\documentclass[aps,prc,onecolumn,floatfix,superscriptaddress]{revtex4-2}
\usepackage{rotating}

\usepackage[pdfencoding=auto,psdextra]{hyperref}
\usepackage{url}

\usepackage{todonotes}

\usepackage{amsmath}
\usepackage{amssymb}
\usepackage{graphicx}

\usepackage{geometry}

\usepackage{cancel}

\usepackage{makecell}

\begin{document}

\def\nuc#1#2{\relax\ifmmode{}^{#1}{\protect\text{#2}}\else${}^{#1}$#2\fi}
\title{Multipole decomposition of tensor interactions of fermionic probes with composite particles and BSM signatures in nuclear reactions}

\author{Ayala Glick-Magid}
\affiliation{Racah Institute of Physics, The Hebrew University, The Edmond J. Safra Campus, Givat Ram, Jerusalem 9190401, Israel}
\author{Doron Gazit}
\email[E-mail:~]{doron.gazit@mail.huji.ac.il}
\affiliation{Racah Institute of Physics, The Hebrew University, The Edmond J. Safra Campus, Givat Ram, Jerusalem 9190401, Israel}

\begin{abstract}
A multipole decomposition of a cross-section is a useful tool to simplify the analysis of reactions due to their symmetry properties. By using a new approach to decompose antisymmetric tensor-type interactions within the multipole analysis, we introduce a general mathematical formalism for working with tensor couplings. This allows us to present a general tensor nuclear response, which is particularly useful for ongoing $\beta$-decay experiments looking for physics beyond the Standard Model, as well as other exotic particle scatterings off nuclei, e.g., in dark matter direct detection experiments.
Using this method, beyond the Standard Model operators identify with the known Standard Model operators, eliminating the need for calculations of additional matrix elements. We present in detail BSM expressions useful for $\beta$-decay experiments and give an exemplary application for $^6$He $\beta$-decay, although the formalism is easily generalizable for calculating other exotic scattering reactions.
\end{abstract}

\maketitle

\section{Introduction}

Tensor interactions have been investigated over the years, with a focus on gravitational radiation~\cite{RevModPhys.52.299}, which introduces a coupling between symmetric tensors – a space-time-metric and the stress-energy-momentum tensor. 
Recently, there has been a renewed interest in the tensor coupling, this time in the search for beyond the Standard Model (BSM) interactions, involving interactions with fermions, and therefore introducing antisymmetric tensors.

A priori, when discussing the weak nuclear interaction of quarks and leptons, the most general Lorentz-invariant form of an interaction Hamiltonian can be written as a linear combination of the five bilinear covariants with specific symmetries, i.e., scalar (S), pseudoscalar (P), polar-vector (V), axial-vector (A) and tensor (T)~\cite{PhysRev.82.531}. However, it was shown experimentally, initially using $\beta$-decays, that the weak interaction between quarks and leptons has a $V-A$ structure, i.e., a polar-vector current and an axial-vector current, with the same amplitude and opposite signs~\cite{1742-6596-196-1-012002}.

In recent years, several experiments~\cite{RevModPhys.78.991,doi:10.1146/annurev-nucl-102010-130410,ANDP:ANDP201300072,1402-4896-2013-T152-014018,RevModPhys.87.1483,gonzalez2019new,Ohayon2018} have focused on $\beta$-decays again, but this time to find deviations from the 
$V-A$ structure of the Standard Model (SM). In particular, these experiments search for minute signatures of interactions with scalar and tensor symmetries. To identify such effects, it is necessary to determine what are the theoretical properties of transitions which have these symmetries.

The theoretical interest in understanding the qualitative behavior of transitions of esoteric character stems additionally from ongoing efforts to directly detect dark matter~\cite{Hoferichter:2015ipa}.
The existence of this material is currently inferred indirectly, as it provides an explanation for certain cosmological gravitational phenomena. Elucidating the nature of dark matter is one of the most pressing challenges in contemporary particle physics and astrophysics.

Among the candidates for dark matter are weakly-interacting massive particles (WIMPs), such as the neutralino in supersymmetric extensions of the Standard Model. This paradigm has spurred the development of detectors on earth, searching for direct interactions of WIMPs from outer space, by measuring the recoil energy of WIMP scattering off nuclei on the detectors. The relevant momentum transfer in such reaction is $q\sim100\mbox{ MeV}$~\cite{Menendez:2012tm}, compared to the typical momentum transfer of $\beta$-decays, just a few $\mbox{MeV}$ ($1-4$). These beyond the Standard Model particles might have many different kinds of couplings to matter, so the overall expression, including the tensor term, will be necessary to interpret the data from these detectors.

In the low energy regime of the weak interaction, one can assume the force-carrying exchange-boson is heavy compared to the momentum transfer. This is particularly a reasonable assumption in $\beta$-decay where the momentum transfer is usually around a few MeV. The weak interaction Hamiltonian between nuclei and light particles is then presumed to be a multiplication of a nuclear current and a probe current of the same kind.
Focusing on the tensor type, the interaction Hamiltonian is expressed in the Schr\"odinger picture as:
\begin{eqnarray}
\label{eq:tensor weak Hamiltonian}
\hat{H}_{\mbox{w}}^{T} & = & \int d^{3}r\hat{j}_{\mu\nu}\left(\vec{r}\right)\hat{\mathcal{J}}^{\mu\nu}\left(\vec{r}\right)\text{,}
\end{eqnarray}
whereas $\hat{\mathcal{J}}^{\mu\nu}\left(\vec{r}\right)$ corresponds to the tensor hadron current, and $\hat{j}_{\mu\nu}\left(\vec{r}\right)$ to the tensor probe current.

As opposed to the vector and axial weak interactions, which have been extensively studied within the Standard Model, and to the scalar and pseudoscalar symmetries, which also have their own formalisms, both for the exotic weak interactions~\cite{Menendez:2012tm,Klos:2013rwa,PhysRevD.94.063505}, and for dark matter~\cite{1475-7516-2013-02-004,PhysRevC.89.065501}, a complete study of cross sections of nuclei with tensor interactions has not been performed.

Here we develop a method of decomposing the tensor coupling within the multipole expansion. In the method we present we do not restrict ourselves to the weak interaction between hadron and lepton currents, but only require a tensor coupling between antisymmetric tensors (i.e., consist of fermions). This work can be viewed as a complimentary to previous works regarding symmetric tensor couplings for the case of gravitational radiation~\cite{RevModPhys.52.299}.
We then use this method to present a general mathematical formalism for the tensor type of interactions with nuclei, applicable to semileptonic interactions like $\beta$-decays.

We note that, particularly in the early days of $\beta$-decay research, there have been several studies~\cite{PhysRev.82.531,PhysRev.104.254,PhysRev.106.517} that aimed to calculate the antisymmetric tensor coupling, during the mission to discover the symmetry nature of the weak nuclear current~\cite{1742-6596-196-1-012002,RevModPhys.78.991}.
These have focused on Fermi and Gamow-Teller decays, and had explicit low momentum transfer approximations.
There is, however, no general formula for non-vanishing momentum transfer of the tensor coupling, depending on the momentum transition. Additionally, there is no general term for non-vanishing momentum transfer of the interference term between the $V-A$ SM symmetry and the BSM tensor symmetry, known as the Fierz term.

The paper is built as follows.
In Sec.~\ref{sec:Tensor-multipole-decomposition}, we present an approach for decomposing a generic coupling of antisymmetric tensor currents within the multipole expansion. This decomposition is suitable for any antisymmetric tensor probe. In the current work we concentrate on the interaction of a tensor probe with a nucleus. In Sec.~\ref{sec:BSM-multipole-operators} we focus on the tensor nuclear single-nucleon current, construct it through our decomposition, and derive from it tensor multipole operators, along with other BSM multipole operators suitable for any semi-leptonic process (a derivation of scalar and pseudoscaler multipole operators is detailed in Appendix~\ref{sec:Appendix Scalar-Completeness}).
Then, in Sec.~\ref{sec:A-General-Expression}, we focus the discussion, present the $\beta$-decay formalism, and write general rate expressions for allowed (Fermi and Gamow-Teller) and forbidden transitions, reviewing how BSM signatures appear in $\beta$-decay observables relevant to contemporary experiments. In Sec.~\ref{sec:Examples} we give an exemplary application for $^6$He $\beta$-decay, of current experimental interest.
We summarize our findings and provide an outlook for future research in Sec.~\ref{sec:Summary}.

\section{Tensor multipole decomposition\label{sec:Tensor-multipole-decomposition}}

Consider a general tensor density of a composite object, e.g., a nucleus, $\hat{\mathcal{J}}^{\mu\nu}$ with a CPT invariant (Lorentz invariant) probe $\hat{j}_{\mu\nu}$, taking the form $\int d^{3}r\hat{j}_{\mu\nu}\left(\vec{r}\right)\hat{\mathcal{J}}^{\mu\nu}\left(\vec{r}\right)$. Assuming the probe has a plane wave character (otherwise one should expand it in plane waves, similarly to what is done in the case of a muon capture from an atomic orbital~\cite{WALECKA1975113}), its general matrix element between its initial and final states can be written as:
\begin{eqnarray}
\left\langle f\left|\hat{j}_{\mu\nu}\left(\vec{r}\right)\right|i\right\rangle  & \equiv & l_{\mu\nu}e^{-i\vec{q}\cdot\vec{r}}\mbox{,}
\end{eqnarray}
where $\vec{q} \equiv \vec{k}_f - \vec{k}_i$ is the momentum transfer between the final and initial probe states, and $l_{\mu\nu}$ depends on all the other physical properties of the probe (a detailed $l_{\mu\nu}$ for a lepton current can be found in Appendix~\ref{sec:Appendix Tensor-Lepton-Traces}).

Typically, the multipole expansion is expressed as a sum of spherical harmonics. For the polar-vector and axial-vector weak interactions in the SM, the traditional way to perform the multipole expansion involves using vector spherical harmonics~\cite{WALECKA1975113}, which are an extension of scalar spherical harmonics. For a multipole expansion of a tensor coupling, we naturally turn to the notion of tensor spherical harmonics. 
The tensor spherical harmonics have been constructed and used in several works on general relativity problems~\cite{RevModPhys.52.299}. Although they were defined in that field only for symmetric representations of ranks 0 and 2 (antisymmetric representations of rank 1 are of no relevance to gravitational wave theory), their completeness for rank 1 stems easily.

However, since rank 1 tensors are actually vectors, we suggest, instead, to simplify the tensor decomposition, taking advantage of its vector nature.
For that, we suggest dismantling the antisymmetric tensors into vector-like objects as follows: first, we decompose $l_{\mu\nu}$ into a temporal scalar $l_{00}$, two mixed spatial-temporal 3-vectors $l_{0i}$ and $l_{i0}$, and an Euclidean (spatial-only) $3\times3$ tensor $l_{ij}$ where $i,j\in\left\{ 1,2,3\right\}$.
Following its antisymmetric nature, we get that $l_{00}=0$ and $l_{i0}=-l_{0i}$. For convenience, we will define a vector $\vec{l}^{T'}$ such that 
\begin{subequations}
\label{eq: tensor current defenitions}
\begin{eqnarray}
l_{i}^{T'} & \equiv & \sqrt{2}l_{0i}.
\end{eqnarray}

Let us now focus on the remaining tensor, $l_{ij}$. It is a Cartesian tensor of the second rank, and therefore can be decomposed into three irreducible spherical tensors of ranks 0, 1 and 2. These will be a scalar, which is the trace of the Cartesian tensor, a vector, which is the antisymmetric part of the Cartesian tensor, and a quadrupole spherical trace-free tensor, which is the remaining symmetric part of the Cartesian tensor.
Using again the fact that $l_{\mu\nu}$ is antisymmetric, it follows that the symmetric scalar and quadrupole spherical tensors vanish, leaving us only with the reduced spherical tensor of rank 1, the spherical vector projector $\vec{l}^{T}\equiv\left[l_{ij}\right]^{\left(1\right)}$. This is a vector that its Cartesian components $i\in\left\{ 1,2,3\right\}$ are defined by
\begin{eqnarray}
l_{i}^{T} & \equiv & -\frac{i}{\sqrt{2}}\epsilon_{ijk}l_{jk}\text{,}
\label{eq: Euclidean components of the reduced spherical vector}
\end{eqnarray}
where $\epsilon_{ijk}$ is the 3-d Levi-Civita symbol (which is $1$ if $\left(i,j,k\right)$ is an even permutation of $\left(1,2,3\right)$, $-1$ if it is an odd permutation, and $0$ if any index is repeated).

The same procedure is done for $\hat{\mathcal{J}}^{\mu\nu}$, which is also antisymmetric, with the definitions of its spatial and spatial-temporal parts as was done to $l_{\mu \nu}$:
\begin{eqnarray}
\left[\vec{\mathcal{J}}^{T}\right]_{i} & \equiv & -\frac{i}{\sqrt{2}}\epsilon_{ijk}\mathcal{\hat{J}}_{jk},\label{eq: tensor current defenition T} \\
\left[\vec{\mathcal{J}}^{T'}\right]_{i} & \equiv & \sqrt{2}\hat{\mathcal{J}}_{0i}\text{.}
\end{eqnarray}
\end{subequations}
We finally conclude the tensor decomposition into vector-like objects, and get to write the tensor product $l_{\mu\nu}\hat{\mathcal{J}}^{\mu\nu}$
as a sum of vectors products, a product of the spatial vector-like parts of the original tensors, and a product of the spatial-temporal vector-like parts of the original tensors:
\begin{eqnarray}
\label{eq:vectors products}
l_{\mu\nu}\hat{\mathcal{J}}^{\mu\nu}\left(\vec{r}\right) & = & -\left[\vec{l}^{T}\cdot\vec{\mathcal{J}}^{T}\left(\vec{r}\right)+\vec{l}^{T'}\cdot\vec{\mathcal{J}}^{T'}\left(\vec{r}\right)\right]\mbox{.}
\end{eqnarray}
While the minus sign of $\vec{l}^{T'}\cdot\vec{\mathcal{J}}^{T'}$ comes from the metric, since it has only one spatial index, the minus sign before $\vec{l}^{T}\cdot\vec{\mathcal{J}}^{T}$ comes from the definitions in Eqs.~\eqref{eq: Euclidean components of the reduced spherical vector} and~\eqref{eq: tensor current defenition T}.

Having gained this vector-like decomposition, all that remains is to carry out the usual vector multipole analysis. For this, we write $l^{T^{\left('\right)}}$ using the circular polarization base unit vector, defined as:
\begin{subequations}
\begin{eqnarray}
\hat{e}_{\pm1} & 
\equiv & \mp\frac{1}{\sqrt{2}}\left(\hat{x}\pm i\hat{y}\right),\\
\hat{e}_0 &
\equiv & \hat{z}\equiv\hat{q},
\end{eqnarray}
\end{subequations}
where we chose the $\hat{z}$ axis to be the direction of the momentum transfer $\hat{q}$.
Now, any vector can be expanded in this set,
$\vec{l} = \sum_{\lambda=-1}^1 l_\lambda \hat{e}^+_\lambda$,
so we can write Eq.~\eqref{eq:vectors products} as
\begin{eqnarray}
l_{\mu\nu}\hat{\mathcal{J}}^{\mu\nu}\left(\vec{r}\right) & = & -\sum_{\lambda=-1}^{1}\left[l_{\lambda}^{T}\hat{e}_{\lambda}^{+}\cdot\vec{\mathcal{J}}^{T}\left(\vec{r}\right)+l_{\lambda}^{T'}\hat{e}_{\lambda}^{+}\cdot\vec{\mathcal{J}}^{T'}\left(\vec{r}\right)\right]\mbox{.}
\end{eqnarray}
Finally, using the identity~\cite{Edmonds:1974:AMQM},
\begin{eqnarray}
\hat{e}_{\lambda}^{+}e^{-i\vec{q}\cdot\vec{r}} & = & \begin{cases}
\frac{i}{q}\sum_{J=0}^{\infty}\sqrt{4\pi\left(2J+1\right)}\left(-i\right)^{J}\vec{\nabla}\left[j_{J}\left(qx\right)Y_{J0}\left(\hat{x}\right)\right] & \lambda=0\\
-\sum_{J=1}^{\infty}\sqrt{2\pi\left(2J+1\right)}\left(-i\right)^{J}\cdot\\
\phantom{----}\cdot\left\{ \lambda j_{J}\left(qx\right)\vec{Y}_{JJ1}^{-\lambda}+\frac{1}{q}\vec{\nabla}\times\left[j_{J}\left(qx\right)\vec{Y}_{JJ1}^{-\lambda}\left(\hat{x}\right)\right]\right\}  & \lambda\in\left\{\pm1\right\}
\end{cases},\label{eq:identity}
\end{eqnarray}
with $j_{J}$ the spherical Bessel functions, $Y_{JM}$ the spherical harmonics, and $\vec{Y}_{Jl1}^{M}$ the vector spherical harmonics defined by the relation~\cite{Edmonds:1974:AMQM}
$\vec{Y}_{Jl1}^{M}\left(\hat{r}\right)\equiv\sum_{\mu=-l}^{l}\sum_{\lambda=-1}^{1}\left\langle l\mu1\lambda|JM\right\rangle Y_{l\mu}\left(\hat{r}\right)\hat{e}_{\lambda}$,
one gets the multipole expansion of the tensor interaction:
\begin{multline}
\left\langle f\left|\int d^{3}r\hat{j}_{\mu\nu}\left(\vec{r}\right)\hat{\mathcal{J}}^{\mu\nu}\left(\vec{r}\right)\right|i\right\rangle =-\sum_{J=0}^{\infty}\sqrt{4\pi\left(2J+1\right)}\left(-i\right)^{J}\left[l_{3}^{T}\left\langle f\left|\hat{L}_{J0}^{T}\right|i\right\rangle +l_{3}^{T'}\left\langle f\left|\hat{L}_{J0}^{T'}\right|i\right\rangle \right]\\
\left.+\sum_{J=1}^{\infty}\sqrt{2\pi\left(2J+1\right)}\left(-i\right)^{J}\sum_{\lambda=\pm1}\left[l_{\lambda}^{T}\left\langle f\left|\hat{E}_{J,-\lambda}^{T}+\lambda\hat{M}_{J,-\lambda}^{T}\right|i\right\rangle +l_{\lambda}^{T'}\left\langle f\left|\hat{E}_{J,-\lambda}^{T'}+\lambda\hat{M}_{J,-\lambda}^{T'}\right|i\right\rangle \right]\vphantom{\sum_{J=0}^{\infty}}\right\}
\label{eq: multipole expansion}
\end{multline}
(for $\hat{\mathcal{J}}^{\mu\nu}$ an hadron current and $\hat{j}_{\mu\nu}$ a lepton current, this is the matrix element of the tensor part of the weak interaction Hamiltonian described in Eq.~\eqref{eq:tensor weak Hamiltonian}, i.e.,
$\left\langle f\left|\hat{H}_{\mbox{w}}^{T}\right|i\right\rangle$).
Here the superscript $T$ ($T'$) denotes a multipole operator calculated
with the spatial (spatial-temporal) vector-like part of the original tensor, $\vec{\mathcal{J}}^{T}$($\vec{\mathcal{J}}^{T'}$). The Coulomb, longitudinal,
electric and magnetic multipole operators are defined by:
\begin{subequations}\label{eq: multipole operators}
\begin{eqnarray}
\hat{C}_{JM}\left(q\right) & \equiv & \int d^{3}r M_{JM}\left(q\vec{r}\right) \mathcal{J}_{0}\left(\vec{r}\right), \\
\hat{L}_{JM}\left(q\right) & \equiv & \frac{i}{q}\int d^{3}r \vec{\nabla}M_{JM}\left(q\vec{r}\right) \cdot\vec{\mathcal{J}}\left(\vec{r}\right), \\
\hat{E}_{JM}\left(q\right) & \equiv & \frac{1}{q}\int d^{3}r\left[ \vec{\nabla}\times \vec{M}_{JJ1}^{M}\left(q\vec{r}\right)
\right] \cdot\vec{\mathcal{J}}\left(\vec{r}\right), \\
\hat{M}_{JM}\left(q\right) & \equiv & \int d^{3}r\vec{M}_{JJ1}^{M}\left(q\vec{r}\right)\cdot\vec{\mathcal{J}}\left(\vec{r}\right)\text{,}
\end{eqnarray}
\end{subequations}
where
\begin{subequations}
\label{eq:M_operators}
\begin{eqnarray}
M_{JM}\left(q\vec{r}\right) & \equiv & j_{J}\left(qr\right)Y_{JM}\left(\hat{r}\right), \\
\vec{M}_{JL1}^{M}\left(q\vec{r}\right) & \equiv & j_{L}\left(qr\right)\vec{Y}_{JL1}^{M}\left(\hat{r}\right)\mbox{.}
\end{eqnarray}
\end{subequations}

Unlike the vector multipole expansion (see, e.g.,~\cite{WALECKA1975113}), the tensor multipole expansion presented in Eq.~\eqref{eq: multipole expansion} does not contain the Coulomb multipole operator, $\hat{C}_{JM}$, which depends on the temperal part $\mathcal{J}_0$ (charge) of a 4-vector current $\mathcal{J}_{\mu}$. It perfectly makes sense, since the tensor is antisymmetric, and therefore, its pure temporal part, $\hat{\mathcal{J}}_{00}$, vanishes. As will be presented in the following, Coulomb multipole operator appears in expressions related to the scalar and pseudoscalar interactions (a detailed discussion about the scalar and pseudoscalar symmetries is presented in Appendix~\ref{sec:Appendix Scalar-Completeness}).

\section{BSM nuclear multipole operators~\label{sec:BSM-multipole-operators}}

For the weak interaction, the multipole expansion of the matrix element of the tensor Hamiltonian (Eq.~\eqref{eq:tensor weak Hamiltonian}), described in Eq.~\eqref{eq: multipole expansion}, depends on the multipole operators (Eq.~\eqref{eq: multipole operators}) calculated with the density of the tensor nuclear current.
In the traditional nuclear physics picture, the nuclear current is constructed from the properties of free nucleons.
In the case of experimental searches, BSM signatures are most likely to be $10^{-3}$ at most~\cite{glick2021formalism}. Thus, we will ignore two-body (and above) currents, leading to a systematic additional uncertainty of $\epsilon_{\text{EFT}}\sim0.3$ in the nuclear model~\cite{cirgiliano2019precision}. 
For dark matter searches, where the couplings to tensor sources need not be smaller than other couplings, the experiments aim at a discovery rather than measuring to high precision a specific coupling. Thus, lower accuracy is needed from the nuclear calculations, a fact that allows neglecting two-body tensor currents at least in the initial stage.
Moreover, chiral perturbation theory with tensor sources suggests that two-body tensor currents are expected at higher order~\cite{Cat__2007}. 

The general form of a single-nucleon matrix element of the tensor part of the charge changing weak current can be written as~\cite{Cirigliano:2013xha}:
\begin{multline}
\label{eq:tensor current single-nucleon matrix element}
\left\langle \vec{p}',\sigma',\rho'\left|\hat{\mathcal{J}}_{\mu\nu}\right|\vec{p},\sigma,\rho\right\rangle =\frac{1}{\Omega}\bar{u}\left(\vec{p}',\sigma'\right)\eta_{\rho'}^{+}\frac{1}{2}\left[g_{T}\left(q^{2}\right)\sigma_{\mu\nu}+g_{T}^{\left(1\right)}\left(q^{2}\right)\left(q_{\mu}\gamma_{\nu}-q_{\nu}\gamma_{\mu}\right)\right.\\
\left.+g_{T}^{\left(2\right)}\left(q^{2}\right)\left(q_{\mu}P_{\nu}-q_{\nu}P_{\mu}\right)+g_{T}^{\left(3\right)}\left(q^{2}\right)\left(\gamma_{\mu}\cancel{q}\gamma_{\nu}-\gamma_{\nu}\cancel{q}\gamma_{\mu}\right)\right]\tau^{\pm}\eta_{\rho}u\left(\vec{p},\sigma\right)\mbox{,}
\end{multline}
with the nowadays conventions, where $\gamma^5\equiv i\gamma^0\gamma^1\gamma^2\gamma^3$ and $\sigma_{\mu\nu}\equiv\frac{i}{2}\left[\gamma_{\mu},\gamma_{\nu}\right]$ the commutator of Dirac gamma matrices.
$u\left(\vec{p},\sigma\right)=\sqrt{\frac{E_{N}+m_{N}}{2E_{N}}}\left(\begin{array}{c}
1\\
\frac{\vec{\sigma}\cdot\vec{p}}{E_{N}+m_{N}}
\end{array}\right)\chi_{\sigma}$
is a Dirac spinor for a free nucleon of mass $m_{N}$ and momentum $p_{\mu}$, $E_{N}=\sqrt{p^{2}+m_{N}^{2}}$ is the energy of the particle, $\chi_{\sigma}$ is a two-component Pauli spinor for a spin up and down along the $\hat{q}$ axis, $\eta_{\rho}$ is a two-component Pauli isospinor, and $\tau^{\pm}\equiv\mp\frac{1}{2}\left(\tau_{x}\pm i\tau_{y}\right)$ are the isospin raising and lowering operators that change a proton into a neutron and vice versa. 
Here, $P_{\mu}\equiv p_{\mu}+p_{\mu}^{'}$, and $q_{\mu}\equiv p_{\mu}-p_{\mu}^{'}$ is the momentum transfer, as before.
$\Omega$ is a normalization volume (we impose periodic boundary conditions on the large volume $\Omega$ and check that its dependence drops subsequently).

Lattice QCD suggests that the tensor nuclear charge $g_{T}$ has a similar magnitude to the SM axial-vector nuclear charge $g_{A}$~\cite{Bhattacharya2016}.
The other tensor form factors $g_{T}^{\left(i\right)}\left(q^{2}\right)$ ($i\in\left\{ 1,2,3\right\}$) are smaller.
In the nomenclature of~\cite{glick2021formalism} that we will use in the following, they are of the order of $\epsilon_{\text{recoil}}\sim\frac{q}{m_{N}}$ ($\approx0.002$ for an endpoint of $\approx2\text{MeV}$)~\cite{Cirigliano:2013xha}.
In addition, $g_{T}^{\left(3\right)}$ is a second class current
and therefore vanishes in the isospin (SU$\left(2\right)_f$) limit~\cite{PhysRev.112.1375}.
Although $g_{T}\sim g_{A}$, the tensor expression is suppressed by a coefficient of the effective theory, $\epsilon_{T}\propto\left(\frac{m_{W}}{\Lambda}\right)^{n}$, which comes from the effective weak interaction Lagrangian, where $m_{W}$ is the mass of the $W$ boson,
$\Lambda$ represents the new physics scale, and $n\geq2$.
For the simplest BSM operator ($n=2$), a TeV scale means $\epsilon_T\sim10^{-3}$. New experiments, looking for BSM signatures, will have this $10^{-3}$ level of precision, making them sensitive to new physics at the TeV scale.

To obtain the vector-like tensor multipole operators used in Eq.~\eqref{eq: multipole expansion}, we extract the tensor current density from Eq.~\eqref{eq:tensor current single-nucleon matrix element}, and separate it into its spatial and spatial-temporal vector-like parts, respectively:
\begin{subequations}
\begin{align}
\vec{\mathcal{J}}^{T}\left(\vec{r}\right) & =-\frac{i}{\sqrt{2}}\sum_{j=1}^{A}\left(g_{T}+2iE_0 g_{T}^{\left(3\right)}\right)\vec{\sigma}_{j}\delta^{\left(3\right)}\left(\vec{r}-\vec{r}_{j}\right)\tau_{j}^{\pm}+\mathcal{O}\left(\epsilon_{\text{NR}}^{2}\right), \label{eq:spatial density}\\
\vec{\mathcal{J}}^{T'}\left(\vec{r}\right) & =\frac{1}{\sqrt{2}}\sum_{j=1}^{A}\left\{ \left(ig_{T}^{\left(1\right)}-\frac{g_{T}}{2m_{N}}\right)\vec{\nabla}\delta^{\left(3\right)}\left(\vec{r}-\vec{r}_{j}\right)-\frac{g_{T}}{2m_{N}}\vec{\sigma}_{j}\times\left\{ \vec{p}_{j},\delta^{\left(3\right)}\left(\vec{r}-\vec{r}_{j}\right)\right\} \vphantom{\frac{1}{2_{N}}}\right.\nonumber \\
 & \left.+\left(2g_{T}^{\left(3\right)}-\frac{E_0}{2m_{N}}g_{T}^{\left(1\right)}\right)\vec{\sigma}_{j}\times\vec{\nabla}\delta^{\left(3\right)}\left(\vec{r}-\vec{r}_{j}\right)\right\} \tau_{j}^{\pm}+\mathcal{O}\left(\epsilon_{\text{NR}}^{2}\right)\text{,}
 \label{eq:spatial-temperal density}
\end{align}
\end{subequations}
where $A$ is the mass number of the nucleus, \(\vec{r}_j\) \((\tau^{+}_j)\) is the \(j\)th nucleon position vector (isospin-raising operator), \(\vec{\sigma}_j\) is the Pauli spin matrices vector associated with nucleon \(j\), and $E_0=q_0$ is the energy transfer.
Here, we used the non-relativistic expansion to expand the currents in powers of $\epsilon_{\text{NR}}\sim\frac{P_{\text{fermi}}}{m_{N}}\approx0.2$ (the calculations are detailed in appendix~\ref{sec:Appendix Tensor-Nuclear-Current}).

In the nuclear tensor current densities we obtained, one can see that the spatial-temperal current terms (Eq.~\eqref{eq:spatial-temperal density}) are suppressed by $\epsilon_{\text{NR}}$ or $\epsilon_{\text{recoil}}$. These suppressions are on top of the small tensor effective theory coefficient, so the spatial-temperal current does not appear in the BSM leading order. That leaves us with the spatial vector-like tensor current. A closer look reveals  that its leading order is the same as the leading order of the 3-vector spatial component of the SM axial-vector current density, i.e.,
\begin{align}
\label{eq:tensor_current_density}
\vec{\mathcal{J}}^{T}\left(\vec{r}\right) &
=-\frac{i}{\sqrt{2}}\frac{g_{T}}{g_{A}}\vec{\mathcal{J}}^{A}\left(\vec{r}\right)+\mathcal{O}\left(\epsilon_{\text{NR}}^{2}, \epsilon_{\text{recoil}}\right)
\end{align}
(A more accurate form will include second class currents:
$\vec{\mathcal{J}}^{T}\left(\vec{r}\right)
=-\frac{i}{\sqrt{2}}\frac{g_{T}+2iE_0 g_{T}^{\left(3\right)}}{g_{A}-\frac{E_0}{2m_{N}}\tilde{g}_{T\left(A\right)}}\vec{\mathcal{J}}^{A}\left(\vec{r}\right)+\mathcal{O}\left(\epsilon_{\text{NR}}^{2}\right)$,
where $g_{T}^{\left(3\right)}$ and
$\tilde{g}_{T\left(A\right)}$, both second class
currents form factors, are themselves $\mathcal{O}\left(\epsilon_{\text{recoil}}\right)$.
For more detail, see Appendix~\ref{sec:Appendix-Second-Class_Nuclear-Currents-and}).
With these current densities, the multipole operators from Eq.~\eqref{eq: multipole operators} can be written as a sum of one-body operators.
Eq.~\eqref{eq:tensor_current_density} clearly shows that the spatial vector-like tensor multipole operators are proportional to the spatial axial-vector multipole operators:
\begin{align}
\hat{O}_{J}^{T}\left(q\right) & 
\approx -\frac{i}{\sqrt{2}}\frac{g_{T}}{g_{A}}\hat{O}_{J}^{A}\left(q\right), &
\hat{O} \in\left\{ \hat{L},\hat{E},\hat{M}\right\}
\label{eq:T-A multipole operators}
\end{align}
(a more accurate form will include second class currents: 
$
\hat{O}_{J}^{T}\left(q\right) =-\frac{i}{\sqrt{2}}\frac{g_{T}+2iE_0 g_{T}^{\left(3\right)}}{g_{A}-\frac{E_0}{2m_{N}}\tilde{g}_{T\left(A\right)}}\hat{O}_{J}^{A}\left(q\right)+\mathcal{O}\left(\epsilon_{\text{NR}}^{2}\right)=
-\frac{i}{\sqrt{2}}\frac{g_{T}}{g_{A}}\left[1+E_0\left(2i\frac{g_{T}^{\left(3\right)}}{g_{T}}+\frac{g_{A}}{g_{T}}\frac{\tilde{g}_{T\left(A\right)}}{2m_{N}}\right)\right]\hat{O}_{J}^{A}\left(q\right)
+\mathcal{O}\left(\epsilon_{\text{NR}}^{2}\right)
$
).
This is a significant result that greatly simplifies the work with the tensor, allowing calculations of BSM tensor interaction using only the well known SM axial-vector multipole operators:
\begin{subequations}
\begin{align}
\hat{L}_{J}^{A}\left(q\right) & =ig_{A}\sum_{j=1}^{A}\tau_{j}^{\pm}\left[\frac{1}{q}\vec{\nabla}M_{J}\left(q\vec{r}_{j}\right)\right]\cdot\vec{\sigma}+\mathcal{O}\left(\epsilon_{qr}^{J-1}\epsilon_{\text{NR}}^{2}\right), \\
\hat{E}_{J}^{A}\left(q\right) & =g_{A}\sum_{j=1}^{A}\tau_{j}^{\pm}\left[\frac{1}{q}\vec{\nabla}\times\vec{M}_{JJ1}\left(q\vec{r}_{j}\right)\right]\cdot\vec{\sigma}+\mathcal{O}\left(\epsilon_{qr}^{J-1}\epsilon_{\text{NR}}^{2}\right), \\
\hat{M}_{J}^{A}\left(q\right) & =g_{A}\sum_{j=1}^{A}\tau_{j}^{\pm}\vec{M}_{JJ1}\left(q\vec{r}_{j}\right)\cdot\vec{\sigma}+\mathcal{O}\left(\epsilon_{qr}^{J}\epsilon_{\text{NR}}^{2}\right),
\end{align}
\end{subequations}
with $\epsilon_{qr}\sim qR$ ($\approx0.01A^{\frac{1}{3}}$ for an endpoint of $\approx2\text{MeV}$. $R$ is the radius of the nucleus).
Eq.~\eqref{eq:T-A multipole operators} here is accurate to  $\mathcal{O}\left(\epsilon_{qr}^{J}\epsilon_{\text{NR}}^{2}\right)$
for $\hat{M}_{J}$, and to $\mathcal{O}\left(\epsilon_{qr}^{J-1}\epsilon_{\text{NR}}^{2}\right)$
for $\hat{E}_{J}$ and $\hat{L}_{J}$ (when $J>0$. For $\hat{L}_{0}$, it is $\mathcal{O}\left(\epsilon_{qr}\epsilon_{\text{NR}}^{2}\right)$).

The Vector-like spatial-temperal tensor current introduces new multipole operators:
\begin{subequations}
\begin{align}
\hat{L}_{J}^{T'}\left(q\right) & =-\frac{1}{\sqrt{2}}\frac{q}{m_{N}}\sum_{j=1}^{A}\left\{ \left(2m_{N}g_{T}^{\left(1\right)}+ig_{T}\right)M_{J}\left(q\vec{r}_{j}\right)+g_{T}\left[\left(\frac{1}{q}\vec{\nabla}M_{J}\left(q\vec{r}_{j}\right)\right)\times\vec{\sigma}_{j}\right]\cdot\frac{1}{q}\vec{\nabla}\right\} \tau_{j}^{\pm}\nonumber\\
&+\mathcal{O}\left(\epsilon_{qr}^{J-1}\epsilon_{\text{NR}}^{2}\right), \\
\hat{E}_{J}^{T'}\left(q\right) & =\frac{1}{\sqrt{2}}\frac{q}{m_{N}}\sum_{j=1}^{A}\left\{ ig_{T}\left[\left(\frac{1}{q}\vec{\nabla}\times\vec{M}_{JJ1}\left(q\vec{r}_{j}\right)\right)\times\vec{\sigma}_{j}\right]\cdot\frac{1}{q}\vec{\nabla}\right.\nonumber \\
 & \left.+\left(\frac{i}{2}g_{T}+\frac{E_0}{2}g_{T}^{\left(1\right)}-2m_{N}g_{T}^{\left(3\right)}\right)\vec{\sigma}_{j}\cdot\vec{M}_{JJ1}\left(q\vec{r}_{j}\right)\right\} \tau_{j}^{\pm}+\mathcal{O}\left(\epsilon_{qr}^{J-1}\epsilon_{\text{NR}}^{2}\right),\\
\hat{M}_{J}^{T'}\left(q\right) & =\frac{1}{\sqrt{2}}\frac{q}{m_{N}}\sum_{j=1}^{A}\left\{ ig_{T}\left[\vec{M}_{JJ1}\left(q\vec{r}_{j}\right)\times\vec{\sigma}_{j}\right]\cdot\frac{1}{q}\vec{\nabla}\right.\nonumber \\
 & \left.+\left(\frac{i}{2}g_{T}+\frac{E_0}{2}g_{T}^{\left(1\right)}-2m_{N}g_{T}^{\left(3\right)}\right)\vec{\sigma}_{j}\cdot\left[\frac{1}{q}\vec{\nabla}\times\vec{M}_{JJ1}\left(q\vec{r}_{j}\right)\right]\right\} \tau_{j}^{\pm}+\mathcal{O}\left(\epsilon_{qr}^{J}\epsilon_{\text{NR}}^{2}\right)\text{,}
\end{align}
\end{subequations}
but, as mentioned above, they do not appear in the BSM leading order. The BSM leading order is controlled only by the multipole operators $\hat{L}_{J}^{T},\hat{E}_{J}^{T}\propto\epsilon_{qr}^{J-1}$ or $\hat{M}_{J}^{T},\hat{C}_{J}^{S}\propto\epsilon_{qr}^{J}$,
depending on the parity of the transition in question.
$\hat{C}_{J}^{S}$ is the Coulomb multipole operator when it is calculated with the BSM scalar nuclear current. 
Similarly to the tensor leading order operators, its form is proportional to a SM multipole operator (for full discussion and derivation of the scalar and pseudoscalar multipole operators, refer to Appendix~\ref{sec:Appendix Scalar-Completeness}):
\begin{eqnarray}
\label{eq:S multipole operators}
\hat{C}_{J}^{S}\left(q\right) & = & \frac{g_{S}}{g_{V}}\hat{C}_{J}^{V}\left(q\right)+\mathcal{O}\left(\epsilon_{qr}^{J}\epsilon_{\text{NR}}^{2}\right),
\end{eqnarray}
where $\hat{C}_{J}^{V}$ is the polar-vector Coulomb multipole operator:
\begin{eqnarray}
\hat{C}_{J}^{V}\left(q\right) &
=g_{V}\sum_{j=1}^{A}M_{J}\left(q\vec{r}_{j}\right)\tau_{j}^{\pm}+\mathcal{O}\left(\epsilon_{qr}^{J}\epsilon_{\text{NR}}^{2}\right).
\end{eqnarray}
Here $g_{V}$ is the vector nuclear charge form factor, which, due to the conservation of the vector current, is $1$ up to second order corrections in isospin breaking~\cite{Ademollo1964,DONOGHUE1990243}. The scalar nuclear charge $g_{S}=g_{V}\frac{M_{n}-M_{p}}{m_{d}-m_{u}}\approx0.8-1.2$, where $M_n$ ($M_p$) is the mass of the neutron (proton) and $m_d$ ($m_u$) is the mass of the down (up) quark.
Since this is a scalar, other multipole operators, associated with the vector type of the current, do not exist.

In order to complete the picture, let us introduce the last BSM multipole operator - the Coulomb multipole operator calculated with the pseudoscalar nuclear current. As with the vector-like spatial-temporal tensor operators, the pseudoscalar multipole operator is suppressed by an additional small parameter, $\epsilon_{\text{recoil}}$:
\begin{eqnarray}
\hat{C}_{J}^{P}\left(q\right) & = & \frac{iq}{2m_{N}}g_{P}\sum_{j=1}^{A}\left[\frac{1}{q}\vec{\nabla}M_{J}\left(q\vec{r}_{j}\right)\right]\cdot\vec{\sigma}_{j}\tau_{j}^{\pm}+\mathcal{O}\left(\epsilon_{qr}^{J}\epsilon_{\text{NR}}^{2}\right),
\end{eqnarray}
with the pseudoscalar charge $g_{P}=g_{V}\frac{M_{n}+M_{p}}{m_{d}+m_{u}}=349(9)$~\cite{Cirigliano:2013xha}.

In summary, we found that for their leading orders, the BSM multipole operators identify with the well-known SM multipole operators. In this way, BSM contributions can be calculated only from the SM phenomena, without calculating new matrix elements for BSM.
For this discussion to be complete, we must make note of another aspect of the BSM signatures in nuclear currents, the second-class currents. These currents do not add any new multipole operators, but correct the existing SM polar-vector and axial-vector operators with some small contributions. The derivation of those corrections can be found in Appendix~\ref{sec:Appendix-Second-Class_Nuclear-Currents-and}.

\section{$\beta$-decay BSM contributions\label{sec:A-General-Expression}}
%
Here we introduce explicitly the use of the tensor decomposition for the experimentally important case of $\beta$-decays. Nuclear beta minus (plus) decay is a weak reaction in which an atomic nucleus transforms into another by changing one of its nuclear neutrons (protons) into a proton (neutron), increasing (decreasing) its charge by one, and emitting an electron (positron) and an antineutrino (neutrino). 

Consider a $\beta^{\mp}$-decay process with $p_{\mu}$ ($p_{\mu}^{'}$) as the initial (final) nucleus momentum, $k_{\mu}$ ($\nu_{\mu}$) as the electron (neutrino) momentum, and $q_{\mu}\equiv p_{\mu}-p_{\mu}^{'}=k_{\mu}+\nu_{\mu}$ as the momentum transfer. The decay rate, which follows from the Golden rule of Fermi, is~\cite{WALECKA1975113}:
\begin{align}
\frac{d^{5}\omega}{dE\frac{d\hat{k}}{4\pi}\frac{d\hat{\nu}}{4\pi}} & 
=\frac{4}{\pi^{2}}\left(E_0-E\right)^{2}kE F^{\mp}\left(Z_{f},E\right)C_{\text{corr}}\frac{1}{2J_{i}+1}\Theta\left(q,\vec{\beta}\cdot\hat{\nu}\right)\text{,}\label{eq:general-decay-rate}
\end{align}
where the function
\begin{align}
\label{eq:Theta}
\Theta\left(q,\vec{\beta}\cdot\hat{\nu}\right) &
\equiv\frac{1}{4\pi}
\frac{\Omega^{2}}{2}\sum_{\mbox{lepton spins}}\sum_{M_{i}}\sum_{M_{f}}
\left|\left\langle f\left|\hat{H}_{\mbox{w}}\right|i\right\rangle \right|^{2},
\end{align}
is the part depending on the nuclear wave functions, represented here as the initial and final states.
This is also the part that is affected by non $V-A$ currents which may be included in the weak interaction Hamiltonian. 
We sum over final target states (spin projection $M_{f}$), and average over initial states ($M_{i}$). $J_{i}$ is the total angular momentum of the initial nucleus, $E_{i}$ ($E_{f}$) is the initial (final) energy of the nuclear system, while $E\equiv k_{0}$ is the electron energy, and $\nu$ is the energy of the neutrino.

To Eq.~\ref{eq:general-decay-rate}, we have added some known corrections. The deformation of the lepton wave function, due to the long-range electromagnetic interaction with the nucleus, is taken into account in the Fermi function $F^{\mp}\left(Z_{f},E\right)$ for a $\beta^{\mp}$-decay, where $Z_{f}$ is the charge of the nucleus after the decay. Other corrections to the nuclear-independent part, such as radiative corrections, finite mass and electrostatic finite size effects, as well as atomic and chemical effects, are represented by $C_{\text{corr}}$. In the literature, these corrections are assumed to be known and do not seem to limit experimental accuracy significantly (for more details see~\cite{RevModPhys.90.015008,HAYEN2019152}).

Jackson, Treiman and Wyld in their paper from 1957~\cite{Jackson1957}, described the $\beta$-decay rate at its leading order, that is, for allowed transitions (Fermi and Gamow-teller), as proportional to 
\begin{align}
\label{eq:Jackson, Treiman and Wyld}
d^5 \omega \propto \xi \left(1+a\vec{\beta}\cdot\hat{\nu}+b \frac{m_e}{E} + \text{...}\right),
\end{align}
where $a$ is the electron-nutrino angular correlation, and $b$ is the Fierz interference term, both observables are important for ongoing BSM searches. $a$ can be extracted from measurements of the angle between the emitted leptons, and $b$ can be extracted from measurements of the energy spectrum of the electron.
The Fierz interference term $b$ do not exists in the Standard Model leading order, and appears when considering the full probe-nucleus interaction Hamiltonian, $\hat{H}_{\mbox{w}}=\hat{H}_{\mbox{w}}^{SM}+\hat{H}_{\mbox{w}}^{BSM}$, which results in an interference term involving both SM and BSM currents (for the derivation of Fierz interference term involving $V-A$ and tensor currents, see Appendix~\ref{sec:Appendix Fierz-Term}. The scalar and pseudoscalar Fierz terms are described in Appendix~\ref{sec:Appendix Scalar-Completeness})

We would like to extend these observable terms to also include forbidden transitions, that are unavailable in their complete form for tensor BSM symmetry.
For that, we will use the notation we developed in~\cite{glick2021formalism}. As outlined there,
a $J_{i}^{\pi_{i}}\rightarrow J_{f}^{\pi_{f}}$ $\beta$-decay transition with $J_{i}$ ($J_{f}$) and $\pi_{i}$ ($\pi_{f}$) the initial (final) angular momentum and parity, 
will include all integer angular momentum changes that satisfy the selection rules $\left|J_{i}-J_{f}\right|\leq J\leq J_{i}+J_{f}$ and $\Delta\pi=\pi_{i}\cdot\pi_{f}$:
\begin{align}
\Theta\left(q,\vec{\beta}\cdot\hat{\nu}\right) & =\sum_{J=\left|J_{i}-J_{f}\right|}^{J_{i}+J_{f}}\Theta^{J^{\Delta\pi}}\left(q,\vec{\beta}\cdot\hat{\nu}\right)\text{.}\label{eq:teta sum}
\end{align}
In the following, we will present the BSM contributions to each $\Theta^{J^{\Delta\pi}}$, arranged as in~\cite{glick2021formalism}, based on the final equations we present in Appendices~\ref{sec:Appendix Tensor-Lepton-Traces}, ~\ref{sec:Appendix Fierz-Term} and ~\ref{sec:Appendix Scalar-Completeness}.

\subsection{Fermi transition}
%
Having $J=0$, BSM contributions to the Fermi transition ($J^{\Delta\pi}=0^{+}$) come from the scalar multipole operator $\hat{C}_{0}^{S}$, which is proportional to the Fermi operator,
$\hat{C}_{0}^{S}\approx\frac{g_{S}}{g_{V}}\hat{C}_{0}^{V}$:
\begin{multline}
\Theta^{0^{+}}\left(q,\vec{\beta}\cdot\hat{\nu}\right)=
\frac{\left|C_{V}\right|^{2}+\left|C_{V}^{'}\right|^{2}}{2\left|g_{V}\right|^{2}}\left|\left\langle \left\Vert \hat{C}_{0}^{V}\right\Vert \right\rangle \right|^{2}
\left(1+\delta_1^{0^{+}}+\frac{\left|C_{S}\right|^{2}+\left|C_{S}^{'}\right|^{2}}{\left|C_{V}\right|^{2}+\left|C_{V}^{'}\right|^{2}}\right)\\
\times\left[1+\vec{\beta}\cdot\hat{\nu}
\left(1+\tilde{\delta}_a^{0^{+}} -2\frac{\left|C_{S}\right|^{2}+\left|C_{S}^{'}\right|^{2}}{\left|C_{V}\right|^{2}+\left|C_{V}^{'}\right|^{2}}\right)
+\frac{m_{e}}{E}
\left(\delta_{b}^{0^{+}} 
\pm2\mathfrak{Re}\frac{C_{V}C_{S}^{*}+C_{V}^{'}C_{S}^{'*}}{\left|C_{V}\right|^{2}+\left|C_{V}^{'}\right|^{2}}
\right)
\right]
\text{,}
\label{eq:Fermi BSM}
\end{multline}
where $\pm$ are for $\beta^{\mp}$-decays, $\left\langle \left\Vert \hat{O}_{J}\right\Vert \right\rangle$ is a short notation for the reduced matrix element $\left\langle f \left\Vert \hat{O}_{J}\right\Vert i \right\rangle$ of a multipole operator $\hat{O}_{J}$ between the final and initial nuclear states,
and $\delta_1^{0^{+}}$, $\tilde{\delta}_a^{0^{+}}$ and $\delta_{b}^{0^{+}}$ are the SM next-to-leading order (NLO) nuclear structure and recoil corrections discussed in Ref.~\cite{glick2021formalism}.
There are two observables of interest here for the search for BSM signatures. The first is the electron-nutrino angular correlation,
\begin{align}
a^{0^{+}}=1+\tilde{\delta}_a^{0^{+}} -2\frac{\left|C_{S}\right|^{2}+\left|C_{S}^{'}\right|^{2}}{\left|C_{V}\right|^{2}+\left|C_{V}^{'}\right|^{2}}\text{,}
\end{align}
which is $a^{0^{+}}=1$ in the SM leading order.
The second is the Fierz interference term,
\begin{align}
\label{eq: Fierz, 0+}
b^{0^{+}}=\delta_{b}^{0^{+}} 
\pm 2\mathfrak{Re}\frac{C_{V}C_{S}^{*}+C_{V}^{'}C_{S}^{'*}}{\left|C_{V}\right|^{2}+\left|C_{V}^{'}\right|^{2}}
\text{,}
\end{align}
which vanishes in the SM leading order.
These results recover the well-known Jackson, Treiman and Wyld results~\cite{Jackson1957} for allowed leading orders. In their formulation (Eq.~\eqref{eq:Jackson, Treiman and Wyld}),
\begin{subequations}
\begin{align}
    \label{eq:xi0+}
    \xi^{0^{+}} &= 
    \frac{\left|\left\langle \left\Vert \hat{C}_{0}^{V}\right\Vert \right\rangle \right|^{2}}
    {\left|g_{V}\right|^{2}}
    \left[\left(\left|C_{V}\right|^{2}+\left|C_{V}^{'}\right|^{2}\right)\left(1+\delta_1^{0^+}\right)+\left|C_{S}\right|^{2}+\left|C_{S}^{'}\right|^{2}\right],\\
    \label{eq:xi_a0+}
    a^{0^{+}} \xi^{0^{+}} &=
    \frac{\left|\left\langle \left\Vert \hat{C}_{0}^{V}\right\Vert \right\rangle \right|^{2}}
    {\left|g_{V}\right|^{2}}
    \left[\left(\left|C_{V}\right|^{2}+\left|C_{V}^{'}\right|^{2}\right)
    \left(1+\delta_a^{0^{+}}\right)
    -\left|C_{S}\right|^{2}-\left|C_{S}^{'}\right|^{2}\right],\\
    \label{eq:xi_b0+}
    b^{0^{+}} \xi^{0^{+}} &=
    \frac{\left|\left\langle \left\Vert \hat{C}_{0}^{V}\right\Vert \right\rangle \right|^{2}}
    {\left|g_{V}\right|^{2}}
    \left[\left(\left|C_{V}\right|^{2}+\left|C_{V}^{'}\right|^{2}\right)
    \delta_{b}^{0^{+}}
    \pm 2\mathfrak{Re}
    \left(C_{V}C_{S}^{*}+C_{V}^{'}C_{S}^{'*}\right)\right],
\end{align}
\end{subequations}
where $\frac{\left\langle \left\Vert \hat{C}_{0}^{V}\right\Vert \right\rangle}{g_V}=M_F$ is the Fermi matrix element used in their paper,
and the NLO SM corrections $\delta_1^{0^{+}}$, $\delta_a^{0^{+}}=\tilde{\delta}_a^{0^{+}}+\delta_1^{0^{+}}$, and $\delta_{b}^{0^{+}}$, are higher order precision corrections not found in the Jackson, Treiman and Wyld paper.

\subsection{Non-unique first-forbidden transition $J^{\Delta\pi}=0^{-}$}
%
For the non-unique first-forbidden transition $J^{\Delta\pi}=0^{-}$, its nuclear structure expression
includes BSM contributions from the tensor multipole operator
$\hat{L}_{0}^{T}\approx-\frac{i}{\sqrt{2}}\frac{g_{T}}{g_{A}}\hat{L}_{0}^{A}$ as follows:
\begin{multline}
\Theta^{0^{-}}\left(q,\vec{\beta}\cdot\hat{\nu}\right)=
\frac{\left|C_{A}\right|^{2}+\left|C_{A}^{'}\right|^{2}}{2\left|g_{A}\right|^{2}}
\left\{ \left|\left\langle \left\Vert \hat{C}_{0}^{A}\right\Vert \right\rangle \right|^{2}
+\left(1+\frac{\left|C_{T}\right|^{2}+\left|C_{T}^{'}\right|^{2}}{\left|C_{A}\right|^{2}+\left|C_{A}^{'}\right|^{2}}\right)\left|\left\langle \left\Vert \hat{L}_{0}^{A}\right\Vert \right\rangle \right|^{2}\right.\\
-2\mathfrak{Re}\left[\left(\frac{E_0}{q}
\mp \frac{m_e}{q}
\frac{C_{A}^{*}C_{T}+C_{A}^{'*}C_{T}^{'}}{\left|C_{A}\right|^{2}+\left|C_{A}^{'}\right|^{2}}\right)
\left\langle \left\Vert \hat{L}_{0}^{A}\right\Vert \right\rangle \left\langle \left\Vert \hat{C}_{0}^{A}\right\Vert \right\rangle ^{*}\right]
\\
+\vec{\beta}\cdot\hat{\nu}
\left[\left|\left\langle \left\Vert \hat{C}_{0}^{A}\right\Vert \right\rangle \right|^{2}
+\left(1-
\frac{\left|C_{T}\right|^{2}+\left|C_{T}^{'}\right|^{2}}{\left|C_{A}\right|^{2}+\left|C_{A}^{'}\right|^{2}}\right)\left|\left\langle \left\Vert \hat{L}_{0}^{A}\right\Vert \right\rangle \right|^{2}\right.\\
\left.-2\mathfrak{Re}\left[\left(\frac{E_0}{q}
\pm \frac{m_e}{q}
\frac{C_{A}^{*}C_{T}+C_{A}^{'*}C_{T}^{'}}{\left|C_{A}\right|^{2}+\left|C_{A}^{'}\right|^{2}}\right)
\left\langle \left\Vert \hat{L}_{0}^{A}\right\Vert \right\rangle \left\langle \left\Vert \hat{C}_{0}^{A}\right\Vert \right\rangle ^{*}\right]
\vphantom{\frac{\left|C_{T}^{'}\right|^{2}}{\left|C_{A}^{'}\right|^{2}}}\right]\\
+\frac{m_{e}}{E}2\mathfrak{Re}\left[
\left(\frac{m_{e}}{q}
\mp\frac{E_0}{q}\frac{C_{A}^{*}C_{T}+C_{A}^{'*}C_{T}^{'}}{\left|C_{A}\right|^{2}+\left|C_{A}^{'}\right|^{2}}\right)
\left\langle \left\Vert \hat{L}_{0}^{A}\right\Vert \right\rangle 
\left\langle \left\Vert \hat{C}_{0}^{A}\right\Vert \right\rangle ^{*}
\pm \frac{C_{A}^{*}C_{T}+C_{A}^{'*}C_{T}^{'}}{\left|C_{A}\right|^{2}+\left|C_{A}^{'}\right|^{2}}
\left|\left\langle \left\Vert \hat{L}_{0}^{A}\right\Vert \right\rangle \right|^{2}
\right]\\
\left.+2\frac{E\left(E_0-E\right)}{q^{2}}\left[\beta^{2}-\left(\vec{\beta}\cdot\hat{\nu}\right)^{2}\right]
\left(1-\frac{\left|C_{T}\right|^{2}+\left|C_{T}^{'}\right|^{2}}{\left|C_{A}\right|^{2}+\left|C_{A}^{'}\right|^{2}}\right)\left|\left\langle \left\Vert \hat{L}_{0}^{A}\right\Vert \right\rangle \right|^{2}
\right\} +\mathcal{O}\left(\epsilon_M\right)
\end{multline}
($\epsilon_M$ presents SM recoiled nucleus corrections, which we will not give explicitly here, since they are relevant only for very light nuclei. For example, for the $\beta$-decay of $^6$He, $\epsilon_M\sim 7\cdot 10^{-4}$~\cite{glick2021formalism}).
It is possible to recognize BSM tensor signatures $\frac{\left|C_{T}\right|^{2}+\left|C_{T}^{'}\right|^{2}}{\left|C_{A}\right|^{2}+\left|C_{A}^{'}\right|^{2}}$ 
and
$\frac{C_{A}^{*}C_{T}+C_{A}^{'*}C_{T}^{'}}{\left|C_{A}\right|^{2}+\left|C_{A}^{'}\right|^{2}}$, 
in this non-unique first-forbidden transition, similarly to the way they are recognized in allowed transitions.
In the notion of Jackson, Treiman and Wyld, these will be:
\begin{subequations}
\begin{align}
    \xi^{0^{-}} &=
    \frac{1}{\left|g_{A}\right|^{2}}\left\{
    \left(\left|C_{A}\right|^{2}+\left|C_{A}^{'}\right|^{2}\right)
    \left[\left|\left\langle \left\Vert \hat{C}_{0}^{A}\right\Vert \right\rangle \right|^{2}
    -\frac{E_0}{q}2\mathfrak{Re}\left(\left\langle \left\Vert \hat{L}_{0}^{A}\right\Vert \right\rangle \left\langle \left\Vert \hat{C}_{0}^{A}\right\Vert \right\rangle ^{*}\right)
    +\left|\left\langle \left\Vert \hat{L}_{0}^{A}\right\Vert \right\rangle \right|^{2}\right]\right.\nonumber\\
    &+ \left.\left(\left|C_{T}\right|^{2}+\left|C_{T}^{'}\right|^{2}\right)
    \left|\left\langle \left\Vert \hat{L}_{0}^{A}\right\Vert \right\rangle \right|^{2}\right\}\\
    a^{0^{-}} \xi^{0^{-}} &=
    \frac{1}{\left|g_{A}\right|^{2}}\left\{
    \left(\left|C_{A}\right|^{2}+\left|C_{A}^{'}\right|^{2}\right)
    \left[\left|\left\langle \left\Vert \hat{C}_{0}^{A}\right\Vert \right\rangle \right|^{2}
    -\frac{E_0}{q}2\mathfrak{Re}\left(\left\langle \left\Vert \hat{L}_{0}^{A}\right\Vert \right\rangle \left\langle \left\Vert \hat{C}_{0}^{A}\right\Vert \right\rangle ^{*}\right)
    +\left|\left\langle \left\Vert \hat{L}_{0}^{A}\right\Vert \right\rangle \right|^{2}\right]\right.\nonumber\\
    &- \left. \left(\left|C_{T}\right|^{2}+\left|C_{T}^{'}\right|^{2}\right)
    \left|\left\langle \left\Vert \hat{L}_{0}^{A}\right\Vert \right\rangle \right|^{2}\right\}\\
    b^{0^{-}} \xi^{0^{-}} &=
    \frac{1}{\left|g_{A}\right|^{2}}
    2\mathfrak{Re}\left[
    \left(\left|C_{A}\right|^{2}+\left|C_{A}^{'}\right|^{2}\right)
    \frac{m_{e}}{q}
    \left\langle \left\Vert \hat{L}_{0}^{A}\right\Vert \right\rangle 
    \left\langle \left\Vert \hat{C}_{0}^{A}\right\Vert \right\rangle ^{*}\right.\nonumber\\
    &\left.\mp \left(C_{A}^{*}C_{T}+C_{A}^{'*}C_{T}^{'}\right)
    \left(\hat{\nu}\cdot\hat{q}
    \left\langle \left\Vert \hat{L}_{0}^{A}\right\Vert \right\rangle 
    \left\langle \left\Vert \hat{C}_{0}^{A}\right\Vert \right\rangle ^{*}
    -\left|\left\langle \left\Vert \hat{L}_{0}^{A}\right\Vert \right\rangle \right|^{2}\right)
    \right],
\end{align}
\end{subequations}
where $\hat{C}_{0}^{A}\propto\epsilon_{\text{NR}}$ and $\hat{L}_{0}^{A}\propto\epsilon_{qr}$ are the SM operators that dominate the $J^{\Delta\pi}=0^{-}$ non-unique first-forbidden transition.

\subsection{Gamow-Teller and unique forbidden transitions}

In discussing the $\Theta^{J^{\Delta\pi}}$ expressions for $J$'s greater than $0$, we distinguish between transitions with two parity types:
$\Delta\pi=\left(-\right)^{J}$, and $\Delta\pi=\left(-\right)^{J-1}$.
$J^{\left(-\right)^{J}}$ angular momentum presents non-unique $J^{\rm th}$ forbidden transitions, while $J^{\left(-\right)^{J-1}}$ presents, for $J=1$, the allowed Gamow-Teller transition, and for $J>1$, unique $(J-1)^{\rm th}$ forbidden transitions (we will refer to them together as unique transitions).

Starting with the unique transitions, using the relation $\hat{L}_{J}^{T}\left(q\right)\approx-\frac{i}{\sqrt{2}}\frac{g_{T}}{g_{A}}\hat{L}_{J}^{A}\left(q\right)$ (Eq.~\eqref{eq:T-A multipole operators}), a general expression which includes the BSM contributions along with the SM NLO corrections can be written as:
\begin{multline}
\Theta^{J^{\left(-\right)^{J-1}}}\left(q,\vec{\beta}\cdot\hat{\nu}\right)=\frac{\left|C_{A}\right|^{2}+\left|C_{A}^{'}\right|^{2}}{2\left|g_{A}\right|^{2}}\left|\left\langle \left\Vert \hat{L}_{J}^{A}\right\Vert \right\rangle \right|^{2}
\frac{2J+1}{J}\left(1+\delta_1^{J^{\left(-\right)^{J-1}}}+\frac{\left|C_{T}\right|^{2}+\left|C_{T}^{'}\right|^{2}}{\left|C_{A}\right|^{2}+\left|C_{A}^{'}\right|^{2}}\right)\\
\times\left\{ 1-\frac{1}{2J+1}
\vec{\beta}\cdot\hat{\nu}
\left(1+\tilde{\delta}_a^{J^{\left(-\right)^{J-1}}}-2\frac{\left|C_{T}\right|^{2}+\left|C_{T}^{'}\right|^{2}}{\left|C_{A}\right|^{2}+\left|C_{A}^{'}\right|^{2}}\right)
+\frac{m_{e}}{E}
\left(\delta_b^{J^{\left(-\right)^{J-1}}}
\pm2\mathfrak{Re}\frac{C_{A}^{*}C_{T}+C_{A}^{'*}C_{T}^{'}}{\left|C_{A}\right|^{2}+\left|C_{A}^{'}\right|^{2}}\right)\right.\\
\left.+\frac{J-1}{2J+1}
\frac{E\left(E_0-E\right)}{q^{2}}
\left[\beta^{2}-\left(\vec{\beta}\cdot\hat{\nu}\right)^{2}\right]
\left(1+\tilde{\delta}_{\beta^2}^{J^{\left(-\right)^{J-1}}}-2\frac{\left|C_{T}\right|^{2}+\left|C_{T}^{'}\right|^{2}}{\left|C_{A}\right|^{2}+\left|C_{A}^{'}\right|^{2}}\right)
\right\} 
+\mathcal{O}\left(\epsilon_M\epsilon_{qr}^{2J-2}\right),
\label{eq:Axial-dominant BSM}
\end{multline}
where the different $\delta^{J^{\left(-\right)^{J-1}}}$ are the NLO SM corrections described in~\cite{glick2021formalism} ($\epsilon_M\epsilon_{qr}^{2J-2}$ presents SM recoiled nucleus corrections, which we will not display here, since they are relevant only for very light nuclei~\cite{glick2021formalism}).
In the Gamow-Teller case ($J=1$), the term
$\frac{J-1}{2J+1}
\left(1+\tilde{\delta}_{\beta^2}^{J^{\left(-\right)^{J-1}}}-2\frac{\left|C_{T}\right|^{2}+\left|C_{T}^{'}\right|^{2}}{\left|C_{A}\right|^{2}+\left|C_{A}^{'}\right|^{2}}\right)$
$\times
\frac{E\left(E_0-E\right)}{q^{2}}
\left[\beta^{2}-\left(\vec{\beta}\cdot\hat{\nu}\right)^{2}\right]$
do not exist, and instead there is an NLO SM correction, $\tilde{\delta}^{1^+}_{\beta^2,\left(\beta\nu\right)^2}$~\cite{glick2021formalism}.

According to the $V-A$ structure of the weak interaction, the $\beta-\nu$ correlation leading order should be $a^{J^{\left(-\right)^{J-1}}}=-\frac{1}{2J+1}$.
When BSM contributions are added, the $\beta-\nu$ correlation becomes
\begin{align}
a^{J^{\left(-\right)^{J-1}}} & =-\frac{1}{2J+1}\left(1+\tilde{\delta}_{a}^{J^{\left(-\right)^{J-1}}}
-2\frac{\left|C_{T}\right|^{2}+\left|C_{T}^{'}\right|^{2}}{\left|C_{A}\right|^{2}+\left|C_{A}^{'}\right|^{2}}
\right)\text{.}
\label{eq: a_bn, -^(J-1) J>0}
\end{align}
As for the Fierz term that vanishes for unique transitions leading order in the $V-A$ structure, its BSM form (including a term with a similar spectral behavior that can be extracted from the NLO SM spectrum) will be:
\begin{align}
\label{eq: delta_b J>0}
b^{J^{\left(-\right)^{J-1}}}=\delta_{b}^{J^{\left(-\right)^{J-1}}} \pm2\mathfrak{Re}\frac{C_{A}^{*}C_{T}+C_{A}^{'*}C_{T}^{'}}{\left|C_{A}\right|^{2}+\left|C_{A}^{'}\right|^{2}}\text{.}
\end{align}

In the notion of Jackson, Treiman and Wyld, one can recognize:
\begin{subequations}
\begin{align}
    \label{eq:xiJ-1}
    \xi^{J^{\left(-\right)^{J-1}}} &= 
    \frac{\left|\left\langle \left\Vert \hat{L}_{J}^{A}\right\Vert \right\rangle \right|^{2}}
    {\left|g_{A}\right|^{2}}
    \left[\left(\left|C_{A}\right|^{2}+\left|C_{A}^{'}\right|^{2}\right)
    \left(1+\delta_1^{J^{\left(-\right)^{J-1}}}\right)
    +\left|C_{T}\right|^{2}+\left|C_{T}^{'}\right|^{2}\right],\\
    \label{eq:xi_aJ-1}
    a^{J^{\left(-\right)^{J-1}}} \xi^{J^{\left(-\right)^{J-1}}} &=
    -\frac{1}{2J+1}
    \frac{\left|\left\langle \left\Vert \hat{L}_{J}^{A}\right\Vert \right\rangle \right|^{2}}{\left|g_{A}\right|^{2}}
    \left[\left(\left|C_{A}\right|^{2}+\left|C_{A}^{'}\right|^{2}\right)
    \left(1+\delta_a^{J^{\left(-\right)^{J-1}}}\right)
    -\left|C_{T}\right|^{2}-\left|C_{T}^{'}\right|^{2}\right],\\
    \label{eq:xi_bJ-1}
    b^{J^{\left(-\right)^{J-1}}} \xi^{J^{\left(-\right)^{J-1}}} &=
    \frac{\left|\left\langle \left\Vert \hat{L}_{J}^{A}\right\Vert \right\rangle \right|^{2}}
    {\left|g_{A}\right|^{2}}
    \left[\left(\left|C_{A}\right|^{2}+\left|C_{A}^{'}\right|^{2}\right)
    \delta_{b}^{J^{\left(-\right)^{J-1}}}
    \pm 2\mathfrak{Re}
    \left(C_{A}^{*}C_{T}+C_{A}^{'*}C_{T}^{'}\right)
    \right],
\end{align}
\end{subequations}
where again the NLO SM corrections $\delta_1$, $\delta_a=\tilde{\delta}_a+\delta_1$ and $\delta_{b}$ are higher order precision corrections not found in the Jackson, Treiman and Wyld paper.

Consider, for example, the allowed Gamow-Teller transition, which is $J^{\Delta\pi}=1^+$. According to Eq.~\eqref{eq:Axial-dominant BSM},
\begin{multline}
\Theta^{1^{+}}\left(q,\vec{\beta}\cdot\hat{\nu}\right)
=\frac{\left|C_{A}\right|^{2}+\left|C_{A}^{'}\right|^{2}}{2\left|g_{A}\right|^{2}}\left|\left\langle \left\Vert \hat{L}_{1}^{A}\right\Vert \right\rangle \right|^{2}3\left(1+\delta_1^{1^{+}}+\frac{\left|C_{T}\right|^{2}+\left|C_{T}^{'}\right|^{2}}{\left|C_{A}\right|^{2}+\left|C_{A}^{'}\right|^{2}}\right)\\
\times\left[1-\frac{1}{3}
\vec{\beta}\cdot\hat{\nu}
\left(1+\tilde{\delta}_a^{1^{+}}
-2\frac{\left|C_{T}\right|^{2}+\left|C_{T}^{'}\right|^{2}}
{\left|C_{A}\right|^{2}+\left|C_{A}^{'}\right|^{2}}\right)
+\frac{m_{e}}{E}
\left(\delta_b^{1^+}\pm 2\mathfrak{Re}\frac{C_{A}^{*}C_{T}+C_{A}^{'*}C_{T}^{'}}
{\left|C_{A}\right|^{2}+\left|C_{A}^{'}\right|^{2}}\right)
+\tilde{\delta}^{1^+}_{\beta^2,\left(\beta\nu\right)^2}\right]
\text{,}
\label{eq:GT BSM}
\end{multline}
where $\frac{\left\langle \left\Vert \hat{L}_{1}^{A}\right\Vert \right\rangle}{g_A} = M_{GT}$ is the Gamow-Teller matrix element used in the Jackson, Treiman and Wyld paper. 
A full allowed (mixed Gamow-Teller and Fermi) transition will be a sum of Eqs.~\eqref{eq:Fermi BSM} and \eqref{eq:GT BSM}, where the full $\xi$ presented in the Jackson, Treiman and Wyld paper is the sum of Eqs.~\eqref{eq:xi0+} and~\eqref{eq:xiJ-1} for $J=1$, $a\xi$ is the sum of Eqs.~\eqref{eq:xi_a0+} and~\eqref{eq:xi_aJ-1} for $J=1$, and $b\xi$ is the sum of Eqs.~\eqref{eq:xi_b0+} and~\eqref{eq:xi_bJ-1} for $J=1$. All are with agreement with their paper.

\subsection{Non-unique forbidden transitions}
%
In the case of non-unique $J^{\rm th}$ forbidden transitions, i.e., decays with angular momentum change $J$ greater than 0, and parity change $\pi = \left(-\right)^{J}$, the $\Theta^{J^{\left(-\right)^{J}}}$
expression can be written as:
\begin{multline}
\Theta^{J^{\left(-\right)^{J}}}\left(q,\vec{\beta}\cdot\hat{\nu}\right)=
\left\{ \frac{\left|C_{V}\right|^{2}+\left|C_{V}^{'}\right|^{2}}{2\left|g_{V}\right|^{2}}\left[1+\frac{1}{J}\frac{E_{0}^{2}}{q^{2}}\left(1-\left(J+1\right)2\mathfrak{Re}\delta^{J^{\left(-\right)^{J}}}\right)\right]+\frac{\left|C_{S}\right|^{2}+\left|C_{S}^{'}\right|^{2}}{2\left|g_{V}\right|^{2}}\right.\\
\left.\mp\frac{m_{e} E_0}{q^2}2\mathfrak{Re}\frac{C_{V}C_{S}^*+C_{V}^{'}C_{S}^{'*}}{2\left|g_{V}\right|^{2}}\vphantom{\frac{\left|C_{V}^{'}\right|^{2}}{2\left|g_{V}\right|^{2}}}\right\} \left|\left\langle \left\Vert \hat{C}_{J}^{V}\right\Vert \right\rangle \right|^{2}
+\left(\frac{\left|C_{A}\right|^{2}+\left|C_{A}^{'}\right|^{2}}{2\left|g_{A}\right|^{2}}+\frac{\left|C_{T}\right|^{2}+\left|C_{T}^{'}\right|^{2}}{2\left|g_{A}\right|^{2}}\right)
\left|\left\langle \left\Vert \hat{M}_{J}^{A}\right\Vert \right\rangle \right|^{2}\\
\pm\sqrt{\frac{J+1}{J}}2\mathfrak{Re}\left[\frac{E_{0}}{q}
\left(\frac{E_{0}-2E}{q}\frac{C_{V}C_{A}^{*}+C_{V}^{'}C_{A}^{'*}}{2g_{V}g_{A}^{*}}\left(1-\delta^{J^{\left(-\right)^{J}}}\right)\right.\right.\\
\left.\left.-\frac{m_{e}}{q}\frac{C_{V}C_{T}^{*}+C_{V}^{'}C_{T}^{'*}}{2g_{V}g_{A}^{*}}\right)
\left\langle \left\Vert \hat{C}_{J}^{V}\right\Vert \right\rangle \left\langle \left\Vert \hat{M}_{J}^{A}\right\Vert \right\rangle ^{*}\right]\\
+\vec{\beta}\cdot\hat{\nu}\left\{ \left[\left[1-\frac{2J+1}{J}\frac{E_{0}^{2}}{q^{2}}\left(1-\frac{J+1}{2J+1}2\mathfrak{Re}\delta^{J^{\left(-\right)^{J}}}\right)\right]
\frac{\left|C_{V}\right|^{2}+\left|C_{V}^{'}\right|^{2}}{2\left|g_{V}\right|^{2}}-\frac{\left|C_{S}\right|^{2}+\left|C_{S}^{'}\right|^{2}}{2\left|g_{V}\right|^{2}}\right.\right.\\
\left.\pm\frac{m_{e} E_0}{q^2}2\mathfrak{Re}\frac{C_{V}C_{S}^{*}+C_{V}^{'}C_{S}^{'*}}{2\left|g_{V}\right|^{2}}\vphantom{\frac{\left|C_{V}^{'}\right|^{2}}{2\left|g_{V}\right|^{2}}}\right]
\left|\left\langle \left\Vert \hat{C}_{J}^{V}\right\Vert \right\rangle \right|^{2}
+\left(-\frac{\left|C_{A}\right|^{2}+\left|C_{A}^{'}\right|^{2}}{2\left|g_{A}\right|^{2}}+\frac{\left|C_{T}\right|^{2}+\left|C_{T}^{'}\right|^{2}}{2\left|g_{A}\right|^{2}}\right)
\left|\left\langle \left\Vert \hat{M}_{J}^{A}\right\Vert \right\rangle \right|^{2}\\
\mp\sqrt{\frac{J+1}{J}}2\mathfrak{Re}\left[\frac{E_{0}}{q}\left(\frac{E_{0}-2E}{q}\frac{C_{V}C_{A}^{*}+C_{V}^{'}C_{A}^{'*}}{2g_{V}g_{A}^{*}}\left(1-\delta^{J^{\left(-\right)^{J}}}\right)\right.\right.\\
\left.\left.\left.-\frac{m_{e}}{q}\frac{C_{V}C_{T}^{*}+C_{V}^{'}C_{T}^{'*}}{2g_{V}g_{A}^{*}}\right)\left\langle \left\Vert \hat{C}_{J}^{V}\right\Vert \right\rangle \left\langle \left\Vert \hat{M}_{J}^{A}\right\Vert \right\rangle ^{*}\right]\vphantom{\frac{\left|C_{V}^{'}\right|^{2}}{2\left|g_{V}\right|^{2}}}\right\} \\
+\frac{m_{e}}{E}\left\{ \left[2\frac{m_{e}E_{0}}{q^{2}}\frac{\left|C_{V}\right|^{2}+\left|C_{V}^{'}\right|^{2}}{2\left|g_{V}\right|^{2}}
\pm\left(1+\frac{E_{0}^2}{q^2}\right)2\mathfrak{Re}\frac{C_{V}C_{S}^{*}+C_{V}^{'}C_{S}^{'*}}{2\left|g_{V}\right|^{2}}\right]\left|\left\langle \left\Vert \hat{C}_{J}^{V}\right\Vert \right\rangle \right|^{2}\right.\\
\mp2\mathfrak{Re}\frac{C_{A}^{*}C_{T}+C_{A}^{'*}C_{T}^{'}}{2\left|g_{A}\right|^{2}}\left|\left\langle \left\Vert \hat{M}_{J}^{A}\right\Vert \right\rangle \right|^{2}\\
\pm\sqrt{\frac{J+1}{J}}2\mathfrak{Re}\left[\left(\frac{m_{e}E_{0}}{q^{2}}
\frac{C_{V}C_{A}^{*}+C_{V}^{'}C_{A}^{'*}}{2g_{V}g_{A}^{*}}
\left(1-\delta^{J^{\left(-\right)^{J}}}\right)\right.\right.\\
\left.\left.\left.+\frac{E_{0}^{2}}{q^{2}}\frac{C_{V}C_{T}^{*}+C_{V}^{'}C_{T}^{'*}}{2g_{V}g_{A}^{*}}\right)\left\langle \left\Vert \hat{C}_{J}^{V}\right\Vert \right\rangle \left\langle \left\Vert \hat{M}_{J}^{A}\right\Vert \right\rangle ^{*}\right]\vphantom{\frac{\left|C_{V}^{'}\right|^{2}}{2\left|g_{V}\right|^{2}}}\right\} \\
+\frac{E\left(E_{0}-E\right)}{q^{2}}\left[\beta^{2}-\left(\vec{\beta}\cdot\hat{\nu}\right)^{2}\right]\left[\frac{J-1}{J}\frac{E_{0}^{2}}{q^{2}}\left(1+\frac{J+1}{J-1}2\mathfrak{Re}\delta^{J^{\left(-\right)^{J}}}\right)\frac{\left|C_{V}\right|^{2}+\left|C_{V}^{'}\right|^{2}}{2\left|g_{V}\right|^{2}}\left|\left\langle \left\Vert \hat{C}_{J}^{V}\right\Vert \right\rangle \right|^{2}\right.\\
\left.
+\left(-\frac{\left|C_{A}\right|^{2}+\left|C_{A}^{'}\right|^{2}}{2\left|g_{A}\right|^{2}}+\frac{\left|C_{T}\right|^{2}+\left|C_{T}^{'}\right|^{2}}{2\left|g_{A}\right|^{2}}\right)
\left|\left\langle \left\Vert \hat{M}_{J}^{A}\right\Vert \right\rangle \right|^{2}\right]
+\mathcal{O}\left(\epsilon_M\epsilon_{qr}^{2J}\right)\text{,}\label{eq:Vector-dominant BSM}
\end{multline}
where $\delta^{J^{\left(-\right)^{J}}}$ is an NLO SM correction described in~\cite{glick2021formalism} ($\epsilon_M\epsilon_{qr}^{2J}$ presents SM recoiled nucleus corrections, which we will not display here, since they are relevant only for very light nuclei~\cite{glick2021formalism}).
The multipole operators involved are $\hat{C}_{J}^{V},\hat{M}_{J}^{A}\propto\epsilon_{qr}^{J}$, and the terms of Jackson, Treiman and Wyld are:
\begin{subequations}
\begin{align}
    \xi^{J^{\left(-\right)^{J}}} &=
    \left\{
    \left[1+\frac{1}{J}\frac{E_{0}^{2}}{q^{2}}\left(1-\left(J+1\right)2\mathfrak{Re}\delta^{J^{\left(-\right)^{J}}}\right)\right]
    \left(\left|C_{V}\right|^{2}+\left|C_{V}^{'}\right|^{2}\right)
    +\left(\left|C_{S}\right|^{2}+\left|C_{S}^{'}\right|^{2}\right)\right.\nonumber\\
    &\left.\mp\frac{m_{e} E_0}{q^2}2\mathfrak{Re}
    \left(C_{V}C_{S}^{*}+C_{V}^{'}C_{S}^{'*}\right)
    \vphantom{\frac{E_{0}^{2}}{q^{2}}}\right\}
    \frac{1}{\left|g_{V}\right|^{2}}
    \left|\left\langle \left\Vert \hat{C}_{J}^{V}\right\Vert \right\rangle \right|^{2}\nonumber\\
    &+\left(\left|C_{A}\right|^{2}+\left|C_{A}^{'}\right|^{2}+\left|C_{T}\right|^{2}+\left|C_{T}^{'}\right|^{2}\right)
    \frac{1}{\left|g_{A}\right|^{2}}
    \left|\left\langle \left\Vert \hat{M}_{J}^{A}\right\Vert \right\rangle \right|^{2}\nonumber\\
    &\pm\sqrt{\frac{J+1}{J}}2\mathfrak{Re}\left\{\left[\frac{E_{0}\left(E_{0}-2E\right)}{q^{2}}
    \left(C_{V}C_{A}^{*}+C_{V}^{'}C_{A}^{'*}\right)
    \left(1-\delta^{J^{\left(-\right)^{J}}}\right)\right.\right.\nonumber\\
    &\left.\left.
    -\frac{m_{e}E_{0}}{q^{2}}
    \left(C_{V}C_{T}^{*}+C_{V}^{'}C_{T}^{'*}\right)\right]
    \frac{1}{g_{V}g_{A}^{*}}
    \left\langle \left\Vert \hat{C}_{J}^{V}\right\Vert \right\rangle 
    \left\langle \left\Vert \hat{M}_{J}^{A}\right\Vert \right\rangle ^{*}\right\},\\
    a^{J^{\left(-\right)^{J}}} \xi^{J^{\left(-\right)^{J}}} &=
    \left\{\left[1-\frac{2J+1}{J}\frac{E_{0}^{2}}{q^{2}}\left(1-\frac{J+1}{2J+1}2\mathfrak{Re}\delta^{J^{\left(-\right)^{J}}}\right)\right]
    \left(\left|C_{V}\right|^{2}+\left|C_{V}^{'}\right|^{2}\right)
    -\left(\left|C_{S}\right|^{2}+\left|C_{S}^{'}\right|^{2}\right)\right.\nonumber\\
    &\left.\pm\frac{m_{e} E_0}{q^2}2\mathfrak{Re}
    \left(C_{V}C_{S}^{*}+C_{V}^{'}C_{S}^{'*}\right)
    \vphantom{\frac{E_{0}^{2}}{q^{2}}}\right\}
    \frac{1}{\left|g_{V}\right|^{2}}
    \left|\left\langle \left\Vert \hat{C}_{J}^{V}\right\Vert \right\rangle \right|^{2}\nonumber\\
    &+\left(-\left|C_{A}\right|^{2}-\left|C_{A}^{'}\right|^{2}+\left|C_{T}\right|^{2}+\left|C_{T}^{'}\right|^{2}\right)
    \frac{1}{\left|g_{A}\right|^{2}}
    \left|\left\langle \left\Vert \hat{M}_{J}^{A}\right\Vert \right\rangle \right|^{2}\nonumber\\
    &\mp\sqrt{\frac{J+1}{J}}2\mathfrak{Re}\left\{\left[\frac{E_{0}\left(E_{0}-2E\right)}{q^{2}}
    \left(C_{V}C_{A}^{*}+C_{V}^{'}C_{A}^{'*}\right)
    \left(1-\delta^{J^{\left(-\right)^{J}}}\right)
    \right.\right.\nonumber\\
    &\left.\left.
    -\frac{m_{e}E_{0}}{q^{2}}
    \left(C_{V}C_{T}^{*}+C_{V}^{'}C_{T}^{'*}\right)\right]
    \frac{1}{g_{V}g_{A}^{*}}
    \left\langle \left\Vert \hat{C}_{J}^{V}\right\Vert \right\rangle
    \left\langle \left\Vert \hat{M}_{J}^{A}\right\Vert \right\rangle ^{*}\right\},
    \\
    b^{J^{\left(-\right)^{J}}} \xi^{J^{\left(-\right)^{J}}} &=
    \left[2\frac{m_{e}E_{0}}{q^{2}}
    \left(\left|C_{V}\right|^{2}+\left|C_{V}^{'}\right|^{2}\right)
    \pm\left(1+\frac{E_{0}^2}{q^2}\right)2\mathfrak{Re}
    \left(C_{V}C_{S}^{*}+C_{V}^{'}C_{S}^{'*}\right)\right]
    \frac{1}{\left|g_{V}\right|^{2}}
    \left|\left\langle \left\Vert \hat{C}_{J}^{V}\right\Vert \right\rangle \right|^{2}\nonumber\\
    &\mp2\mathfrak{Re}
    \left(C_{A}^{*}C_{T}+C_{A}^{'*}C_{T}^{'}\right)
    \frac{1}{\left|g_{A}\right|^{2}}
    \left|\left\langle \left\Vert \hat{M}_{J}^{A}\right\Vert \right\rangle \right|^{2}\nonumber\\
    &
    \pm\sqrt{\frac{J+1}{J}}2\mathfrak{Re}
    \left\{\left[\frac{m_{e}E_{0}}{q^{2}}
    \left(C_{V}C_{A}^{*}+C_{V}^{'}C_{A}^{'*}\right)
    \left(1-\delta^{J^{\left(-\right)^{J}}}\right)\right.\right.\nonumber\\
    &
    \left.\left.+\frac{E_{0}^{2}}{q^{2}}
    \left(C_{V}C_{T}^{*}+C_{V}^{'}C_{T}^{'*}\right)\right]
    \frac{1}{g_{V}g_{A}^{*}}
    \left\langle \left\Vert \hat{C}_{J}^{V}\right\Vert \right\rangle
    \left\langle \left\Vert \hat{M}_{J}^{A}\right\Vert \right\rangle ^{*}\right\}.
\end{align}
\end{subequations}

\section{Sensitivity to BSM signatures in $^6$He as an exemplary nucleus of current experimental interest 
\label{sec:Examples}}

$^6$He decays into $^6$Li in a pure GT transition. This is a light nucleus, amendable to state-of-the-art ab-initio calculations, and its half life is about $\approx 1\, sec$, making it ideal for experimental study using traps. For these reasons, it has a prominent role in several ongoing precision experiments (see Ref.~\cite{glickmagid2021nuclear}). We thus use it here as a case study to demonstrate the application of the theory presented here.

Identifying theoretically a BSM signal relies on correctly evaluating the theoretical prediction of the $\beta$-decay observables. We thus plot the ratio, $R\left(\text{BSM}/\text{SM}\right)$, of BSM signal to the size of associated nuclear structure related SM corrections. If the ratio $R\left(\text{BSM}/\text{SM}\right)$ is of the order of $1$, then the corrections should be calculated explicitly. The limit of theoretical uncertainty consequently occurs when the ratio $R\left(\text{BSM}/\text{SM}\right)$ is about the size of the theoretical uncertainty in calculating the SM corrections.  

\begin{figure}[htb]
  \centering
  \includegraphics[width=\columnwidth]{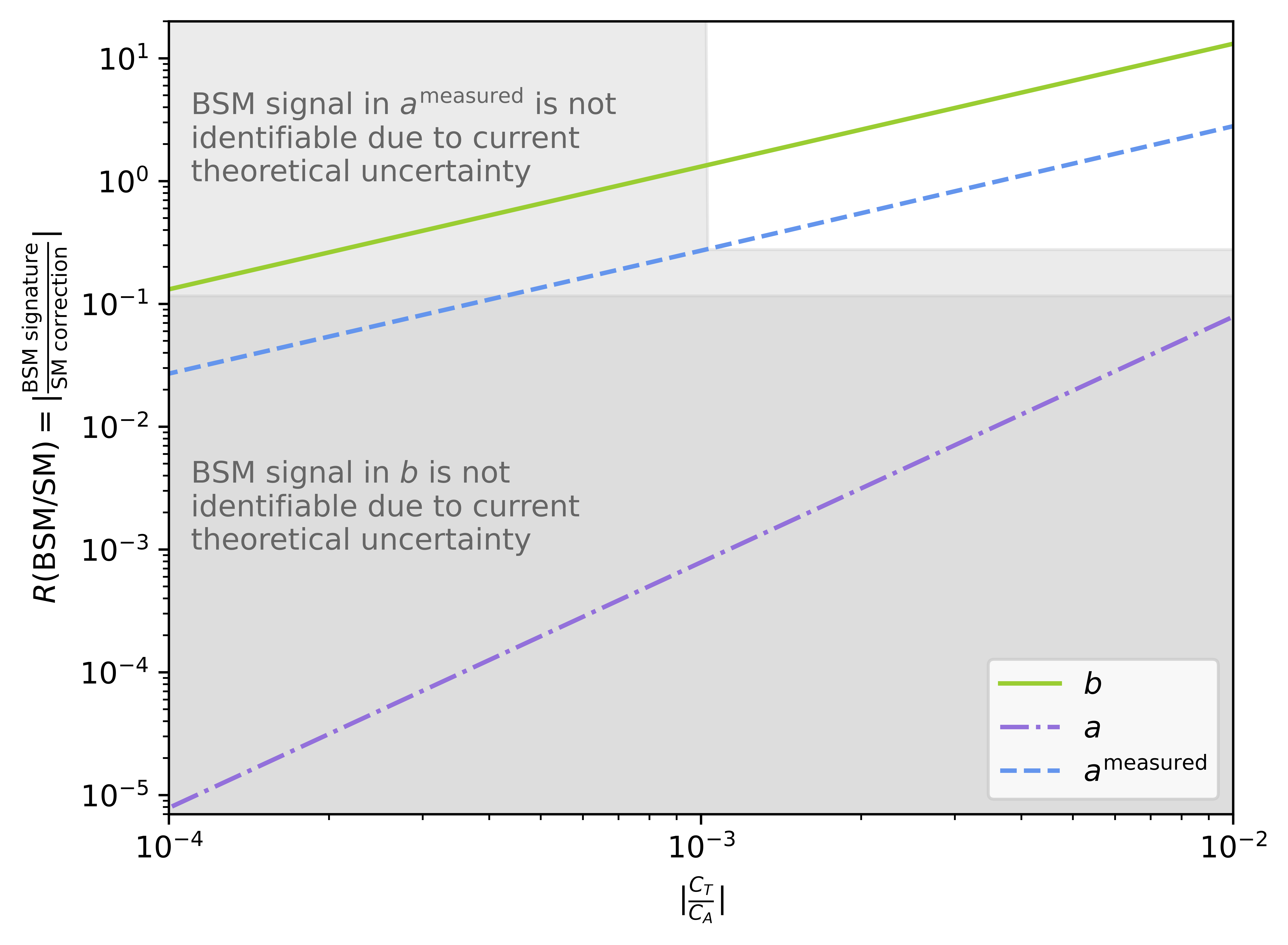}
  \caption{(Color online) The ratio $R\left(\text{BSM}/\text{SM}\right)$ of BSM signatures in $\beta$-decay observables to the associated SM correction calculated for $^6$He $\beta$-decay in~\cite{glickmagid2021nuclear}, for different values of the BSM coupling constant. For visualization simplicity, we assume $C_{A}^{'}=C_A$ and $C_{T}^{'}=C_T$.
   Solid green line is the ratio for Fierz term $b$.
   Dashed-dotted purple line is the ratio for the angular correlation $a$.
   Dashed blue line is the ratio for the measured value of the angular correlation, $a^{\text{measured}}$.
   In the white domain, considering the theoretical uncertainty from ~\cite{glickmagid2021nuclear}, separating the BSM signal from the SM corrections in both $b$ and $a^{\text{measured}}$ is possible. In the other domains separation is limited by theoretical uncertainties.
\label{fig:BSM_signatures}}
\end{figure}

We focus on tensor couplings of the order of $\left|\frac{C_T}{C_A}\right| = \left|\frac{C_{T}^{'}}{C_A}\right| \sim10^{-3}$, corresponding to new physics at a few TeV scale. 
Fig.~\ref{fig:BSM_signatures} compares the ratio $R\left(\text{BSM}/\text{SM}\right)$ of BSM signatures in the aforementioned $\beta$-decay observables,  $a$ and $b$, to the associated SM correction calculated for $^6$He $\beta$-decay in~\cite{glickmagid2021nuclear}.

As  $^6$He $\beta$-decay is a pure Gamow-Teller transition, we use Eq.~\eqref{eq: delta_b J>0} and compare its Fierz term to the correction $\delta_b$  in the spectrum, originating in nuclear structure corrections that has a spectral behavior similar to the Fierz term, i.e., 
$R^{b}\left(\text{BSM}/\text{SM}\right)=\left|2\mathfrak{Re}\frac{C_{A}^{*}C_{T}+C_{A}^{'*}C_{T}^{'}} {\left|C_{A}\right|^{2}+\left|C_{A}^{'}\right|^{2}} / \delta_b\right|$.
In Fig.~\ref{fig:BSM_signatures}, domains where current theory enables separation of the BSM signal from nuclear structure related SM corrections appearing in Fierz term, i.e., domains where $R^{b}\left(\text{BSM}/\text{SM}\right) > \left|\left(\delta_b\text{ uncertainty} \right) / \delta_b \right|$ are shown in white and light gray. 
As apparent in Fig.~\ref{fig:BSM_signatures}, a BSM Fierz signal is identifiable already tensor couplings as small as $\left|\frac{C_T}{C_A}\right| \sim10^{-4}$.

In contrast, a similar approach for the angular correlation, i.e, $R^{a}\left(\text{BSM}/\text{SM}\right)=\left|2\frac{\left|C_{T}\right|^{2}+\left|C_{T}^{'}\right|^{2}} {\left|C_{A}\right|^{2}+\left|C_{A}^{'}\right|^{2}} / \left<\tilde{\delta}_{a}\right>\right|$ (see Eq.~\eqref{eq: a_bn, -^(J-1) J>0}), where the angle brackets represent an average weighted by the spectrum,
shows that the theory cannot identify the naive BSM signal of the angular correlation from the SM corrections even for $\left|\frac{C_T}{C_A}\right| \sim 10^{-2}$.

However, when taking into account the way that $a$ is extracted from measurements,  
the spectral shape suggests that $a^{\text{measured}}=\frac{\left<a\right>} {1+b\left<\frac{m_{e}}{E}\right>}$~\cite{PhysRevC.94.035503},
making the measured value of $a$ sensitive also to the Fierz term, as specified in the following relation for Gamow-Teller and unique forbidden transitions:
\begin{equation}
\label{eq:a_measured}
  \begin{split}
    a^{\text{measured}} &=
    -\frac{1}{2J+1}
    \left(1+\left<\tilde{\delta}_a^{J^{\left(-\right)^{J-1}}}\right> - \delta_b^{J^{\left(-\right)^{J-1}}}\left<\frac{m_{e}}{E}\right> 
    \vphantom{\frac{\left|C_{T}^{'}\right|^{2}}{\left|C_{A}^{'}\right|^{2}}}
    \right.\\
     &\left.
    \mp 2\mathfrak{Re}\frac{C_{A}^{*}C_{T}+C_{A}^{'*}C_{T}^{'}}
    {\left|C_{A}\right|^{2}+\left|C_{A}^{'}\right|^{2}}
    \left<\frac{m_{e}}{E}\right>
    -2\frac{\left|C_{T}\right|^{2}+\left|C_{T}^{'}\right|^{2}}
    {\left|C_{A}\right|^{2}+\left|C_{A}^{'}\right|^{2}}
    \right).
    \end{split}
  \end{equation}
This results in a relative size of the BSM signal, 
\begin{equation}
R^{a^{\text{measured}}}\left(\text{BSM}/\text{SM}\right)=\left|\frac{ 2\mathfrak{Re}\frac{C_{A}^{*}C_{T}+C_{A}^{'*}C_{T}^{'}}
{\left|C_{A}\right|^{2}+\left|C_{A}^{'}\right|^{2}}
\left<\frac{m_{e}}{E}\right>
+2\frac{\left|C_{T}\right|^{2}+\left|C_{T}^{'}\right|^{2}}
{\left|C_{A}\right|^{2}+\left|C_{A}^{'}\right|^{2}} }{
\left<\tilde{\delta}_{a}\right> - \delta_b\left<\frac{m_{e}}{E}\right>}\right|,    
\end{equation}
which enables a separation between the BSM signal and SM corrections for $\left|\frac{C_T}{C_A}\right| = \left|\frac{C_{T}^{'}}{C_A}\right| \sim10^{-3}$, as shown in Fig.~\ref{fig:BSM_signatures} (the white domain).

To understand the ability of experiments to identify these signals, we notice that the SM nuclear structure related corrections are of the order of $10^{-3}$ for $b$ and for $a^{\text{measured}}$. Thus, experimental accuracy of about $10^{-3}$ in the measurement of both these observables is needed, as is aimed in current and planned experimental campaigns.

A complete application of the theory presented, extracting both the angular correlation and Fierz term from measurements of the recoiled ion energy of $^{23}$Ne as well as $^6$He $\beta$-decays, and combining the theory and experiment sensitivities to present new bounds on BSM tensor coupling constants, can be found in~\cite{Mishnayot-23Ne}.

The above discussion concentrated on allowed transitions, as they are the focus of many experimental campaigns. However, theoretical considerations that arise also from the formalism presented here, show that studying $\beta$-decays beyond allowed transitions can be of advantage. For unique forbidden transitions, there appears an additional term in the spectrum (see Eq.~\eqref{eq:Axial-dominant BSM}),
\begin{equation}
\frac{J-1}{2J+1}
\frac{E\left(E_0-E\right)}{q^{2}}
\left[\beta^{2}-\left(\vec{\beta}\cdot\hat{\nu}\right)^{2}\right] 
\times \left(1+\tilde{\delta}_{\beta^2}^{J^{\left(-\right)^{J-1}}}-2\frac{\left|C_{T}\right|^{2}+\left|C_{T}^{'}\right|^{2}}{\left|C_{A}\right|^{2}+\left|C_{A}^{'}\right|^{2}}\right),
\end{equation}
which does not appear in Gamow-Teller transitions. This fact makes the energy spectrum of the electron sensitive to both angular correlation and Fierz term, as we detailed in~\cite{GLICKMAGID2017285} for unique first forbidden transitions. Consequently, two experimental campaigns were initiated to measure unique first-forbidden transitions~\cite{refId0,Ohayon2018}.

\section{Conclusions and outlook\label{sec:Summary}}

In this paper, we introduced a general mathematical formalism for calculating the interaction of a Lorentz invariant probe with a nucleus. As we demonstrated, this general formalism is useful for various types of BSM physics analysis, from exotic interactions with standard particles to interactions with new particles, such as those expected in the astrophysical dark matter scenario, and thus completing the theoretical parts needed for analyzing ongoing and future experiments looking for BSM physics.

This paper results in three main findings:
the first is the multipole expansion of tensor interactions with a composite particle, presented in Eq.~\eqref{eq: multipole expansion}. 
The second is the expression for the tensor nuclear current, in Eq.~\eqref{eq:tensor_current_density}.
The third are BSM expressions for $\beta$-decay, useful for precision experiments searching for BSM signatures, described in detail in Sec.~\ref{sec:A-General-Expression}, with an exemplary application given in Sec.\ref{sec:Examples}. The latter shows the usefulness of the theoretical analysis we presented in analysis of the potential of experimental campaigns in identifying BSM signals.

In addition, our multipole expansion technique reveals that the BSM multipole operators are identical to operators appearing in the SM, to the needed approximation. Consequently, in order to compute BSM contributions to semi-leptonic processes, such as $\beta$-decays, which are frequently used in BSM searches today, it is not necessary to compute any new matrix elements, but to use the well-established SM matrix elements, as shown in Eqs.~\eqref{eq:T-A multipole operators} and~\eqref{eq:S multipole operators}.

Moreover, the additional terms we found, which complete the ingredients for $\beta$-decays, are crucial for accurately identifying the expected size of the BSM effect, as we demonstrate for $^6$He in Sec.~\ref{sec:Examples}. This can assist in the design of future experiments to study BSM effects, as we pointed out in~\cite{GLICKMAGID2017285}, where, supported by this formalism, we showed that the unique first forbidden decay spectrum is more sensitive to BSM signatures. In light of that, experiments are underway at the SARAF accelerator, Israel, and the Oak Ridge National Laboratory, Tennessee, to measure the spectrum of unique first forbidden $\beta$-decays~\cite{refId0,Ohayon2018}.

\appendix

\section*{Appendices}

\section{Tensor lepton traces\label{sec:Appendix Tensor-Lepton-Traces}}

In the low energy frame, the weak interaction Hamiltonian between nuclei and a lepton probe is presumed to be a multiplication of a nuclear current and the lepton current of the same kind.
We describe the probes using relativistic quantum fields. In the interaction representation, fermion fields take the following form~\cite{WALECKA1975113}:
\begin{eqnarray}
\psi\left(\vec{r}\right) & \equiv & \frac{1}{\sqrt{\Omega}}\sum_{\vec{k}}\left[a_{\vec{k}}u\left(\vec{k}\right)e^{i\vec{k}\cdot\vec{r}}+b_{\vec{k}}^{+}v\left(\vec{k}\right)e^{-i\vec{k}\cdot\vec{r}}\right]\mbox{.}\label{eq:Fremion fields}
\end{eqnarray}
In this expression, $a_{\vec{k}}$ destroys a lepton with momentum $\vec{k}$ and $b_{\vec{k}}^{+}$ creates an antilepton with the same momentum; $u$ ($v$) is the free-particle (antiparticle) Dirac spinor.
We write the tensor lepton current in its most general form~\cite{PhysRev.104.254,PhysRev.106.517}, adjusted to the nowadays convention~\cite{gonzalez2019new}:
\begin{eqnarray}
\hat{j}_{\mu\nu}\left(\vec{r}\right) & = & \bar{\psi}'\left(\vec{r}\right)\sigma_{\mu\nu}\left(C_{T}-C_{T}^{'}\gamma_{5}\right)\psi\left(\vec{r}\right)\mbox{,}\label{eq: j_mu,nu}
\end{eqnarray}
where $\bar{\psi}\equiv{\psi}^{+}\gamma_{0}$. The coupling constants $C$ and $C'$ (these of the tensor coupling (T), as well as these of scalar (S), pseudoscalar (P), polar-vector (V) and axial-vector (A) couplings that will be mentioned in the following Appendices) are real if time reversal invariance is preserved in the process, but this will not be assumed in the following.
Assuming the leptons have a plane wave character (interaction with the nucleus will be inserted perturbatively), the general matrix element can be written as
\begin{eqnarray}
\left\langle f\left|\hat{j}_{\mu\nu}\left(\vec{r}\right)\right|i\right\rangle  & \equiv & l_{\mu\nu}e^{-i\vec{q}\cdot\vec{r}}\mbox{,}
\end{eqnarray}
where $\vec{q} \equiv \vec{k}_f - \vec{k}_i$ is the momentum transfer, and $l_{\mu\nu}$ is defined as
$l_{\mu\nu}=\frac{1}{\Omega}\bar{l}'\left(\vec{k}'\right)\sigma_{\mu\nu} \left(C_{T}-C_{T}^{'}\gamma_{5}\right)$ $\times l\left(\vec{k}\right)$, where $l,l'\in\left\{ u,v\right\} $.

Using Wigner-Eckart theorem~\cite{Edmonds:1974:AMQM}:
\begin{eqnarray}
\label{eq: Wigner-Eckart}
\left\langle J_f M_f \left|\hat{O}_{J M}\right|J_i M_i\right\rangle = \left(-1\right)^{J_f-M_f}
\left(\begin{array}{ccc}
J_f & J & J_i \\
-M_f & M & M_i
\end{array}\right)
\left\langle J_f \left\Vert\hat{O}_{J}\right\Vert J_i\right\rangle,
\end{eqnarray}
where $\hat{O}_{J M}$ is a spherical tensor with rank $J$ 
and projection $M \in \left\{-J,-J+1,...,J \right\}$, $\left\langle J_f \left\Vert\hat{T}_{J}\right\Vert J_i\right\rangle$ 
is its reduced matrix element, and
$\left(\begin{array}{ccc}
J_f & J & J_i \\
-M_f & M & M_i
\end{array}\right)$ 
is a 3-j coefficient; with the orthonormality of the 3-j coefficients:
\begin{eqnarray}
\sum_{M_{i}}\sum_{M_{f}}
\left(\begin{array}{ccc}
J_f & J & J_i \\
-M_f & M & M_i
\end{array}\right)
\left(\begin{array}{ccc}
J_f & J^{'} & J_i \\
-M_f & M^{'} & M_i
\end{array}\right)
=\frac{1}{2J+1}\delta_{J J^{'}} \delta_{M M^{'}},
\end{eqnarray}
and some identities of complex numbers,
we distract from Eq.~\eqref{eq: multipole expansion} in the main text, a general result for any semileptonic nuclear tensor process in terms of reduced matrix elements of the multipole operators:
\begin{multline}
\sum_{M_{i}}\sum_{M_{f}}\left|\left\langle f\left|\hat{H}_{\mbox{w}}^{T}\right|i\right\rangle \right|^{2}=
4\pi\left\{ \sum_{J=0}^{\infty}\left[l_{3}^{T}l_{3}^{T*}\left|\left\langle \left\Vert \hat{L}_{J}^{T}\right\Vert \right\rangle \right|^{2}+l_{3}^{T'}l_{3}^{T'*}\left|\left\langle \left\Vert \hat{L}_{J}^{T'}\right\Vert \right\rangle \right|^{2}\right.\right.\\
\left.+2\mathfrak{Re}\left(l_{3}^{T}l_{3}^{T'*}\left\langle \left\Vert \hat{L}_{J}^{T}\right\Vert \right\rangle \left\langle \left\Vert \hat{L}_{J}^{T'}\right\Vert \right\rangle ^{*}\right)\right]\\
+\frac{1}{2}\sum_{J=1}^{\infty}\left[\left(\vec{l}^{T}\cdot\vec{l}^{T*}-l_{3}^{T}l_{3}^{T*}\right)\left(\left|\left\langle \left\Vert \hat{E}_{J}^{T}\right\Vert \right\rangle \right|^{2}+\left|\left\langle \left\Vert \hat{M}_{J}^{T}\right\Vert \right\rangle \right|^{2}\right)\right.\\
+\left(\vec{l}^{T'}\cdot\vec{l}^{T'*}-l_{3}^{T'}l_{3}^{T'*}\right)\left(\left|\left\langle \left\Vert \hat{E}_{J}^{T'}\right\Vert \right\rangle \right|^{2}+\left|\left\langle \left\Vert \hat{M}_{J}^{T'}\right\Vert \right\rangle \right|^{2}\right)\\
\left.\left.+2\mathfrak{Re}\left[\left(\vec{l}^{T}\cdot\vec{l}^{T'*}-l_{3}^{T}l_{3}^{T'*}\right)\left(\left\langle \left\Vert \hat{E}_{J}^{T}\right\Vert \right\rangle \left\langle \left\Vert \hat{E}_{J}^{T'}\right\Vert \right\rangle ^{*}+\left\langle \left\Vert \hat{M}_{J}^{T}\right\Vert \right\rangle \left\langle \left\Vert \hat{M}_{J}^{T'}\right\Vert \right\rangle ^{*}\right)\right]\right.\right.\\
\left.\left.\hphantom{+\frac{1}{2}\sum_{J=1}^{\infty}}-2\mathfrak{Re}\left[i\left(\vec{l}^{T}\times\vec{l}^{T*}\right)_{3}\left(\left\langle \left\Vert \hat{E}_{J}^{T}\right\Vert \right\rangle \left\langle \left\Vert \hat{M}_{J}^{T}\right\Vert \right\rangle ^{*}\right)+i\left(\vec{l}^{T'}\times\vec{l}^{T'*}\right)_{3}\left(\left\langle \left\Vert \hat{E}_{J}^{T'}\right\Vert \right\rangle \left\langle \left\Vert \hat{M}_{J}^{T'}\right\Vert \right\rangle ^{*}\right)\right.\right.\right.\\
\left.\left.\left.+i\left(\vec{l}^{T}\times\vec{l}^{T'*}\right)_{3}\left(\left\langle \left\Vert \hat{E}_{J}^{T}\right\Vert \right\rangle \left\langle \left\Vert \hat{M}_{J}^{T'}\right\Vert \right\rangle ^{*}+\left\langle \left\Vert \hat{M}_{J}^{T}\right\Vert \right\rangle \left\langle \left\Vert \hat{E}_{J}^{T'}\right\Vert \right\rangle ^{*}\right)\right]\vphantom{\left(\vec{J}^{2}\right)^{2}}\right]\right\} .\label{eq: rate with l*l}
\end{multline}

To find $\Theta$ (Eq.~\eqref{eq:Theta} in the main text), we need to calculate the different lepton traces $\sum_{\mbox{lepton spins}}l_{\mu\nu}l_{\rho\sigma}^{*}$ from the lepton matrix elements.
These are the coefficients of the multipole expansion and therefore are essential for any specific calculation.
Firstly, we will use the $\gamma_{0}$ conjugate characteristics:
\begin{subequations}
\begin{align}
\gamma_{0}^{2} &=1,\\
\gamma_{0}\sigma_{\mu\nu}\gamma_{0} &=\sigma_{\mu\nu}^{+},\\
\gamma_{0}\gamma_{5}\gamma_{0} &=-\gamma_{5}^{+}=-\gamma_{5},
\end{align}
and the fact that for any two spinors $l_{1}$ and $l_{2}$, and any $4\times4$ matrix $\Gamma$,
\begin{align}
\left(\bar{l}_{1}\Gamma l_{2}\right)^{*} & 
=\bar{l}_{2}\left(\gamma_{0}\Gamma^{+}\gamma_{0}\right)l_{1}, 
\end{align}
to find that the Hermitian conjugate of $l_{\mu\nu}$ is
$l_{\mu\nu}^{*}=\frac{1}{\Omega}\bar{l}\left(\vec{k}\right)\left(C_{T}^{*}+C_{T}^{'*}\gamma_{5}\right)\sigma_{\mu\nu} l^{'}\left(\vec{k^{'}}\right)$.
Secondly, using
\begin{align}
\mbox{Tr}\left(\gamma_{\mu}\right) &= 0,\\
\gamma_{5}^{2} &= 1,\\
\left\{ \gamma_{5},\gamma_{\mu}\right\} &= 0,
\end{align}
one can show that a trace of any product of an odd number of $\gamma_{\mu}$ is zero, and so is a trace of $\gamma_{5}$ times a product of an odd number of $\gamma_{\mu}$. This, along with the following identities~\cite{Itzykson:1980:QTF}:
\begin{eqnarray}
\gamma_{5}\sigma_{\mu\nu} & = & \frac{i}{2}\epsilon_{\mu\nu\rho\sigma}\sigma^{\rho\sigma}\mbox{,} \\
\mbox{Tr}\left(\gamma_{\mu}\gamma_{\nu}\gamma_{\rho}\gamma_{\sigma}\right) & = & 4\left(g_{\mu\nu}g_{\rho\sigma}-g_{\mu\rho}g_{\nu\sigma}+g_{\mu\sigma}g_{\nu\rho}\right)\mbox{,}
\end{eqnarray}
\end{subequations}
and the invariant of the trace under cyclic permutations, leads to
the features below: 
\begin{subequations}
\begin{eqnarray}
\mbox{Tr}\left(\gamma_{a}\gamma_{b}\gamma_{c}\gamma_{d}\gamma_{e}\gamma_{f}\right) & = & 4\left(g_{ab}g_{cd}g_{ef}-g_{ac}g_{bd}g_{ef}+g_{ad}g_{bc}g_{ef}-g_{ab}g_{ce}g_{df}+g_{ac}g_{be}g_{df}\right.\nonumber \\
 &  & \left.-g_{ae}g_{bc}g_{df}+g_{ab}g_{cf}g_{de}-g_{ad}g_{be}g_{cf}+g_{ae}g_{bd}g_{cf}-g_{ac}g_{bf}g_{de}\right.\nonumber \\
 &  & \left.+g_{ad}g_{bf}g_{ce}-g_{ae}g_{bf}g_{cd}+g_{af}g_{bc}g_{de}-g_{af}g_{bd}g_{ce}+g_{af}g_{be}g_{cd}\right)\mbox{,}\\
\mbox{Tr}\left(\sigma_{ef}\gamma_{a}\sigma_{bc}\gamma_{d}\right) & = & 4\left(g_{ab}g_{ce}g_{df}-g_{ab}g_{cf}g_{de}+g_{ac}g_{bf}g_{de}-g_{ac}g_{be}g_{df}+g_{ad}g_{be}g_{cf}\right.\nonumber \\
 &  & \left.-g_{ad}g_{bf}g_{ce}+g_{ae}g_{bf}g_{cd}-g_{ae}g_{bd}g_{cf}+g_{af}g_{bd}g_{ce}-g_{af}g_{be}g_{cd}\right)\mbox{,}\\
\mbox{Tr}\left(\gamma_{5}\sigma_{ef}\gamma_{a}\sigma_{bc}\gamma_{d}\right) & = & 4i\epsilon_{efgh}\left(g_{ab}g_{c}^{g}g_{d}^{h}+g_{ac}g_{d}^{g}g_{b}^{h}+g_{ad}g_{b}^{g}g_{c}^{h}+g_{bd}g_{c}^{g}g_{a}^{h}+g_{cd}g_{a}^{g}g_{b}^{h}\right)\mbox{.}
\end{eqnarray}
\end{subequations}
Thirdly, we notice that a sum over Dirac spinors of mass $m$, momentum $k_{\beta}$ and potential energy $E$, is
$\sum_{\mbox{lepton spins}}u\left(\vec{k}\right)\bar{u}\left(\vec{k}\right)=\frac{\gamma_{\beta}k^{\beta}+m}{2E}$.
While for a massless spinor, as a neutrino or an anti-neutrino, with momentum $\nu_{\alpha}$ and energy $\nu=\nu_{0}$, the sum is
$\sum_{\mbox{lepton spins}}u\left(\vec{\nu}\right)\bar{u}\left(\vec{\nu}\right)=\frac{\gamma_{\alpha}\nu^{\alpha}}{2\nu}$.
All these facts, along with the commutator $\left[\gamma_{5},\sigma^{\mu\nu}\right]=0$,
allow us to calculate generic lepton traces for basic semileptonic weak nuclear processes involving a neutrino or an antineutrino (as neutrino/antineutrino reaction, charged lepton capture, and $\beta$-decay):
\begin{multline}
\frac{\Omega^{2}}{2}\sum_{\mbox{lepton spins}}l_{\mu\nu}l_{\rho\sigma}^{*}
=\frac{1}{2}\mbox{Tr}\left[\sigma_{\mu\nu}\left(C_{T}-C_{T}^{'}\gamma_{5}\right)\left(\frac{\gamma_{\alpha}\nu^{\alpha}}{2\nu}\right)\left(C_{T}^{*}+C_{T}^{'*}\gamma_{5}\right)\sigma_{\rho\sigma}\left(\frac{\gamma_{\beta}k^{\beta}+m}{2E}\right)\right]\\
=\frac{\left|C_{T}\right|^{2}+\left|C_{T}^{'}\right|^{2}}{2}\left[\left(g_{\rho\mu}g_{\sigma\nu}-g_{\rho\nu}g_{\sigma\mu}\right)\frac{\nu^{\alpha}k_{\alpha}}{\nu E}+g_{\sigma\mu}\left(\frac{\nu_{\rho}k_{\nu}}{\nu E}+\frac{\nu_{\nu}k_{\rho}}{\nu E}\right)\right.\\
\left.-g_{\sigma\nu}\left(\frac{\nu_{\rho}k_{\mu}}{\nu E}+\frac{\nu_{\mu}k_{\rho}}{\nu E}\right)+g_{\rho\nu}\left(\frac{\nu_{\sigma}k_{\mu}}{\nu E}+\frac{\nu_{\mu}k_{\sigma}}{\nu E}\right)-g_{\rho\mu}\left(\frac{\nu_{\sigma}k_{\nu}}{\nu E}+\frac{\nu_{\nu}k_{\sigma}}{\nu E}\right)\right]\\
+i\frac{C_{T}C_{T}^{'*}+C_{T}^{'}C_{T}^{*}}{2}\epsilon_{\mu\nu\gamma\delta}\left[g_{\rho}^{\gamma}g_{\sigma}^{\delta}\frac{\nu^{\alpha}k_{\alpha}}{\nu E}+g_{\sigma}^{\gamma}\left(\frac{\nu_{\rho}k^{\delta}}{\nu E}+\frac{\nu^{\delta}k_{\rho}}{\nu E}\right)+g_{\rho}^{\delta}\left(\frac{\nu_{\sigma}k^{\gamma}}{\nu E}+\frac{\nu^{\gamma}k_{\sigma}}{\nu E}\right)\right]\mbox{.}
\end{multline}
Using the definitions of $l_{i}^{T^{\left('\right)}}$ from Eq.~\eqref{eq: tensor current defenitions} in the main text, we find the required tensor lepton traces (notice that $\hat{q}$
is the 3$^{\text{rd}}$ direction $\hat{z}\equiv\frac{\vec{q}}{\left|q\right|}$):
\begin{subequations}
\begin{align}
\frac{\Omega^{2}}{2}\sum_{\mbox{lepton spins}}l_{3}^{T}l_{3}^{T*} & =\left(\left|C_{T}\right|^{2}+\left|C_{T}^{'}\right|^{2}\right)\left[1+\vec{\beta}\cdot\hat{\nu}-2\left(\hat{\nu}\cdot\hat{q}\right)\left(\vec{\beta}\cdot\hat{q}\right)\right], \\
\frac{\Omega^{2}}{2}\sum_{\mbox{lepton spins}}l_{3}^{T'}l_{3}^{T'*} & =\left(\left|C_{T}\right|^{2}+\left|C_{T}^{'}\right|^{2}\right)\left[1+\vec{\beta}\cdot\hat{\nu}-2\left(\hat{\nu}\cdot\hat{q}\right)\left(\vec{\beta}\cdot\hat{q}\right)\right], \\
\frac{\Omega^{2}}{2}\sum l_{3}^{T}l_{3}^{T'*} & =\left(C_{T}C_{T}^{'*}+C_{T}^{'}C_{T}^{*}\right)\left[1+\vec{\beta}\cdot\hat{\nu}-2\left(\hat{\nu}\cdot\hat{q}\right)\left(\vec{\beta}\cdot\hat{q}\right)\right], \\
\frac{1}{2}\frac{\Omega^{2}}{2}\sum_{\mbox{lepton spins}}\left(\vec{l}^{T}\cdot\vec{l}^{T*}-l_{3}^{T}l_{3}^{T*}\right) & =\left(\left|C_{T}\right|^{2}+\left|C_{T}^{'}\right|^{2}\right)\left[1+\left(\hat{\nu}\cdot\hat{q}\right)\left(\vec{\beta}\cdot\hat{q}\right)\right],\\
\frac{1}{2}\frac{\Omega^{2}}{2}\sum_{\mbox{lepton spins}}\left(\vec{l}^{T'}\cdot\vec{l}^{T'*}-l_{3}^{T'}l_{3}^{T'*}\right) & =\left(\left|C_{T}\right|^{2}+\left|C_{T}^{'}\right|^{2}\right)\left[1+\left(\hat{\nu}\cdot\hat{q}\right)\left(\vec{\beta}\cdot\hat{q}\right)\right], \\
\frac{1}{2}\frac{\Omega^{2}}{2}\sum_{\mbox{lepton spins}}\left(\vec{l}^{T}\cdot\vec{l}^{T'*}-l_{3}^{T}l_{3}^{T'*}\right) & =\left(C_{T}C_{T}^{'*}+C_{T}^{'}C_{T}^{*}\right)\left[1+\left(\hat{\nu}\cdot\hat{q}\right)\left(\vec{\beta}\cdot\hat{q}\right)\right], \\
-\frac{i}{2}\frac{\Omega^{2}}{2}\sum_{\mbox{lepton spins}}\left(\vec{l}^{T}\times\vec{l}^{T*}\right)_{3} & =\left(C_{T}C_{T}^{'*}+C_{T}^{'}C_{T}^{*}\right)\hat{q}\cdot\left(\hat{\nu}+\vec{\beta}\right),\\
-\frac{i}{2}\frac{\Omega^{2}}{2}\sum_{\mbox{lepton spins}}\left(\vec{l}^{T'}\times\vec{l}^{T'*}\right)_{3} & =\left(C_{T}C_{T}^{'*}+C_{T}^{'}C_{T}^{*}\right)\hat{q}\cdot\left(\hat{\nu}+\vec{\beta}\right), \\
-\frac{i}{2}\frac{\Omega^{2}}{2}\sum_{\mbox{lepton spins}}\left(\vec{l}\times\vec{l}^{T'*}\right)_{3} & =\left(\left|C_{T}\right|^{2}+\left|C_{T}^{'}\right|^{2}\right)\hat{q}\cdot\left(\hat{\nu}+\vec{\beta}\right)\mbox{.}
\end{align}
\end{subequations}
These tensor lepton traces are suitable for all semileptonic weak nuclear processes.
Note that:
\begin{subequations}
\label{eq: q identities}
\begin{align}
\hat{\nu}\cdot\hat{q}
&=\frac{E}{q}\vec{\beta}\cdot\hat{\nu}+\frac{E_0-E}{q},\\
\hat{q}\cdot\left(\hat{\nu}+\vec{\beta}\right)
&=\frac{E_0}{q}\left(1+\vec{\beta}\cdot\hat{\nu}\right)-\frac{m_{e}}{E}\frac{m_e}{q},\\
\left(\hat{\nu}\cdot\hat{q}\right)\left(\vec{\beta}\cdot\hat{q}\right)
&=\vec{\beta}\cdot\hat{\nu}+\frac{E\left(E_0-E\right)}{q^{2}}\left[\beta^{2}-\left(\vec{\beta}\cdot\hat{\nu}\right)^{2}\right]. 
\end{align}
\end{subequations}

The symmetry coefficients we used here to obtain the lepton traces terms are $C_{\text{sym}}\sim g_{\text{sym}}\cdot\epsilon_{\text{sym}}$ ($\text{sym}\in\left\{ S,P,V,A,T\right\}$). These are nucleon-level coefficients. Since the quark-level matrix elements already contain the $g_{\text{sym}}$ form factors, when coming to use the lepton traces with the currents quark-level matrix elements we discussed in Sec.~\ref{sec:BSM-multipole-operators}, there is a need to make a small adjustment. A simple replacement of the obtained lepton coefficients $C_{\text{sym}}^{\left('\right)}$, with the adjust coefficients $\frac{C_{\text{sym}}^{\left('\right)}}{g_{\text{sym}}}$, will serve our needs.

From Eq.~\eqref{eq: rate with l*l}, we get a general expression for the $\beta$-decay rate of tensor symmetry transitions between any two nuclear states: 
\begin{multline}
\Theta\left(q,\vec{\beta}\cdot\hat{\nu}\right)
=\sum_{J=0}^{\infty}\left(1+\vec{\beta}\cdot\hat{\nu}-2\left(\hat{\nu}\cdot\hat{q}\right)\left(\vec{\beta}\cdot\hat{q}\right)\right)\left[\frac{\left|C_{T}\right|^{2}+\left|C_{T}^{'}\right|^{2}}{g^2_T}\left(\left|\left\langle \left\Vert \hat{L}_{J}^{T}\right\Vert \right\rangle \right|^{2}+\left|\left\langle \left\Vert \hat{L}_{J}^{T'}\right\Vert \right\rangle \right|^{2}\right)\right.\\
\left.\left.+2\mathfrak{Re}\left(\frac{C_{T}C_{T}^{'*}+C_{T}^{'}C_{T}^{*}}{g^2_T}\left\langle \left\Vert \hat{L}_{J}^{T}\right\Vert \right\rangle \left\langle \left\Vert \hat{L}_{J}^{T'}\right\Vert \right\rangle ^{*}\right)\vphantom{\left(\vec{J}^{2}\right)^{2}}\right]\right.\\
+\sum_{J=1}^{\infty}\left.\left\{\left(1+\left(\hat{\nu}\cdot\hat{q}\right)\left(\vec{\beta}\cdot\hat{q}\right)\right)\left[\frac{\left|C_{T}\right|^{2}+\left|C_{T}^{'}\right|^{2}}{g^2_T}\right.\right.\right.\\
\times
\left(\left|\left\langle \left\Vert \hat{E}_{J}^{T}\right\Vert \right\rangle \right|^{2}+\left|\left\langle \left\Vert \hat{M}_{J}^{T}\right\Vert \right\rangle \right|^{2}+\left|\left\langle \left\Vert \hat{E}_{J}^{T'}\right\Vert \right\rangle \right|^{2}+\left|\left\langle \left\Vert \hat{M}_{J}^{T'}\right\Vert \right\rangle \right|^{2}\right)\hphantom{hhhhhh}\\
\left.\left.\left.+2\mathfrak{Re}\left(\frac{C_{T}C_{T}^{'*}+C_{T}^{'}C_{T}^{*}}{g^2_T}\left(\left\langle \left\Vert \hat{E}_{J}^{T}\right\Vert \right\rangle \left\langle \left\Vert \hat{E}_{J}^{T'}\right\Vert \right\rangle ^{*}+\left\langle \left\Vert \hat{M}_{J}^{T}\right\Vert \right\rangle \left\langle \left\Vert \hat{M}_{J}^{T'}\right\Vert \right\rangle ^{*}\right)\right)\right]\right.\right.\\
\left.\left.+\hat{q}\cdot\left(\hat{\nu}+\vec{\beta}\right)2\mathfrak{Re}\left[\frac{C_{T}C_{T}^{'*}+C_{T}^{'}C_{T}^{*}}{g^2_T}\left(\left\langle \left\Vert \hat{E}_{J}^{T}\right\Vert \right\rangle \left\langle \left\Vert \hat{M}_{J}^{T}\right\Vert \right\rangle ^{*}+\left\langle \left\Vert \hat{E}_{J}^{T'}\right\Vert \right\rangle \left\langle \left\Vert \hat{M}_{J}^{T'}\right\Vert \right\rangle ^{*}\right)\right.\right.\right.\\
\left.\left.+\frac{\left|C_{T}\right|^{2}+\left|C_{T}^{'}\right|^{2}}{g^2_T}\left(\left\langle \left\Vert \hat{E}_{J}^{T}\right\Vert \right\rangle \left\langle \left\Vert \hat{M}_{J}^{T'}\right\Vert \right\rangle ^{*}+\left\langle \left\Vert \hat{M}_{J}^{T}\right\Vert \right\rangle \left\langle \left\Vert \hat{E}_{J}^{T'}\right\Vert \right\rangle ^{*}\right)\right]\vphantom{\sum_{J=0}^{\infty}}\right\} \mbox{,}
\end{multline}
and after taking into account also the parity selection rules, as well as the the relation $\hat{E}_{J}=\sqrt{\frac{J+1}{J}}\hat{L}_{J}+\mathcal{O}\left(\left(qR\right)^{J+1}\right)$,
where $\hat{L}_{J}$ is $\mathcal{O}\left(\left(qR\right)^{J-1}\right)$
for $J>0$~\cite{WALECKA1975113}, the general expression is reduced to:
\begin{multline}
\Theta\left(q,\vec{\beta}\cdot\hat{\nu}\right)
=\frac{\left|C_{T}\right|^{2}+\left|C_{T}^{'}\right|^{2}}{g^2_T}
\left\{ \left(1+\vec{\beta}\cdot\hat{\nu}-2\left(\hat{\nu}\cdot\hat{q}\right)\left(\vec{\beta}\cdot\hat{q}\right)\right)\left(\left|\left\langle \left\Vert \hat{L}_{0}^{T}\right\Vert \right\rangle \right|^{2}+\left|\left\langle \left\Vert \hat{L}_{0}^{T'}\right\Vert \right\rangle \right|^{2}\right)\right.\\
+\sum_{J=1}^{\infty}\left[\frac{2J+1}{J}\left(1+\frac{J}{2J+1}\vec{\beta}\cdot\hat{\nu}-\frac{J-1}{2J+1}\left(\hat{\nu}\cdot\hat{q}\right)\left(\vec{\beta}\cdot\hat{q}\right)\right)\left(\left|\left\langle \left\Vert \hat{L}_{J}^{T}\right\Vert \right\rangle \right|^{2}+\left|\left\langle \left\Vert \hat{L}_{J}^{T'}\right\Vert \right\rangle \right|^{2}\right)\right.\\
+\left.\left(1+\left(\hat{\nu}\cdot\hat{q}\right)\left(\vec{\beta}\cdot\hat{q}\right)\right)\left(\left|\left\langle \left\Vert \hat{M}_{J}^{T}\right\Vert \right\rangle \right|^{2}+\left|\left\langle \left\Vert \hat{M}_{J}^{T'}\right\Vert \right\rangle \right|^{2}\right)\right.\hphantom{hhhhhh}\\
\left.\left.+\sqrt{\frac{J+1}{J}}\hat{q}\cdot\left(\hat{\nu}+\vec{\beta}\right)2\mathfrak{Re}\left(\left\langle \left\Vert \hat{L}_{J}^{T}\right\Vert \right\rangle \left\langle \left\Vert \hat{M}_{J}^{T'}\right\Vert \right\rangle ^{*}+\left\langle \left\Vert \hat{M}_{J}^{T}\right\Vert \right\rangle \left\langle \left\Vert \hat{L}_{J}^{T'}\right\Vert \right\rangle ^{*}\right)\right]\right\} \text{.}
\end{multline}
Finally, using the connection from Eq.~\eqref{eq:T-A multipole operators} in the main text, we find the leading order BSM expression:
\begin{multline}
\Theta\left(q,\vec{\beta}\cdot\hat{\nu}\right)
=\frac{\left|C_{T}\right|^{2}+\left|C_{T}^{'}\right|^{2}}{2g_{A}^{2}}
\left\{ \left(1+\vec{\beta}\cdot\hat{\nu}-2\left(\hat{\nu}\cdot\hat{q}\right)\left(\vec{\beta}\cdot\hat{q}\right)\right)\left|\left\langle \left\Vert \hat{L}_{0}^{A}\right\Vert \right\rangle \right|^{2}\right.\\
+\sum_{J=1}^{\infty}\left[\frac{2J+1}{J}\left(1+\frac{J}{2J+1}\vec{\beta}\cdot\hat{\nu}-\frac{J-1}{2J+1}\left(\hat{\nu}\cdot\hat{q}\right)\left(\vec{\beta}\cdot\hat{q}\right)\right)\left|\left\langle \left\Vert \hat{L}_{J}^{A}\right\Vert \right\rangle \right|^{2}\right.\\
\left.\left.\left(1+\left(\hat{\nu}\cdot\hat{q}\right)\left(\vec{\beta}\cdot\hat{q}\right)\right)\left|\left\langle \left\Vert \hat{M}_{J}^{A}\right\Vert \right\rangle \right|^{2}\right]\right\} \text{.}
\end{multline}
This is a general result which holds for any semileptonic nuclear process, including different types of beyond the Standard Model physics. After substituting Eq.~\eqref{eq: q identities}, it yields the tensor terms presented in Sec.~\ref{sec:A-General-Expression}.

\section{Tensor nuclear currents\label{sec:Appendix Tensor-Nuclear-Current}}

Since the expected signatures of BSM physics is small enough, we will
neglect the two-body (and above) currents, which leads to a systematic
uncertainty of $\epsilon_{\text{EFT}}\approx0.3$ in the nuclear model.
We would like to construct the tensor nuclear current operator, 
\begin{eqnarray}
\hat{\mathcal{J}}_{\mu\nu}\left(\vec{r}\right) & = & \frac{1}{2}\bar{\phi}\left(\vec{r}\right)\sigma_{\mu\nu}\phi'\left(\vec{r}\right)\mbox{.}
\end{eqnarray}
In the traditional nuclear physics picture, the electroweak current
is constructed from the properties of free nucleons, and with this
approach, the general form of the single-nucleon matrix element of
the tensor part of the charge changing weak current is~\cite{Cirigliano:2013xha}:
\begin{multline}
\left\langle \vec{p}',\sigma',\rho'\left|\hat{\mathcal{J}}_{\mu\nu}\right|\vec{p},\sigma,\rho\right\rangle =\frac{1}{\Omega}\bar{u}\left(\vec{p'},\sigma'\right)\eta_{\rho'}^{+}\frac{1}{2}\left[g_{T}\left(q^{2}\right)\sigma_{\mu\nu}+g_{T}^{\left(1\right)}\left(q^{2}\right)\left(q_{\mu}\gamma_{\nu}-q_{\nu}\gamma_{\mu}\right)\right.\\
\left.+g_{T}^{\left(2\right)}\left(q^{2}\right)\left(q_{\mu}P_{\nu}-q_{\nu}P_{\mu}\right)+g_{T}^{\left(3\right)}\left(q^{2}\right)\left(\gamma_{\mu}\cancel{q}\gamma_{\nu}-\gamma_{\nu}\cancel{q}\gamma_{\mu}\right)\right]\tau_{\pm}\eta_{\rho}u\left(\vec{p},\sigma\right)\mbox{.}
\end{multline}

After substituting the explicit form of Dirac spinors,
$u\left(\vec{p},\sigma\right)=\sqrt{\frac{E_{N}+m_{N}}{2E_{N}}}\left(\begin{array}{c}
1\\
\frac{\vec{\sigma}\cdot\vec{p}}{E_{N}+m_{N}}
\end{array}\right)\chi_{\sigma}$ (we use the convention $\bar{u}=u^{+}\gamma_{0}$, so that $u^{+}u=1$),
we make a non-relativistic expansion, expanding the matrix element consistently in powers of $\epsilon_{\text{NR}}\sim\frac{P_{\text{fermi}}}{m_{N}}\approx0.2$, as momenta are assumed here up to few hundred $\mbox{MeV}$'s.
For any tensor $T_{\mu\nu}$ one can write the expansion as: 
\begin{eqnarray}
\bar{u'}\left(\vec{p}',\sigma'\right)T_{\mu\nu}u\left(p,\sigma\right) & = & \chi_{\sigma'}^{+}\left(1,\frac{1}{2m_{N}}\vec{\sigma}\cdot\vec{p'}\right)\gamma_{0}T_{\mu\nu}\left(\begin{array}{c}
1\\
\frac{1}{2m_{N}}\vec{\sigma}\cdot\vec{p}
\end{array}\right)\chi_{\sigma}+\mathcal{O}\left(\epsilon_{\text{NR}}^{2}\right)\mbox{.}
\end{eqnarray}
Using the Dirac representation of the gamma matrices: 
\begin{subequations}
\begin{eqnarray}
\gamma^{0} & = & \left(\begin{array}{cc}
1 & 0\\
0 & -1
\end{array}\right), \\
\vec{\gamma} & = & \gamma^{0}\left(\begin{array}{cc}
0 & \vec{\sigma}\\
\vec{\sigma} & 0
\end{array}\right), \\
\left\{ \gamma_{\mu},\gamma_{\nu}\right\}  & = & 2g_{\mu\nu},
\end{eqnarray}
with the metric $g_{\mu\nu}=\left(\begin{array}{cccc}
1 & 0 & 0 & 0\\
0 & -1 & 0 & 0\\
0 & 0 & -1 & 0\\
0 & 0 & 0 & -1
\end{array}\right)$,
and the above identities for Pauli matrices $\vec{\sigma}$ and
Levi-Civita symbol $\epsilon_{ijk}$: 
\begin{eqnarray}
\vec{\sigma}^{+} & = & \vec{\sigma}, \\
\sigma_{i}\sigma_{j} & = & \delta_{ij}+i\epsilon_{ijk}\sigma_{k}, \\
\epsilon_{ijk}\epsilon_{imn} & = & \delta_{jm}\delta_{kn}-\delta_{jn}\delta_{km}, \\
\epsilon_{ijk}\epsilon_{ijl} & = & 2\delta_{kl}\mbox{,}
\end{eqnarray}
\end{subequations}
one can expand the needed matrix elements:
\begin{subequations}
\begin{eqnarray}
\bar{u'}\left(p',\sigma'\right)\sigma_{ij}u\left(p,\sigma\right) & = & \chi_{\sigma'}^{+}\left(\epsilon_{ijk}\sigma_{k}\right)\chi_{\sigma}+\mathcal{O}\left(\epsilon_{\text{NR}}^{2}\right), \\
\bar{u'}\left(p',\sigma'\right)\frac{i}{2}\left(\gamma_{0}\vec{\gamma}-\vec{\gamma}\gamma_{0}\right)u\left(p,\sigma\right) & = & -\frac{i}{2m_{N}}\chi_{\sigma'}^{+}\left(\vec{q}-i\vec{\sigma}\times\vec{P}\right)\chi_{\sigma}+\mathcal{O}\left(\epsilon_{\text{NR}}^{2}\right), \\
\bar{u'}\left(p',\sigma'\right)\left(q_{i}\gamma_{j}-q_{j}\gamma_{i}\right)u\left(p,\sigma\right) & = & \frac{1}{2m_{N}}\chi_{\sigma'}^{+}\left(P_{j}q_{i}-P_{i}q_{j}\right.\nonumber \\
 &  & \left.+i\epsilon_{jkl}\sigma_{l}q_{i}q_{k}-i\epsilon_{ikl}\sigma_{l}q_{j}q_{k}\right)\chi_{\sigma}+\mathcal{O}\left(\epsilon_{\text{NR}}^{2}\right), \\
\bar{u'}\left(p',\sigma'\right)\left(q_{0}\vec{\gamma}-\vec{q}\gamma_{0}\right)u\left(p,\sigma\right) & = & \chi_{\sigma'}^{+}\left[-\vec{q}+\frac{q_{0}}{2m_{N}}\left(\vec{P}-i\vec{\sigma}\times\vec{q}\right)\right]\chi_{\sigma}+\mathcal{O}\left(\epsilon_{\text{NR}}^{2}\right), \\
\bar{u'}\left(p',\sigma'\right)\left(q_{\mu}P_{\nu}-q_{\nu}P_{\mu}\right)u\left(p,\sigma\right) & = & \chi_{\sigma'}^{+}\left(q_{\mu}P_{\nu}-q_{\nu}P_{\mu}\right)\chi_{\sigma}+\mathcal{O}\left(\epsilon_{\text{NR}}^{2}\right), \\
\bar{u'}\left(p',\sigma'\right)\left(\gamma_{i}\cancel{q}\gamma_{j}-\gamma_{j}\cancel{q}\gamma_{i}\right)u\left(p,\sigma\right) & = & 2i\chi_{\sigma'}^{+}\left(q_{0}\epsilon_{ijk}\sigma_{k}-\frac{1}{2m_{N}}\epsilon_{ijk}q_{k}\sigma_{l}P_{l}\right)\chi_{\sigma}+\mathcal{O}\left(\epsilon_{\text{NR}}^{2}\right), \\
\bar{u'}\left(p',\sigma'\right)\left(\gamma_{0}\cancel{q}\vec{\gamma}-\vec{\gamma}\cancel{q}\gamma_{0}\right)u\left(p,\sigma\right) & = & 2i\chi_{\sigma'}^{+}\left(\vec{\sigma}\times\vec{q}\right)\chi_{\sigma}+\mathcal{O}\left(\epsilon_{\text{NR}}^{2}\right)\mbox{.}
\end{eqnarray}
\end{subequations}

Finally, we use the definitions of the spatial and spatial-temporal parts of the tensor current from Eq.~\eqref{eq: tensor current defenitions} in the main text,
to find the following expansions for the matrix elements of the spatial-temporal and spatial vector-like parts of the tensor current:
\begin{subequations}
\label{eq: hadronic tensor currents}
\begin{align}
\left\langle \vec{p}',\sigma',\rho'\left|\vec{\mathcal{J}}^{T}\right|\vec{p},\sigma,\rho\right\rangle  & =-\frac{i}{\sqrt{2}}\frac{1}{\Omega}\chi_{\sigma'}^{+}\eta_{\rho'}^{+}\left\{ \left(g_{T}+2iq_{0}g_{T}^{\left(3\right)}\right)\vec{\sigma}+g_{T}^{\left(2\right)}\left(\vec{q}\times\vec{P}\right)\right.\nonumber \\
 & \left.+\frac{1}{2m_{N}}\left[g_{T}^{\left(1\right)}\left(\vec{q}\times\vec{P}+i\left(\vec{q}\cdot\vec{\sigma}\right)\vec{q}-iq^{2}\vec{\sigma}\right)-2ig_{T}^{\left(3\right)}\left(\vec{P}\cdot\vec{\sigma}\right)\vec{q}\right]\right\} \tau_{\pm}\eta_{\rho}\chi_{\sigma}\nonumber\\
 &+\mathcal{O}\left(\epsilon_{\text{NR}}^{2}\right),\\
\left\langle \vec{p}',\sigma',\rho'\left|\vec{\mathcal{J}}^{T'}\right|\vec{p},\sigma,\rho\right\rangle  & =\frac{1}{\sqrt{2}}\frac{1}{\Omega}\chi_{\sigma'}^{+}\eta_{\rho'}^{+}\left\{ -g_{T}^{\left(1\right)}\vec{q}+g_{T}^{\left(2\right)}\left(q_{0}\vec{P}-\vec{q}P_{0}\right)+2ig_{T}^{\left(3\right)}\left(\vec{\sigma}\times\vec{q}\right)\vphantom{\frac{1}{2_{N}}}\right.\nonumber \\
 & \left.+\frac{1}{2m_{N}}\left[q_{0}g_{T}^{\left(1\right)}\left(\vec{P}-i\vec{\sigma}\times\vec{q}\right)-ig_{T}\left(\vec{q}-i\vec{\sigma}\times\vec{P}\right)\right]\right\} \tau_{\pm}\eta_{\rho}\chi_{\sigma}+\mathcal{O}\left(\epsilon_{\text{NR}}^{2}\right)\text{.}
\end{align}
\end{subequations}

In order to find the multipole operators, we need to extract the current densities from the currents above.
For that, we proceed as follows: first, we use the definition of the (second quantization)
$\mathcal{J}\left(\vec{r}\right)$ current matrix element as a sum over first
quantization currents $\hat{\mathcal{J}}^{\left(1\right)}$: 
\begin{align}
\left\langle \vec{p}',\sigma',\rho'\left|\mathcal{J}\left(\vec{r}\right)\right|\vec{p},\sigma,\rho\right\rangle  & =\int d^{3}x\phi_{\vec{p}'\sigma'\rho'}^{+}\left(\vec{x}\right)\left[\hat{\mathcal{J}}^{\left(1\right)}\left(\vec{x}\right)\delta^{\left(3\right)}\left(\vec{r}-\vec{x}\right)\right]\phi_{\vec{p}\sigma\rho}\left(\vec{x}\right)\mbox{.}
\end{align}
Evaluated at $\vec{r}=0$, we find out that $\left\langle \vec{p}',\sigma',\rho'\left|\mathcal{J}\left(0\right)\right|\vec{p},\sigma,\rho\right\rangle =\phi_{\vec{p}'\sigma'\rho'}^{+}\left(0\right)\hat{\mathcal{J}}^{\left(1\right)}\left(0\right)\phi_{\vec{p}\sigma\rho}\left(0\right)$,
what permits the identification of the nuclear density operators in
first quantization (from Eq.~\eqref{eq: hadronic tensor currents} which is also evaluated at $\vec{r}=0$):
\begin{subequations}
\begin{align}
\hat{\mathcal{J}}^{T\left(1\right)}\left(0\right) & =-\frac{i}{\sqrt{2}}\left\{ \left(g_{T}+2iE_0 g_{T}^{\left(3\right)}\right)\vec{\sigma}+g_{T}^{\left(2\right)}\left(\vec{q}\times\vec{P}\right)\vphantom{\frac{1}{2_{N}}}\right.\nonumber \\
 & \left.+\frac{1}{2m_{N}}\left[-2ig_{T}^{\left(3\right)}\left(\vec{P}\cdot\vec{\sigma}\right)\vec{q}+g_{T}^{\left(1\right)}\left(\vec{q}\times\vec{P}+i\left(\vec{q}\cdot\vec{\sigma}\right)\vec{q}-iq^{2}\vec{\sigma}\right)\right]\right\} \tau_{\pm}+\mathcal{O}\left(\epsilon_{\text{NR}}^{2}\right), \\
\hat{\mathcal{J}}^{T'\left(1\right)}\left(0\right) & =\frac{1}{\sqrt{2}}\left\{ -\frac{g_{T}}{2m_{N}}\left(\vec{\sigma}\times\vec{P}\right)-\left(g_{T}^{\left(1\right)}+i\frac{g_{T}}{2m_{N}}\right)\vec{q}+2ig_{T}^{\left(3\right)}\left(\vec{\sigma}\times\vec{q}\right)\vphantom{\frac{1}{2_{N}}}\right.\nonumber \\
 & \left.+g_{T}^{\left(1\right)}\frac{E_0}{2m_{N}}\left(\vec{P}-i\vec{\sigma}\times\vec{q}\right)+g_{T}^{\left(2\right)}\left(E_0\vec{P}-\vec{q}P_{0}\right)\right\} \tau_{\pm}+\mathcal{O}\left(\epsilon_{\text{NR}}^{2}\right)\mbox{.}
\end{align}
\end{subequations}

Now, using the current density operator in the first quantization,
\begin{eqnarray}
\mathcal{J}\left(\vec{r}\right) & = & \sum_{j=1}^{A}\hat{\mathcal{J}}^{\left(1\right)}\left(j\right)\delta^{\left(3\right)}\left(\vec{r}-\vec{r}_{j}\right)\text{,}
\end{eqnarray}
and under the assumption of the first quantization, that there is no dependency
on the location, so $\hat{\mathcal{J}}^{\left(1\right)}\left(j\right)=\hat{\mathcal{J}}^{\left(1\right)}\left(0\right)\left(j\right)$,
one gets the following tensor current densities:
\begin{subequations}
\begin{align}
\vec{\mathcal{J}}^{T}\left(\vec{r}\right) & =-\frac{i}{\sqrt{2}}\sum_{j=1}^{A}\left(g_{T}+2iE_0 g_{T}^{\left(3\right)}\right)\vec{\sigma}_{j}\delta^{\left(3\right)}\left(\vec{r}-\vec{r}_{j}\right)\tau_{j}^{\pm}+\mathcal{O}\left(\epsilon_{\text{NR}}^{2}\right), \\
\vec{\mathcal{J}}^{T'}\left(\vec{r}\right) & =\frac{1}{\sqrt{2}}\sum_{j=1}^{A}\left\{ \left(ig_{T}^{\left(1\right)}-\frac{g_{T}}{2m_{N}}\right)\vec{\nabla}\delta^{\left(3\right)}\left(\vec{r}-\vec{r}_{j}\right)-\frac{g_{T}}{2m_{N}}\vec{\sigma}_{j}\times\left\{ \vec{p}_{j},\delta^{\left(3\right)}\left(\vec{r}-\vec{r}_{j}\right)\right\} \vphantom{\frac{1}{2_{N}}}\right.\nonumber \\
 & \left.+\left(2g_{T}^{\left(3\right)}-\frac{E_0}{2m_{N}}g_{T}^{\left(1\right)}\right)\vec{\sigma}_{j}\times\vec{\nabla}\delta^{\left(3\right)}\left(\vec{r}-\vec{r}_{j}\right)\right\} \tau_{j}^{\pm}+\mathcal{O}\left(\epsilon_{\text{NR}}^{2}\right)\text{.}
\end{align}
\end{subequations}
Here we made the operator replacements $\vec{P}\rightarrow\left\{ \vec{p}_{j},\delta^{\left(3\right)}\left(\vec{r}-\vec{r}_{j}\right)\right\} $, and $\vec{q}\rightarrow-i\vec{\nabla}$, the last one based on a
partial integration of Fourier transform of the transition matrix
element of the current, $\int e^{-i\vec{q}\cdot\vec{r}}\left\langle f\left|\mathcal{J}\left(\vec{r}\right)\right|i\right\rangle $,
with localized densities. 

\section{Fierz term and its tensor lepton traces\label{sec:Appendix Fierz-Term}}

To complete the discussion, considering the full probe-nucleus interaction
Hamiltonian, $\mathcal{\hat{H}_{\mbox{w}}}=\hat{H}_{\mbox{w}}^{SM}+\hat{H}_{\mbox{w}}^{BSM}$, results in an interference term, known as Fierz term, involving both SM and BSM currents.
Consider the SM Hamiltonian $\hat{H}_{\mbox{w}}^{SM}=\hat{H}_{\mbox{w}}^{V}+\hat{H}_{\mbox{w}}^{A}$ where $\hat{H}_{\mbox{w}}^{V\left(A\right)} = \int d^{3}r\hat{j}^{V\left(A\right)}_{\mu}\left(\vec{r}\right)\hat{\mathcal{J}}^{V\left(A\right)\mu}\left(\vec{r}\right)$ is its polar (axial)-vector part.
The matrix element of the SM Hamiltonian can be written as~\cite{WALECKA1975113}:
\begin{multline}
\left\langle f\left|\hat{H}_{\mbox{w}}^{V\left(A\right)}\right|i\right\rangle =\sum_{J=0}^{\infty}\sqrt{4\pi\left(2J+1\right)}\left(-i\right)^{J}\left[l_{0}^{V\left(A\right)}\left\langle f\left|\hat{C}_{J0}^{V\left(A\right)}\right|i\right\rangle 
-l_{3}^{V\left(A\right)}\left\langle f\left|\hat{L}_{J0}^{V\left(A\right)}\right|i\right\rangle \right]\\
\left.+\sum_{J=1}^{\infty}\sqrt{2\pi\left(2J+1\right)}\left(-i\right)^{J}\sum_{\lambda=\pm1}l_{\lambda}^{V\left(A\right)}\left\langle f\left|\hat{E}_{J,-\lambda}^{V\left(A\right)}+\lambda\hat{M}_{J,-\lambda}^{V\left(A\right)}\right|i\right\rangle \vphantom{\sum_{J=0}^{\infty}}\right\} \mbox{.}
\label{eq: vector multipole expansion}
\end{multline}
where the superscript $V\left(A\right)$ denotes a multipole operator (Eq.~\eqref{eq: multipole operators} in the main text)
calculated with the polar (axial)-vector nuclear current (described in Appendix~\ref{sec:Appendix-Second-Class_Nuclear-Currents-and}), and
$l^V_{\mu}=\frac{1}{\Omega}\bar{l}'\left(\vec{k}'\right)\gamma_{\mu}\left(C_{V}-C_{V}^{'}\gamma_{5}\right)$ $\times l\left(\vec{k}\right)$ 
($l^A_{\mu}=\frac{1}{\Omega}\bar{l}'\left(\vec{k}'\right)\gamma_{\mu}\left(C_{A}^{'}-C_{A}\gamma_{5}\right) l\left(\vec{k}\right)$).

For the tensor BSM case, $\hat{H}_{\mbox{w}}=\hat{H}_{\mbox{w}}^{SM}+\hat{H}_{\mbox{w}}^{T}$,
the Fierz interference term, involving both SM currents and BSM tensor currents, following from both
the multipole expansion for $V-A$ (Eq.~\ref{eq: vector multipole expansion}) and tensor couplings (Eq.~\eqref{eq: multipole expansion} in the main text), will be:
\begin{multline}
\sum_{M_{i}}\sum_{M_{f}}2\mathfrak{Re}\left(\left\langle f\left|\hat{H}_{\mbox{w}}^{V\left(A\right)}\right|i\right\rangle \left\langle f\left|\hat{H}_{\mbox{w}}^{T}\right|i\right\rangle ^{*}\right)\\
=8\pi\mathfrak{Re}\left\{ \sum_{J=0}^{\infty}\left[l_{3}^{V\left(A\right)}l_{3}^{T*}\left\langle \left\Vert \hat{L}_{J}^{V\left(A\right)}\right\Vert \right\rangle \left\langle \left\Vert \hat{L}_{J}^{T}\right\Vert \right\rangle ^{*}+l_{3}^{V\left(A\right)}l_{3}^{T'*}\left\langle \left\Vert \hat{L}_{J}^{V\left(A\right)}\right\Vert \right\rangle \left\langle \left\Vert \hat{L}_{J}^{T'}\right\Vert \right\rangle ^{*}\right.\right.\\
\left.\hphantom{--\frac{4\pi}{2J_{i}+1}\mathfrak{Re}\left\{ \sum_{J=0}^{\infty}\right.}-l_{0}^{V\left(A\right)}l_{3}^{T*}\left\langle \left\Vert \hat{C}_{J}^{V\left(A\right)}\right\Vert \right\rangle \left\langle \left\Vert \hat{L}_{J}^{T}\right\Vert \right\rangle ^{*}-l_{0}^{V\left(A\right)}l_{3}^{T'*}\left\langle \left\Vert \hat{C}_{J}^{V\left(A\right)}\right\Vert \right\rangle \left\langle \left\Vert \hat{L}_{J}^{T'}\right\Vert \right\rangle ^{*}\right]\\
\hphantom{--\frac{4\pi}{2J_{i}+1}\mathfrak{Re}\left\{ \right.}+\frac{1}{2}\sum_{J=1}^{\infty}\sum_{\lambda=\pm1}\left[l_{\lambda}^{V\left(A\right)}l_{\lambda}^{T*}\left\langle \left\Vert \hat{E}_{J}^{V\left(A\right)}+\lambda\hat{M}_{J}^{V\left(A\right)}\right\Vert \right\rangle \left\langle \left\Vert \hat{E}_{J}^{T}+\lambda\hat{M}_{J}^{T}\right\Vert \right\rangle ^{*}\right.\\
\left.\left.+l_{\lambda}^{V\left(A\right)}l_{\lambda}^{T'*}\left\langle \left\Vert \hat{E}_{J}^{V\left(A\right)}+\lambda\hat{M}_{J}^{V\left(A\right)}\right\Vert \right\rangle \left\langle \left\Vert \hat{E}_{J}^{T'}+\lambda\hat{M}_{J}^{T'}\right\Vert \right\rangle ^{*}\right]\vphantom{\sum_{J=0}^{\infty}}\right\} \mbox{.}
\end{multline}
Using 
\begin{subequations}
\begin{eqnarray}
\sum_{\lambda=\pm1}l_{\lambda}l_{\lambda}^{'*}\left(a+\lambda b\right)\left(a'+\lambda b'\right)^{*} & = & \left(\vec{l}\cdot\vec{l'}^{*}-l_{3}l_{3}^{'*}\right)\left(a\bar{a'}+b\bar{b'}\right)-i\left(\vec{l}\times\vec{l'}^{*}\right)_{3}\left(a\bar{b'}+b\bar{a'}\right), \\
i\left(\vec{l}\times\vec{l'}^{*}\right)_{3} & \in & \mathbb{R}, \\
a\bar{b}+\bar{a}b & = & 2\mathfrak{Re}\left(a\bar{b}\right)=2\mathfrak{Re}\left(\bar{a}b\right),
\end{eqnarray}
\end{subequations}
we get a general result for any semileptonic nuclear Fierz term: 
\begin{multline}
\sum_{M_{i}}\sum_{M_{f}}2\mathfrak{Re}\left(\left\langle f\left|\hat{H}_{\mbox{w}}^{V\left(A\right)}\right|i\right\rangle \left\langle f\left|\hat{H}_{\mbox{w}}^{T}\right|i\right\rangle ^{*}\right)\\
=8\pi\mathfrak{Re}\left\{ \sum_{J=0}^{\infty}
\left[l_{3}^{V\left(A\right)} l_{3}^{T*}
\left\langle \left\Vert \hat{L}_{J}^{V\left(A\right)}\right\Vert \right\rangle 
\left\langle \left\Vert \hat{L}_{J}^{T}\right\Vert \right\rangle ^{*}
+l_{3}^{V\left(A\right)} l_{3}^{T'*}
\left\langle \left\Vert \hat{L}_{J}^{V\left(A\right)}\right\Vert \right\rangle \left\langle \left\Vert \hat{L}_{J}^{T'}\right\Vert \right\rangle ^{*}\right.\right.\\
\left.\left.\hphantom{=\frac{4\pi}{2J_{i}+1}\left\{ \sum_{J=0}^{\infty}hhh\right.}
-l_{0}^{V\left(A\right)} l_{3}^{T*}\left\langle \left\Vert \hat{C}_{J}^{V\left(A\right)}\right\Vert \right\rangle \left\langle \left\Vert \hat{L}_{J}^{T}\right\Vert \right\rangle ^{*}
-l_{0}^{V\left(A\right)} l_{3}^{T'*}\left\langle \left\Vert \hat{C}_{J}^{V\left(A\right)}\right\Vert \right\rangle \left\langle \left\Vert \hat{L}_{J}^{T'}\right\Vert \right\rangle ^{*}\right]\right.\\
+\frac{1}{2}\sum_{J=1}^{\infty}\left.\left[
\left(\vec{l}^{V\left(A\right)}\cdot\vec{l}^{T*}-l_{3}^{V\left(A\right)}\cdot l_{3}^{T*}\right)\left(\left\langle \left\Vert \hat{E}_{J}^{V\left(A\right)}\right\Vert \right\rangle \left\langle \left\Vert \hat{E}_{J}^{T}\right\Vert \right\rangle ^{*}
+\left\langle \left\Vert \hat{M}_{J}^{V\left(A\right)}\right\Vert \right\rangle \left\langle \left\Vert \hat{M}_{J}^{T}\right\Vert \right\rangle ^{*}\right)\right.\right.\\
\left.\left.\hphantom{\left\{\frac{1}{2}\sum_{J=1}^{\infty}\right.}
+\left(\vec{l}^{V\left(A\right)} \cdot\vec{l}^{T'*}-l_{3}^{V\left(A\right)} l_{3}^{T'*}\right)
\left(\left\langle \left\Vert \hat{E}_{J}^{V\left(A\right)}\right\Vert \right\rangle \left\langle \left\Vert \hat{E}_{J}^{T'}\right\Vert \right\rangle ^{*}
+\left\langle \left\Vert \hat{M}_{J}^{V\left(A\right)}\right\Vert \right\rangle \left\langle \left\Vert \hat{M}_{J}^{T'}\right\Vert \right\rangle ^{*}\right)\right.\right.\\
\left.\left.\hphantom{\left\{ \frac{1}{2}\sum_{J=1}^{\infty}\right.}
-i\left(\vec{l}^{V\left(A\right)}\times\vec{l}^{T*}\right)_{3}\left(\left\langle \left\Vert \hat{E}_{J}^{V\left(A\right)}\right\Vert \right\rangle \left\langle \left\Vert \hat{M}_{J}^{T}\right\Vert \right\rangle ^{*}+\left\langle \left\Vert \hat{M}_{J}^{V\left(A\right)}\right\Vert \right\rangle \left\langle \left\Vert \hat{E}_{J}^{T}\right\Vert \right\rangle ^{*}\right)\right.\right.\\
\left.\left.
-i\left(\vec{l}^{V\left(A\right)}\times\vec{l}^{T'*}\right)_{3}\left(\left\langle \left\Vert \hat{E}_{J}^{V\left(A\right)}\right\Vert \right\rangle \left\langle \left\Vert \hat{M}_{J}^{T'}\right\Vert \right\rangle ^{*}+\left\langle \left\Vert \hat{M}_{J}^{V\left(A\right)}\right\Vert \right\rangle \left\langle \left\Vert \hat{E}_{J}^{T'}\right\Vert \right\rangle ^{*}\right)\right]\vphantom{\sum_{J=0}^{\infty}}\right\} \mbox{.}
\end{multline}
Now, to calculate its lepton traces, we find out that for a $\beta^{\mp}$-decay:
\begin{multline}
\frac{\Omega^{2}}{2}\sum_{\mbox{lepton spins}}l_{\mu}^{A}l_{\rho\sigma}^{*}
=\frac{1}{2}\mbox{Tr}\left[\gamma_{\mu}\left(C_{A}^{'}-C_{A}\gamma_{5}\right)\left(\frac{\gamma_{\alpha}\nu^{\alpha}}{2\nu}\right)\left(C_{T}^{*}+C_{T}^{'*}\gamma_{5}\right)\sigma_{\rho\sigma}\left(\frac{\gamma_{\beta}k^{\beta} \pm m}{2E}\right)\right]\\
=\pm i\frac{m}{E}\frac{C_{A}^{'}C_{T}^{*}+C_{A}C_{T}^{'*}}{2}\left(g_{\mu\sigma}\frac{\nu_{\rho}}{\nu}-g_{\mu\rho}\frac{\nu_{\sigma}}{\nu}\right)
\pm \frac{m}{E}\frac{C_{A}C_{T}^{*}+C_{A}^{'}C_{T}^{'*}}{2}\epsilon_{\mu\alpha\rho\sigma}\frac{\nu^{\alpha}}{\nu}\mbox{,}
\end{multline}
so the Fierz lepton traces are:
\begin{subequations}
\begin{eqnarray}
\frac{\Omega^{2}}{2}\sum_{\mbox{lepton spins}}l_{3}^{A}l_{3}^{T*} & = & \mp\frac{i}{\sqrt{2}}\frac{m_{e}}{E}\left(C_{A}C_{T}^{*}+C_{A}^{'}C_{T}^{'*}\right), \\
\frac{\Omega^{2}}{2}\sum_{\mbox{lepton spins}}l_{3}^{A}l_{3}^{T'*} & = & \mp\frac{i}{\sqrt{2}}\frac{m_{e}}{E}\left(C_{A}^{'}C_{T}^{*}+C_{A}C_{T}^{'*}\right), \\
-\frac{\Omega^{2}}{2}\sum_{\mbox{lepton spins}}l_{0}^{A}l_{3}^{T*} & = & \pm\frac{i}{\sqrt{2}}\frac{m_{e}}{E}\left(C_{A}C_{T}^{*}+C_{A}^{'}C_{T}^{'*}\right)\left(\hat{\nu}\cdot\hat{q}\right), \\
-\frac{\Omega^{2}}{2}\sum_{\mbox{lepton spins}}l_{0}^{A}l_{3}^{T'*} & = & \pm\frac{i}{\sqrt{2}}\frac{m_{e}}{E}\left(C_{A}^{'}C_{T}^{*}+C_{A}C_{T}^{'*}\right)\left(\hat{\nu}\cdot\hat{q}\right), \\
\frac{1}{2}\frac{\Omega^{2}}{2}\sum_{\mbox{lepton spins}}\left(\vec{l}^{A}\cdot\vec{l}^{T*}-l_{3}^{A}l_{3}^{T*}\right) & = & \mp\frac{i}{\sqrt{2}}\frac{m_{e}}{E}\left(C_{A}C_{T}^{*}+C_{A}^{'}C_{T}^{'*}\right), \\
\frac{1}{2}\frac{\Omega^{2}}{2}\sum_{\mbox{lepton spins}}\left(\vec{l}^{A}\cdot\vec{l}^{T'*}-l^{A}l_{3}^{T'*}\right) & = & \mp\frac{i}{\sqrt{2}}\frac{m_{e}}{E}\left(C_{A}^{'}C_{T}^{*}+C_{A}C_{T}^{'*}\right), \\
-\frac{i}{2}\frac{\Omega^{2}}{2}\sum_{\mbox{lepton spins}}\left(\vec{l}^{A}\times\vec{l}^{T*}\right)_{3} & = & \mp\frac{i}{\sqrt{2}}\frac{m_{e}}{E}\left(C_{A}^{'}C_{T}^{*}+C_{A}C_{T}^{'*}\right)\left(\hat{\nu}\cdot\hat{q}\right), \\
-\frac{i}{2}\frac{\Omega^{2}}{2}\sum_{\mbox{lepton spins}}\left(\vec{l}^{A}\times\vec{l}^{T'*}\right)_{3} & = & \mp\frac{i}{\sqrt{2}}\frac{m_{e}}{E}\left(C_{A}C_{T}^{*}+C_{A}^{'}C_{T}^{'*}\right)\left(\hat{\nu}\cdot\hat{q}\right)\mbox{,}
\end{eqnarray}
\end{subequations}
where $\pm$ and $\mp$ are for $\beta^{\mp}$-decays. In order to match these lepton traces terms to the quark-level effective theory one-nucleon matrix elements used in Sec.~\ref{sec:BSM-multipole-operators}, we again replace $C_{\text{sym}}^{\left('\right)}$, with the adjust coefficients $\frac{C_{\text{sym}}^{\left('\right)}}{g_{\text{sym}}}$, where $\text{sym}\in\left\{ S,P,V,A,T\right\}$, as we did in Appendix~\ref{sec:Appendix Tensor-Lepton-Traces}.
Now, for the Fierz axial-tensor interference term we get the following
$\beta$-decay rate: 
\begin{multline}
\Theta\left(q,\vec{\beta}\cdot\hat{\nu}\right)
^{AT}=\mp\frac{m_{e}}{E}\\
\times\sqrt{2}\mathfrak{Re}\left\{ \sum_{J=0}^{\infty}i\right.\left[\frac{C_{A}C_{T}^{*}+C_{A}^{'}C_{T}^{'*}}{g_A g_T^*}\left\langle \left\Vert \hat{L}_{J}^{A}\right\Vert \right\rangle \left\langle \left\Vert \hat{L}_{J}^{T}\right\Vert \right\rangle ^{*}+\frac{C_{A}^{'}C_{T}^{*}+C_{A}C_{T}^{'*}}{g_A g_T^*}\left\langle \left\Vert \hat{L}_{J}^{A}\right\Vert \right\rangle \left\langle \left\Vert \hat{L}_{J}^{T'}\right\Vert \right\rangle ^{*}\right.\\
\left.\left.-\left(\hat{\nu}\cdot\hat{q}\right)\frac{C_{A}C_{T}^{*}+C_{A}^{'}C_{T}^{'*}}{g_A g_T^*}\left\langle \left\Vert \hat{C}_{J}^{A}\right\Vert \right\rangle \left\langle \left\Vert \hat{L}_{J}^{T}\right\Vert \right\rangle ^{*}-\left(\hat{\nu}\cdot\hat{q}\right)\frac{C_{A}^{'}C_{T}^{*}+C_{A}C_{T}^{'*}}{g_A g_T^*}\left\langle \left\Vert \hat{C}_{J}^{A}\right\Vert \right\rangle \left\langle \left\Vert \hat{L}_{J}^{T'}\right\Vert \right\rangle ^{*}\vphantom{\frac{i}{\sqrt{2}}}\right]\right.\\
+\sum_{J=1}^{\infty}i\left.\left[\frac{C_{A}C_{T}^{*}+C_{A}^{'}C_{T}^{'*}}{g_A g_T^*}\left(\left\langle \left\Vert \hat{E}_{J}^{A}\right\Vert \right\rangle \left\langle \left\Vert \hat{E}_{J}^{T}\right\Vert \right\rangle ^{*}+\left\langle \left\Vert \hat{M}_{J}^{A}\right\Vert \right\rangle \left\langle \left\Vert \hat{M}_{J}^{T}\right\Vert \right\rangle ^{*}\right)\right.\right.\\
\hphantom{\cdot\left\{ \sum_{J=0}^{\infty}\right.}\left.\left.+\frac{C_{A}^{'}C_{T}^{*}+C_{A}C_{T}^{'*}}{g_A g_T^*}\left(\left\langle \left\Vert \hat{E}_{J}^{A}\right\Vert \right\rangle \left\langle \left\Vert \hat{E}_{J}^{T'}\right\Vert \right\rangle ^{*}+\left\langle \left\Vert \hat{M}_{J}^{A}\right\Vert \right\rangle \left\langle \left\Vert \hat{M}_{J}^{T'}\right\Vert \right\rangle ^{*}\right)\right.\right.\\
\left.\left.\hphantom{\cdot\left\{ \sum_{J=0}^{\infty}\right.}+\left(\hat{\nu}\cdot\hat{q}\right)\frac{C_{A}^{'}C_{T}^{*}+C_{A}C_{T}^{'*}}{g_A g_T^*}\left(\left\langle \left\Vert \hat{E}_{J}^{A}\right\Vert \right\rangle \left\langle \left\Vert \hat{M}_{J}^{T}\right\Vert \right\rangle ^{*}+\left\langle \left\Vert \hat{M}_{J}^{A}\right\Vert \right\rangle \left\langle \left\Vert \hat{E}_{J}^{T}\right\Vert \right\rangle ^{*}\right)\right.\right.\\
\left.\left.+\left(\hat{\nu}\cdot\hat{q}\right)\frac{C_{A}C_{T}^{*}+C_{A}^{'}C_{T}^{'*}}{g_A g_T^*}\left(\left\langle \left\Vert \hat{E}_{J}^{A}\right\Vert \right\rangle \left\langle \left\Vert \hat{M}_{J}^{T'}\right\Vert \right\rangle ^{*}+\left\langle \left\Vert \hat{M}_{J}^{A}\right\Vert \right\rangle \left\langle \left\Vert \hat{E}_{J}^{T'}\right\Vert \right\rangle ^{*}\right)\vphantom{\frac{i}{\sqrt{2}}}\right]\right\} \mbox{.}
\end{multline}
The interference with the polar-vector current will have the same expression, only with the superscript $V$, $g_V$, $C_{V}^{'}$ and $C_{V}$ instead of the superscript $A$, $g_A$, $C_{A}$ and $C_{A}^{'}$, respectively (note the replacement of $^{\left('\right)}$).

We have not made any approximations up to this
stage, and the results for the lepton traces are still correct for
any semileptonic process. After taking into account parity selection
rules, as well as the relation $\hat{E}_{J}\approx\sqrt{\frac{J+1}{J}}\hat{L}_{J}$ (see Appendix~\ref{sec:Appendix Tensor-Lepton-Traces}),
one stays with the full tensor Fierz term:
\begin{multline}
\Theta\left(q,\vec{\beta}\cdot\hat{\nu}\right)^{VT,AT}
=\mp
\frac{m_{e}}{E}\\
\times\sqrt{2}\mathfrak{Re}\left\{ i\left[\frac{C_{A}C_{T}^{*}+C_{A}^{'}C_{T}^{'*}}{g_A g_T^*}\left\langle \left\Vert \hat{L}_{0}^{A}\right\Vert \right\rangle \left\langle \left\Vert \hat{L}_{0}^{T}\right\Vert \right\rangle ^{*}+\frac{C_{V}C_{T}^{*}+C_{V}^{'}C_{T}^{'*}}{g_A g_T^*}\left\langle \left\Vert \hat{L}_{0}^{V}\right\Vert \right\rangle \left\langle \left\Vert \hat{L}_{0}^{T'}\right\Vert \right\rangle ^{*}\right.\right.\\
\left.\left.-\frac{C_{A}C_{T}^{*}+C_{A}^{'}C_{T}^{'*}}{g_A g_T^*}\left(\hat{\nu}\cdot\hat{q}\right)\left\langle \left\Vert \hat{C}_{0}^{A}\right\Vert \right\rangle \left\langle \left\Vert \hat{L}_{0}^{T}\right\Vert \right\rangle ^{*}-\frac{C_{V}C_{T}^{*}+C_{V}^{'}C_{T}^{'*}}{g_A g_T^*}\left(\hat{\nu}\cdot\hat{q}\right)\left\langle \left\Vert \hat{C}_{0}^{V}\right\Vert \right\rangle \left\langle \left\Vert \hat{L}_{0}^{T'}\right\Vert \right\rangle ^{*}\vphantom{\frac{i}{\sqrt{2}}}\right]\right.\\
+i\sum_{J=1}^{\infty}\left[\frac{C_{A}C_{T}^{*}+C_{A}^{'}C_{T}^{'*}}{g_A g_T^*}\frac{2J+1}{J}\left\langle \left\Vert \hat{L}_{J}^{A}\right\Vert \right\rangle \left\langle \left\Vert \hat{L}_{J}^{T}\right\Vert \right\rangle ^{*}+\frac{C_{V}C_{T}^{*}+C_{V}^{'}C_{T}^{'*}}{g_A g_T^*}\frac{2J+1}{J}\left\langle \left\Vert \hat{L}_{J}^{V}\right\Vert \right\rangle \left\langle \left\Vert \hat{L}_{J}^{T'}\right\Vert \right\rangle ^{*}\right.\\
\left.\left.-\frac{C_{A}C_{T}^{*}+C_{A}^{'}C_{T}^{'*}}{g_A g_T^*}\left(\hat{\nu}\cdot\hat{q}\right)\left\langle \left\Vert \hat{C}_{J}^{A}\right\Vert \right\rangle \left\langle \left\Vert \hat{L}_{J}^{T}\right\Vert \right\rangle ^{*}-\frac{C_{V}C_{T}^{*}+C_{V}^{'}C_{T}^{'*}}{g_A g_T^*}\left(\hat{\nu}\cdot\hat{q}\right)\left\langle \left\Vert \hat{C}_{J}^{V}\right\Vert \right\rangle \left\langle \left\Vert \hat{L}_{J}^{T'}\right\Vert \right\rangle ^{*}\vphantom{\frac{i}{\sqrt{2}}}\right]\right.\\
+\frac{C_{A}C_{T}^{*}+C_{A}^{'}C_{T}^{'*}}{g_A g_T^*}\left\langle \left\Vert \hat{M}_{J}^{A}\right\Vert \right\rangle \left\langle \left\Vert \hat{M}_{J}^{T}\right\Vert \right\rangle ^{*}+\frac{C_{V}C_{T}^{*}+C_{V}^{'}C_{T}^{'*}}{g_A g_T^*}\left\langle \left\Vert \hat{M}_{J}^{V}\right\Vert \right\rangle \left\langle \left\Vert \hat{M}_{J}^{T'}\right\Vert \right\rangle ^{*}\\
\left.\left.+\sqrt{\frac{J+1}{J}}\left(\hat{\nu}\cdot\hat{q}\right)\frac{C_{V}C_{T}^{*}+C_{V}^{'}C_{T}^{'*}}{g_A g_T^*}\left(\left\langle \left\Vert \hat{L}_{J}^{V}\right\Vert \right\rangle \left\langle \left\Vert \hat{M}_{J}^{T}\right\Vert \right\rangle ^{*}+\left\langle \left\Vert \hat{M}_{J}^{V}\right\Vert \right\rangle \left\langle \left\Vert \hat{L}_{J}^{T}\right\Vert \right\rangle ^{*}\right)\right.\right.\\
\left.\left.+\sqrt{\frac{J+1}{J}}\left(\hat{\nu}\cdot\hat{q}\right)\frac{C_{A}C_{T}^{*}+C_{A}^{'}C_{T}^{'*}}{g_A g_T^*}\left(\left\langle \left\Vert \hat{L}_{J}^{A}\right\Vert \right\rangle \left\langle \left\Vert \hat{M}_{J}^{T'}\right\Vert \right\rangle ^{*}+\left\langle \left\Vert \hat{M}_{J}^{A}\right\Vert \right\rangle \left\langle \left\Vert \hat{L}_{J}^{T'}\right\Vert \right\rangle ^{*}\right)\vphantom{\frac{i}{\sqrt{2}}}\right]\right\} \text{,}
\end{multline}
or in BSM LO (using Eq.~\eqref{eq:T-A multipole operators} in the main text):
\begin{multline}
\Theta\left(q,\vec{\beta}\cdot\hat{\nu}\right)^{VT,AT}=
\pm
\frac{m_{e}}{E}
2\mathfrak{Re}\left\{
\frac{C_{A}^{*}C_{T}+C_{A}^{'*}C_{T}^{'}}{2\left|g_{A}\right|^{2}}\left[
\left|\left\langle \left\Vert \hat{L}_{0}^{A}\right\Vert \right\rangle \right|^{2}-\left(\hat{\nu}\cdot\hat{q}\right)\left\langle \left\Vert \hat{C}_{0}^{A}\right\Vert \right\rangle ^{*} \left\langle \left\Vert \hat{L}_{0}^{A}\right\Vert \right\rangle \right]\right.\\
+\sum_{J=1}^{\infty}\left[
\frac{C_{A}^{*}C_{T}+C_{A}^{'*}C_{T}^{'}}{2\left|g_{A}\right|^{2}}\left(
\frac{2J+1}{J}\left|\left\langle \left\Vert \hat{L}_{J}^{A}\right\Vert \right\rangle \right|^{2}-\left(\hat{\nu}\cdot\hat{q}\right)\left\langle \left\Vert \hat{C}_{J}^{A}\right\Vert \right\rangle ^{*}\left\langle \left\Vert \hat{L}_{J}^{A}\right\Vert \right\rangle 
+\left|\left\langle \left\Vert \hat{M}_{J}^{A}\right\Vert \right\rangle \right|^{2}
\right)\right.\\
\left.\left.+\sqrt{\frac{J+1}{J}}\left(\hat{\nu}\cdot\hat{q}\right)
\frac{C_{V}^{*}C_{T}+C_{V}^{'*}C_{T}^{'}}{2\left|g_{V}\right|^{2}}
\left\langle \left\Vert \hat{L}_{J}^{V}\right\Vert \right\rangle ^{*}
\left\langle \left\Vert \hat{M}_{J}^{A}\right\Vert \right\rangle \vphantom{\frac{i}{\sqrt{2}}}\right]\right\}.
\end{multline}
This is a general result which hold for any semileptonic nuclear process, including different types of beyond the Standard Model physics. After substituting Eq.~\eqref{eq: q identities}, it yields the interference Fierz tensor-vector terms presented in Sec.~\ref{sec:A-General-Expression}.
A complete beyond the Standard Model discussion, affecting the full Fierz term, will include also the scalar and pseudoscalar terms, mentioned at appendix \ref{sec:Appendix Scalar-Completeness}.

\section{Scalar and pseudoscalar completeness\label{sec:Appendix Scalar-Completeness}}

Starting from the scalar Hamiltonian,
\begin{eqnarray}
\hat{H}_{\mbox{w}}^{S} & = & \int d^{3}r\hat{j}^{S}\left(\vec{r}\right)\hat{\mathcal{J}}^{S}\left(\vec{r}\right)\text{,}
\end{eqnarray}
we write the scalar lepton current in its most general way, as was
customary prior to any Standard Model experimental-related assumptions~\cite{PhysRev.104.254,PhysRev.106.517}:
\begin{eqnarray}
\hat{j}^{S}\left(\vec{r}\right) & = & \bar{\psi}'\left(\vec{r}\right)\left(C_{S}-C_{S}^{'}\gamma_{5}\right)\psi\left(\vec{r}\right).
\label{eq:scalar_lepton_current}
\end{eqnarray}
Here $\psi^{\left('\right)}\left(\vec{r}\right)$ are fermion fildes, as defined in
Eq.~\eqref{eq:Fremion fields}. Assuming the leptons have a plane
wave character (interaction with the nucleus will be inserted perturbatively),
the general matrix element can be written as
\begin{eqnarray}
\left\langle f\left|\hat{j}^{S}\left(\vec{r}\right)\right|i\right\rangle  & \equiv & l^{S}e^{-i\vec{q}\cdot\vec{r}},
\end{eqnarray}
where $l^{S}=\frac{1}{\Omega}\bar{l}'\left(\vec{k}'\right)\left(C_{S}-C_{S}^{'}\gamma_{5}\right)l\left(\vec{k}\right)$. Using the plane wave expansion~\cite{Edmonds:1974:AMQM},
\begin{eqnarray}
e^{-i\vec{q}\cdot\vec{r}} & = & \sum_{J=0}^{\infty}\sqrt{4\pi\left(2J+1\right)}\left(-i\right)^{J}
j_{J}\left(qr\right)Y_{J0}\left(\hat{r}\right)\text{,}
\end{eqnarray}
one can find the multipole expansion of the scalar Hamiltonian:
\begin{align}
\left\langle f\left|\hat{H}_{\mbox{w}}^{S}\right|i\right\rangle & =\sum_{J=0}^{\infty}\sqrt{4\pi\left(2J+1\right)}\left(-i\right)^{J}
l^{S}\left\langle f\left| \hat{C}_{J}^{S}\left(q\right)\right| i\right\rangle,
\end{align}
and distract from it, using Wigner-Eckart theorem (Eq.~\eqref{eq: Wigner-Eckart}), the term
\begin{align}
\sum_{M_{i}}\sum_{M_{f}}\left|\left\langle f\left|\hat{H}_{\mbox{w}}^{S}\right|i\right\rangle \right|^{2} & =4\pi\sum_{J=0}^{\infty}l^{S}l^{S*}\left|\left\langle J_{f}\left\Vert \hat{C}_{J}^{S}\left(q\right)\right\Vert J_{i}\right\rangle \right|^{2},
\end{align}
as well as the Fierz interference terms:
\begin{align}
\sum_{M_{i}}\sum_{M_{f}}2\mathfrak{Re}\left(\left\langle f\left|\hat{H}_{\mbox{w}}^{V\left(A\right)}\right|i\right\rangle \left\langle f\left|\hat{H}_{\mbox{w}}^{S}\right|i\right\rangle ^{*}\right)
&=8\pi \sum_{J=0}^{\infty}\mathfrak{Re}\left[
l_{0}^{V\left(A\right)}l^{S*}\left\langle J_{f}\left\Vert \hat{C}_{J}^{V\left(A\right)}\right\Vert J_{i}\right\rangle \left\langle J_{f}\left\Vert \hat{C}_{J}^{S}\right\Vert J_{i}\right\rangle ^{*}
\right.\nonumber\\
&\left.
-l_{3}^{V\left(A\right)}l^{S*}\left\langle J_{f}\left\Vert \hat{L}_{J}^{V\left(A\right)}\right\Vert J_{i}\right\rangle \left\langle J_{f}\left\Vert \hat{C}_{J}^{S}\right\Vert J_{i}\right\rangle ^{*}\right],\\
\sum_{M_{i}}\sum_{M_{f}}2\mathfrak{Re}\left(\left\langle f\left|\hat{H}_{\mbox{w}}^{T}\right|i\right\rangle \left\langle f\left|\hat{H}_{\mbox{w}}^{S}\right|i\right\rangle ^{*}\right)
&=8\pi \sum_{J=0}^{\infty}\mathfrak{Re}\left[-l_{3}^{T}l^{S*}\left\langle J_{f}\left\Vert \hat{L}_{J}^{T}\right\Vert J_{i}\right\rangle \left\langle J_{f}\left\Vert \hat{C}_{J}^{S}\right\Vert J_{i}\right\rangle ^{*}\right.\nonumber\\
&\left.-l_{3}^{T'}l^{S*}\left\langle J_{f}\left\Vert \hat{L}_{J}^{T'}\right\Vert J_{i}\right\rangle \left\langle J_{f}\left\Vert \hat{C}_{J}^{S}\right\Vert J_{i}\right\rangle ^{*}\right],
\end{align}
where $\hat{C}_{J}^{S}$ is the Coulomb multipole operator, defined
in Eq.~\eqref{eq: multipole operators} in the main text, calculated with the scalar nuclear current.

The pseudoscalar coupling will have the same expansion, only with
$\hat{H}_{\mbox{w}}^{P}$, $\hat{C}_{J}^{P}$
and $l^{P}$, instead of $\hat{H}_{\mbox{w}}^{S}$, $\hat{C}_{J}^{S}$
and $l^{S}$. 
Note that although it is possible to calculate a scalar-pseudoscalar interference term, according to parity selection rules, there will not be any transition that will involve this kind of term.

Using the definition of $l^{S}$ , one can easily calculate the scalar traces as required for semileptonic weak nuclear processes:
\begin{subequations}
\begin{eqnarray}
\frac{\Omega^{2}}{2}\sum_{\mbox{lepton spins}}l^{S} l^{S*} & = & \frac{\left|C_{S}\right|^{2}+\left|C_{S}^{'}\right|^{2}}{2}\left(1-\vec{\beta}\cdot\hat{\nu}\right), \\
\frac{\Omega^{2}}{2}\sum_{\mbox{lepton spins}}l_{0}^{V}l^{S*} & = & \pm\frac{C_{V}C_{S}^{*}+C_{V}^{'}C_{S}^{'*}}{2}\frac{m_{e}}{E},\\
\frac{\Omega^{2}}{2}\sum_{\mbox{lepton spins}}l_{0}^{A}l^{S*} & = & \pm\frac{C_{A}^{'}C_{S}^{*}+C_{A}C_{S}^{'*}}{2}\frac{m_{e}}{E},\\
-\frac{\Omega^{2}}{2}\sum_{\mbox{lepton spins}}l_{3}^{V}l^{S*} & = & \pm\frac{C_{V}C_{S}^{*}+C_{V}^{'}C_{S}^{'*}}{2}\left(\hat{\nu}\cdot\hat{q}\right)\frac{m_{e}}{E}, \\
-\frac{\Omega^{2}}{2}\sum_{\mbox{lepton spins}}l_{3}^{A}l^{S*} & = & \pm\frac{C_{A}^{'}C_{S}^{*}+C_{A}C_{S}^{'*}}{2}\left(\hat{\nu}\cdot\hat{q}\right)\frac{m_{e}}{E}, \\
-\frac{\Omega^{2}}{2}\sum_{\mbox{lepton spins}}l_{3}^{T}l^{S*} & = & 
\mp i\sqrt{2}\frac{C_{T}^{'}C_{S}^{*}+C_{T}C_{S}^{'*}}{2}\hat{q}\cdot\left(\hat{\nu}-\vec{\beta}\right),\\
-\frac{\Omega^{2}}{2}\sum_{\mbox{lepton spins}}l_{3}^{T'}l^{S*} & = & \pm i\sqrt{2}\frac{C_{T}C_{S}^{*}+C_{T}^{'}C_{S}^{'*}}{2}\hat{q}\cdot\left(\hat{\nu}-\vec{\beta}\right).
\end{eqnarray}
\end{subequations}

Pseudoscalar traces, originating from the pseudoscalar leptonic
current, $j^{P}\equiv\bar{\psi}'\left(\vec{r}\right) \left(C_{P}\gamma_{5}-C_{P}^{'}\right)$ $\times \psi\left(\vec{r}\right)$
(compare to the scalar leptonic current in Eq.~\eqref{eq:scalar_lepton_current}), with  $l^{P}=\frac{1}{\Omega}\bar{l}'\left(\vec{k}'\right)\left(C_{P}\gamma_{5}-C_{P}^{'}\right)$ $\times l\left(\vec{k}\right)$,
will be similar, replacing the coefficients $C_{S}$ and $C_{S}^{'}$
with the coefficients $-C_{P}^{'}$ and $-C_{P}$, respectfully. 

In order to match the lepton traces terms to the quark-level effective theory one-nucleon matrix elements, we replace $C_{\text{sym}}^{\left('\right)}$ with the adjust coefficients $\frac{C_{\text{sym}}^{\left('\right)}}{g_{\text{sym}}}$ ($\text{sym}\in\left\{ S,P,V,A,T\right\}$), as we did before.
Taking into account also the parity selection rules, one can finally write a scalar and pseudoscalar expression:
\begin{multline}
\Theta\left(q,\vec{\beta}\cdot\hat{\nu}\right)^{S,P}
=\sum_{J=0}^{\infty}\left\{ \left(1-\vec{\beta}\cdot\hat{\nu}\right)\left[\frac{\left|C_{S}\right|^{2}+\left|C_{S}^{'}\right|^{2}}{2g_S^2}\left|\left\langle \left\Vert \hat{C}_{J}^{S}\right\Vert \right\rangle \right|^{2}+\frac{\left|C_{P}\right|^{2}+\left|C_{P}^{'}\right|^{2}}{2g_P^2}\left|\left\langle \left\Vert \hat{C}_{J}^{P}\right\Vert \right\rangle \right|^{2}\right]\right.\\
\pm2\frac{m_{e}}{E}\mathfrak{Re}\left[\frac{C_{V}C_{S}^{*}+C_{V}^{'}C_{S}^{'*}}{2g_V g_S^*}\left\langle \left\Vert \hat{C}_{J}^{V}\right\Vert \right\rangle \left\langle \left\Vert \hat{C}_{J}^{S}\right\Vert \right\rangle ^{*}-\frac{C_{A}^{'}C_{P}^{'*}+C_{A}C_{P}^{*}}{2g_A g_P^*}\left\langle \left\Vert \hat{C}_{J}^{A}\right\Vert \right\rangle \left\langle \left\Vert \hat{C}_{J}^{P}\right\Vert \right\rangle ^{*}\right]\\
\pm2\frac{m_{e}}{E}\left(\hat{\nu}\cdot\hat{q}\right)\mathfrak{Re}\left[\frac{C_{V}C_{S}^{*}+C_{V}^{'}C_{S}^{'*}}{2g_V g_S^*}\left\langle \left\Vert \hat{L}_{J}^{V}\right\Vert \right\rangle \left\langle \left\Vert \hat{C}_{J}^{S}\right\Vert \right\rangle ^{*}-\frac{C_{A}^{'}C_{P}^{'*}+C_{A}C_{P}^{*}}{2g_A g_P^*}\left\langle \left\Vert \hat{L}_{J}^{A}\right\Vert \right\rangle \left\langle \left\Vert \hat{C}_{J}^{P}\right\Vert \right\rangle ^{*}\right]\\
\pm2\sqrt{2}\hat{q}\cdot\left(\hat{\nu}-\vec{\beta}\right)\mathfrak{Re}\left[i\frac{C_{T}C_{S}^{*}+C_{T}^{'}C_{S}^{'*}}{2g_T g_S^*}\left\langle \left\Vert \hat{L}_{J}^{T'}\right\Vert \right\rangle \left\langle \left\Vert \hat{C}_{J}^{S}\right\Vert \right\rangle ^{*}\right.\\
\left.\left.+i\frac{C_{T}^{'}C_{P}^{'*}+C_{T}C_{P}^{*}}{2g_T g_P^*}\left\langle \left\Vert \hat{L}_{J}^{T}\right\Vert \right\rangle \left\langle \left\Vert \hat{C}_{J}^{P}\right\Vert \right\rangle ^{*}\right]\right\}.
\end{multline}

In order to complete this discussion, one needs to calculate the scalar and pseudoscalar multipole operators. Starting from the hadronic currents:
\begin{subequations}
\begin{eqnarray}
\hat{\mathcal{J}}^{S}\left(\vec{r}\right) & = & \bar{\phi}'\left(\vec{r}\right)\phi\left(\vec{r}\right), \\
\hat{\mathcal{J}}^{P}\left(\vec{r}\right) & = & \bar{\phi}'\left(\vec{r}\right)\gamma_{5}\phi\left(\vec{r}\right),
\end{eqnarray}
\end{subequations}
the general form of the single-nucleon matrix element of the scalar
and pseudoscalar parts of the charge changing weak current, are~\cite{Cirigliano:2013xha}:
\begin{subequations}
\begin{eqnarray}
\left\langle \vec{p}',\sigma',\rho'\left|\hat{\mathcal{J}}^{S}\right|\vec{p},\sigma,\rho\right\rangle  & = & \frac{1}{\Omega}g_{S}\left(q^{2}\right)\bar{u}\left(\vec{p'},\sigma'\right)\eta_{\rho'}^{+}\tau^{\pm}\eta_{\rho}u\left(\vec{p},\sigma\right)+\mathcal{O}\left(\epsilon_{\text{recoil}}^{2}\right), \\
\left\langle \vec{p}',\sigma',\rho'\left|\hat{\mathcal{J}}^{P}\right|\vec{p},\sigma,\rho\right\rangle  & = & \frac{1}{\Omega}g_{P}\left(q^{2}\right)\bar{u}\left(\vec{p'},\sigma'\right)\eta_{\rho'}^{+}\gamma_{5}\tau^{\pm}\eta_{\rho}u\left(\vec{p},\sigma\right)+\mathcal{O}\left(\epsilon_{\text{recoil}}^{2}\right),
\end{eqnarray}
\end{subequations}
with $\epsilon_{\text{recoil}}\sim\frac{q}{m_{N}}$ ($\approx0.002$
for an endpoint of $\approx2\text{MeV}$). Expanding the needed matrix
elements in the inverse mass (following what we did to the tensor
matrix element at appendix \ref{sec:Appendix Tensor-Nuclear-Current}),
one can find the following non-relativistic expansion for the matrix
elements of the scalar and pseudoscalar nuclear currents:
\begin{subequations}
\begin{eqnarray}
\left\langle \vec{p}',\sigma',\rho'\left|\mathcal{J}^{S}\right|\vec{p},\sigma,\rho\right\rangle  & = & \frac{1}{\Omega}g_{S}\left(q^{2}\right)\chi_{\sigma'}^{+}\eta_{\rho'}^{+}\tau_{\pm}\eta_{\rho}\chi_{\sigma}+\mathcal{O}\left(\epsilon_{\text{NR}}^{2}\right),\\
\left\langle \vec{p}',\sigma',\rho'\left|\mathcal{J}^{P}\right|\vec{p},\sigma,\rho\right\rangle  & = & \frac{1}{\Omega}g_{P}\left(q^{2}\right)\chi_{\sigma'}^{+}\eta_{\rho'}^{+}\frac{\vec{\sigma}\cdot\vec{q}}{2m_{N}}\tau^{\pm}\eta_{\rho}\chi_{\sigma}+\mathcal{O}\left(\epsilon_{\text{NR}}^{2}\right),
\end{eqnarray}
\end{subequations}
with $\epsilon_{\text{NR}}\sim\frac{P_{\text{fermi}}}{m_{N}}\approx0.2$.
As we did in Appendix~\ref{sec:Appendix Tensor-Nuclear-Current}, we use the definition of the (second quantization) $\mathcal{J}\left(\vec{r}\right)$
current matrix element as a sum over first-quantization currents $\hat{\mathcal{J}}^{\left(1\right)}$,
identify the nuclear density operators in first quantization:
\begin{subequations}
\begin{eqnarray}
\hat{\mathcal{J}}^{S \left(1\right)}\left(0\right) & = & g_{S}\left(q^{2}\right)\tau^{\pm}+\mathcal{O}\left(\epsilon_{\text{NR}}^{2}\right) ,\\
\hat{\mathcal{J}}^{P \left(1\right)}\left(0\right) & = & g_{P}\left(q^{2}\right)\frac{\vec{\sigma}\cdot\vec{q}}{2m_{N}}\tau^{\pm}+\mathcal{O}\left(\epsilon_{\text{NR}}^{2}\right)\text{,}
\end{eqnarray}
\end{subequations}
and get the (second quantization) current densities:
\begin{subequations}
\begin{eqnarray}
\mathcal{J}^{S}\left(\vec{r}\right) & = & g_{S}\sum_{j=1}^{A}\tau_{j}^{\pm}\delta^{\left(3\right)}\left(\vec{r}-\vec{r}_{j}\right)+\mathcal{O}\left(\epsilon_{\text{NR}}^{2}\right), \\
\mathcal{J}^{P}\left(\vec{r}\right) & = & -\frac{i}{2m_{N}}g_{P}\sum_{j=1}^{A}\vec{\nabla}\delta^{\left(3\right)}\left(\vec{r}-\vec{r}_{j}\right)\cdot\vec{\sigma}_{j}\tau_{j}^{\pm}+\mathcal{O}\left(\epsilon_{\text{NR}}^{2}\right)\text{.}
\end{eqnarray}
\end{subequations}

Using the current densities, the multipole operators (Eq.~\eqref{eq: multipole operators} in the main text) can be written as a sum of one-body
operators. The multipole operators, calculated with the scalar and
pseudoscalar symmetry contributions to the weak nuclear current, will be:
\begin{subequations}
\begin{eqnarray}
\hat{C}_{J}^{S}\left(q\right) & = & \frac{g_{S}}{g_{V}}\hat{C}_{J}^{V}\left(q\right)+\mathcal{O}\left(\epsilon_{qr}^{J}\epsilon_{\text{NR}}^{2}\right), \\
\hat{C}_{J}^{P}\left(q\right) & = & \frac{iq}{2m_{N}}g_{P}\sum_{j=1}^{A}\left[\frac{1}{q}\vec{\nabla}M_{J}\left(q\vec{r}_{j}\right)\right]\cdot\vec{\sigma}_{j}\tau_{j}^{\pm}+\mathcal{O}\left(\epsilon_{qr}^{J}\epsilon_{\text{NR}}^{2}\right),
\end{eqnarray}
\end{subequations}
where $\hat{C}_{J}^{V}\left(q\right)=g_{V}\sum_{j=1}^{A}M_{J}\left(q\vec{r}_{j}\right)\tau_{j}^{\pm}+\mathcal{O}\left(\epsilon_{qr}^{J}\epsilon_{\text{NR}}^{2}\right)$ is the SM polar-vector Coulomb multipole operator.
The BSM LO expression will be:
\begin{multline}
\Theta\left(q,\vec{\beta}\cdot\hat{\nu}\right)
^{S,P}=\sum_{J=0}^{\infty}\left\{ \left(1-\vec{\beta}\cdot\hat{\nu}\right)\frac{\left|C_{S}\right|^{2}+\left|C_{S}^{'}\right|^{2}}{2\left|g_{V}\right|^{2}}\left|\left\langle \left\Vert \hat{C}_{J}^{V}\right\Vert \right\rangle \right|^{2}\right.\\
\left.\pm\frac{m_{e}}{E}2\mathfrak{Re}\left[
\frac{C_{V}C_{S}^{*}+C_{V}^{'}C_{S}^{'*}}{2\left|g_{V}\right|^{2}}
\left(
\left|\left\langle \left\Vert \hat{C}_{J}^{V}\right\Vert \right\rangle \right|^{2}
+\left(\hat{\nu}\cdot\hat{q}\right)
\left\langle \left\Vert \hat{L}_{J}^{V}\right\Vert \right\rangle
\left\langle \left\Vert \hat{C}_{J}^{V}\right\Vert \right\rangle ^{*} 
\right)\right]\right\}.
\end{multline}
This is a general result which hold for any semileptonic nuclear process, including different types of beyond the Standard Model physics. After substituting Eq.~\eqref{eq: q identities}, it yields the scalar and pseudoscalar terms (including their interference Fierz terms) presented in Sec.~\ref{sec:A-General-Expression}.

\section{Second class nuclear currents and multipole operators\label{sec:Appendix-Second-Class_Nuclear-Currents-and}}

The general form of the single-nucleon matrix
element, of the vector and axial parts of the charge changing weak current are (respectively)~\cite{Cirigliano:2013xha}:
\begin{small}
\begin{subequations}
\begin{align}
\left\langle \vec{p}',\sigma',\rho'\left|\mathcal{J}_{\mu}^{V}\left(0\right)\right|\vec{p},\sigma,\rho\right\rangle  & =\frac{1}{\Omega}\bar{u}\left(\vec{p'},\sigma'\right)\eta_{\rho'}^{+}\left[g_{V}\left(q^{2}\right)\gamma_{\mu}-i\frac{\tilde{g}_{T\left(V\right)}\left(q^{2}\right)}{2m_{N}}\sigma_{\mu\nu}q^{\nu}+\frac{\tilde{g}_{S}\left(q^{2}\right)}{2m_{N}}q_{\mu}\right]\tau^{\pm}\eta_{\rho}u\left(\vec{p},\sigma\right),\\
\left\langle \vec{p}',\sigma',\rho'\left|\mathcal{J}_{\mu}^{A}\left(0\right)\right|\vec{p},\sigma,\rho\right\rangle  & =\frac{1}{\Omega}\bar{u}\left(\vec{p'},\sigma'\right)\eta_{\rho'}^{+}\left[g_{A}\left(q^{2}\right)\gamma_{\mu}-i\frac{\tilde{g}_{T\left(A\right)}\left(q^{2}\right)}{2m_{N}}\sigma_{\mu\nu}q^{\nu}+\frac{\tilde{g}_{P}\left(q^{2}\right)}{2m_{N}}q_{\mu}\right]\gamma_{5}\tau^{\pm}\eta_{\rho}u\left(\vec{p},\sigma\right).
\end{align}
\end{subequations}
\end{small}
All the form factors $g_{i}\left(q^{2}\right)$
are functions of $q^{2}$. In the Standard Model, $g_{V}=1$, up
to second-order corrections in isospin breaking~\cite{Ademollo1964,DONOGHUE1990243},
as a result of the conservation of the vector current, and $g_{A}\approx1.276g_{V}$~\cite{Mendenhall2013,Mund2013}. The induced charges, $\tilde{g}_{T\left(V\right)}$, $\tilde{g}_{S}$,
$\tilde{g}_{T\left(A\right)}$ and $\tilde{g}_{P}$ (not to confuse with the actual BSM charges, $g_{S}$, $g_{P}$ and $g_{T}$, which appear in the scalar, pseudoscalar and tensor currents), are all proportional
to $\epsilon_{\text{recoil}}\equiv\frac{q}{m_{N}}$~\cite{Cirigliano:2013xha}.
$\tilde{g}_{S}$ and $\tilde{g}_{T\left(A\right)}$, known as second class currents, do not exist in the Standard Model, $\tilde{g}_{S}$
due to current conservation, and $\tilde{g}_{T\left(A\right)}$ because
of G-parity considerations~\cite{PhysRev.112.1375}.

As before, we substitute the explicit form of Dirac spinors and make a non-relativistic expansion, to find the required matrix elements:
\begin{subequations}
\begin{align}
\left\langle \vec{p}',\sigma',\rho'\left|\mathcal{J}_{0}^{V}\left(0\right)\right|\vec{p},\sigma,\rho\right\rangle 
&=\frac{1}{\Omega}\chi_{\sigma'}^{+}\eta_{\rho'}^{+}\left(g_{V}+\frac{E_0}{2m_{N}}\tilde{g}_{S}\right)\tau^{\pm}\eta_{\rho}\chi_{\sigma}+\mathcal{O}\mathrm{\left(\epsilon_{\text{NR}}^{2}\right)}, \\
\left\langle \vec{p}',\sigma',\rho'\left|\vec{\mathcal{J}}^{V}\left(0\right)\right|\vec{p},\sigma,\rho\right\rangle 
&=\frac{1}{\Omega}\chi_{\sigma'}^{+}\eta_{\rho'}^{+}\frac{1}{2m_{N}}\left[g_{V}\vec{P}+\left(g_{V}+\tilde{g}_{T\left(V\right)}\right)i\vec{q}\times\vec{\sigma}+\tilde{g}_{S}\vec{q}\right]\tau^{\pm}\eta_{\rho}\chi_{\sigma}+\mathcal{O}\mathrm{\left(\epsilon_{\text{NR}}^{2}\right)}\label{eq: Hadron J_V},\\
\left\langle \vec{p}',\sigma',\rho'\left|\mathcal{J}_{0}^{A}\left(0\right)\right|\vec{p},\sigma,\rho\right\rangle 
&=\frac{1}{\Omega}\bar{u}\left(\vec{p'},\sigma'\right)\eta_{\rho'}^{+}\frac{1}{2m_{N}}\left[g_{A}\vec{P}\cdot\vec{\sigma}-\left(\tilde{g}_{T\left(A\right)}-\frac{E_0}{2m_{N}}\tilde{g}_{P}\right)\vec{q}\cdot\vec{\sigma}\right]\tau^{\pm}\eta_{\rho}u\left(\vec{p},\sigma\right)\nonumber\\
&+\mathcal{O}\mathrm{\left(\epsilon_{\text{NR}}^{2}\right)}, \\
\left\langle \vec{p}',\sigma',\rho'\left|\vec{\mathcal{J}}^{A}\left(0\right)\right|\vec{p},\sigma,\rho\right\rangle 
&=\frac{1}{\Omega}\bar{u}\left(\vec{p'},\sigma'\right)\eta_{\rho'}^{+}\left(g_{A}\vec{\sigma}-\tilde{g}_{T\left(A\right)}\frac{E_0}{2m_{N}}\vec{\sigma}\right)\tau^{\pm}\eta_{\rho}u\left(\vec{p},\sigma\right)+\mathcal{O}\mathrm{\left(\epsilon_{\text{NR}}^{2}\right)}\text{.}\label{eq: hadron J_A}
\end{align}
\end{subequations}
We use the definition of the second quantization current matrix element to identify the following currents:
\begin{subequations}
\label{eq: V-A nuclear currents}
\begin{align}
\mathcal{J}_{0}^{V}\left(\vec{r}\right) & =\sum_{j=1}^{A}\left[g_{V}+\frac{E_0}{2m_{N}}\tilde{g}_{S}\right]\tau_{j}^{\pm}\delta^{\left(3\right)}\left(\vec{r}-\vec{r}_{j}\right)+\mathcal{O}\mathrm{\left(\epsilon_{\text{NR}}^{2}\right)}, \\
\vec{\mathcal{J}}^{V}\left(\vec{r}\right) & =\frac{1}{2m_{N}}\sum_{j=1}^{A}\left[g_{V}\left\{ \vec{p}_{j},\delta^{\left(3\right)}\left(\vec{r}-\vec{r}_{j}\right)\right\} +\left(g_{V}+\tilde{g}_{T\left(V\right)}\right)\vec{\nabla}\times\vec{\sigma}_{j}\delta^{\left(3\right)}\left(\vec{r}-\vec{r}_{j}\right)\right.\nonumber \\
 & \left.-i\tilde{g}_{S}\vec{\nabla}\delta^{\left(3\right)}\left(\vec{r}-\vec{r}_{j}\right)\right]\tau_{j}^{\pm}+\mathcal{O}\mathrm{\left(\epsilon_{\text{NR}}^{2}\right)}, \\
\mathcal{J}_{0}^{A}\left(\vec{r}\right) & =\frac{1}{2m_{N}}\sum_{j=1}^{A}\left[g_{A}\left\{ \vec{p}_{j},\delta^{\left(3\right)}\left(\vec{r}-\vec{r}_{j}\right)\right\} +i\left(\tilde{g}_{T\left(A\right)}-\frac{E_0}{2m_{N}}\tilde{g}_{P}\right)\vec{\nabla}\delta^{\left(3\right)}\left(\vec{r}-\vec{r}_{j}\right)\right]\cdot\vec{\sigma}_{j}\tau_{j}^{\pm}+\mathcal{O}\mathrm{\left(\epsilon_{\text{NR}}^{2}\right)}, \\
\vec{\mathcal{J}}^{A}\left(\vec{r}\right) & =\sum_{j=1}^{A}\left[g_{A}-\frac{E_0}{2m_{N}}\tilde{g}_{T\left(A\right)}\right]\vec{\sigma}_{j}\tau_{j}^{\pm}\delta^{\left(3\right)}\left(\vec{r}-\vec{r}_{j}\right)+\mathcal{O}\mathrm{\left(\epsilon_{\text{NR}}^{2}\right)}.
\end{align}
\end{subequations}
Positioning Eq. \eqref{eq: V-A nuclear currents} into the multipole
operators definition, leads to the explicit expressions for the vector
and axial currents multipole operators:
\begin{subequations}
\begin{align}
\hat{C}_{J}^{V\text{[2c]}}\left(q\right) & =\left(g_{V}+\frac{E_0}{2m_{N}}\tilde{g}_{S}\right)\sum_{j=1}^{A}M_{J}\left(q\vec{r}_{j}\right)\tau_{j}^{\pm}+\mathcal{O}\left(\epsilon_{qr}^{J}\epsilon_{\text{NR}}^{2}\right), \\
\hat{L}_{J}^{V\text{[2c]}}\left(q\right) & =-\frac{q}{2m_{N}}\sum_{j=1}^{A}\left\{ \left(g_{V}-\tilde{g}_{S}\right)M_{J}\left(q\vec{r}_{j}\right)-2g_{V}\left[\frac{1}{q}\vec{\nabla}M_{J}\left(q\vec{r}_{j}\right)\right]\cdot\frac{1}{q}\vec{\nabla}\right\} \tau_{j}^{\pm}+\mathcal{O}\left(\epsilon_{qr}^{J-1}\epsilon_{\text{NR}}^{2}\right), \\
\hat{E}_{J}^{V\text{[2c]}}\left(q\right) & =\frac{q}{m_{N}}\sum_{j=1}^{A}\left\{ -ig_{V}\left[\frac{1}{q}\vec{\nabla}\times\vec{M}_{JJ1}\left(q\vec{r}_{j}\right)\right]\cdot\frac{1}{q}\vec{\nabla}+\frac{g_{V}+\tilde{g}_{T\left(V\right)}}{2}\vec{M}_{JJ1}\left(q\vec{r}_{j}\right)\cdot\vec{\sigma}\right\} \tau_{j}^{\pm}\nonumber\\
&+\mathcal{O}\left(\epsilon_{qr}^{J-1}\epsilon_{\text{NR}}^{2}\right), \\
\hat{M}_{J}^{V\text{[2c]}}\left(q\right) & =-\frac{iq}{m_{N}}\sum_{j=1}^{A}\left\{ g_{V}\vec{M}_{JJ1}\left(q\vec{r}_{j}\right)\cdot\frac{1}{q}\vec{\nabla}+i\frac{g_{V}+\tilde{g}_{T\left(V\right)}}{2}\left[\frac{1}{q}\vec{\nabla}\times\vec{M}_{JJ1}\left(q\vec{r}_{j}\right)\right]\cdot\vec{\sigma}\right\} \tau_{j}^{\pm}\nonumber\\
&+\mathcal{O}\left(\epsilon_{qr}^{J}\epsilon_{\text{NR}}^{2}\right),\\
\hat{C}_{J}^{A\text{[2c]}}\left(q\right) & =-\frac{iq}{m_{N}}\sum_{j=1}^{A}\left\{ g_{A}M_{J}\left(q\vec{r}_{j}\right)\vec{\sigma}\cdot\frac{1}{q}\vec{\nabla}+\frac{1}{2}\left(g_{A}+\tilde{g}_{T\left(A\right)}-\frac{E_0}{2m_{N}}\tilde{g}_{P}\right)\left[\frac{1}{q}\vec{\nabla}M_{J}\left(q\vec{r}\right)\right]\cdot\vec{\sigma}\right\} \tau_{j}^{\pm}\nonumber\\
&+\mathcal{O}\left(\epsilon_{qr}^{J}\epsilon_{\text{NR}}^{2}\right), \\
\hat{L}_{J}^{A\text{[2c]}}\left(q\right) & =i\left(g_{A}-\frac{E_0}{2m_{N}}\tilde{g}_{T\left(A\right)}\right)\sum_{j=1}^{A}\left[\frac{1}{q}\vec{\nabla}M_{J}\left(q\vec{r}_{j}\right)\right]\cdot\vec{\sigma}\tau_{j}^{\pm}+\mathcal{O}\left(\epsilon_{qr}^{J-1}\epsilon_{\text{NR}}^{2}\right), \\
\hat{E}_{J}^{A\text{[2c]}}\left(q\right) & =\left(g_{A}-\frac{E_0}{2m_{N}}\tilde{g}_{T\left(A\right)}\right)\sum_{j=1}^{A}\left[\frac{1}{q}\vec{\nabla}\times\vec{M}_{JJ1}\left(q\vec{r}_{j}\right)\right]\cdot\vec{\sigma}\tau_{j}^{\pm}+\mathcal{O}\left(\epsilon_{qr}^{J-1}\epsilon_{\text{NR}}^{2}\right), \\
\hat{M}_{J}^{A\text{[2c]}}\left(q\right) & =\left(g_{A}-\frac{E_0}{2m_{N}}\tilde{g}_{T\left(A\right)}\right)\sum_{j=1}^{A}\vec{M}_{JJ1}\left(q\vec{r}_{j}\right)\cdot\vec{\sigma}\tau_{j}^{\pm}+\mathcal{O}\left(\epsilon_{qr}^{J}\epsilon_{\text{NR}}^{2}\right)\text{,}
\end{align}
\end{subequations}
with $M_{J}$ and $\vec{M}_{JL1}$ defined in Eq.~\eqref{eq:M_operators} in the main text.
One can recognize that these multipoles operators that include the second class currents are actually the SM multipoles operators with small changes:
\begin{subequations}
\begin{align}
\hat{C}_{J}^{V\text{[2c]}}\left(q\right) & =\frac{g_{V}+\frac{E_0}{2m_{N}}\tilde{g}_{S}}{g_{V}}\hat{C}_{J}^{V}\left(q\right), \\
\hat{L}_{J}^{V\text{[2c]}}\left(q\right) & =\hat{L}_{J}^{V}\left(q\right)+\frac{q}{2m_{N}}\frac{\tilde{g}_{S}}{g_{V}}\hat{C}_{J}^{V}\left(q\right), \\
\hat{C}_{J}^{A\text{[2c]}}\left(q\right) & =\hat{C}_{J}^{A}\left(q\right)-\frac{q}{2m_{N}}\frac{\tilde{g}_{T\left(A\right)}}{g_{A}}\hat{L}_{J}^{A}\left(q\right), \\
\hat{O}_{J}^{A\text{[2c]}}\left(q\right) & =\frac{g_{A}-\frac{E_0}{2m_{N}}\tilde{g}_{T\left(A\right)}}{g_{A}}\hat{O}_{J}^{A}\left(q\right), &
\hat{O} \in\left\{ \hat{L},\hat{E},\hat{M}\right\} \text{.}
\end{align}
\end{subequations}
The multipole operator $\hat{E}_{J}^{V}$ and $\hat{M}_{J}^{V}$
stay with no change for their leading orders when including second class currents.

\bibliographystyle{unsrt}
\addcontentsline{toc}{section}{\refname}\bibliography{bib}

\begin{thebibliography}{10}

\bibitem{RevModPhys.52.299}
Kip~S. Thorne.
\newblock Multipole expansions of gravitational radiation.
\newblock {\em Rev. Mod. Phys.}, 52:299--339, Apr 1980.

\bibitem{PhysRev.82.531}
E.~Greuling and M.~L. Meeks.
\newblock Electron-neutrino angular correlation.
\newblock {\em Phys. Rev.}, 82:531--537, May 1951.

\bibitem{1742-6596-196-1-012002}
Steven Weinberg.
\newblock V-a was the key.
\newblock {\em Journal of Physics: Conference Series}, 196(1):012002, 2009.

\bibitem{RevModPhys.78.991}
Nathal Severijns, Marcus Beck, and Oscar Naviliat-Cuncic.
\newblock Tests of the standard electroweak model in nuclear beta decay.
\newblock {\em Rev. Mod. Phys.}, 78:991--1040, Sep 2006.

\bibitem{doi:10.1146/annurev-nucl-102010-130410}
Nathal Severijns and Oscar Naviliat-Cuncic.
\newblock Symmetry tests in nuclear beta decay.
\newblock {\em Annual Review of Nuclear and Particle Science}, 61(1):23--46,
  2011.

\bibitem{ANDP:ANDP201300072}
Oscar Naviliat-Cuncic and Mart{\'i}n Gonz{\'a}lez-Alonso.
\newblock Prospects for precision measurements in nuclear $\ensuremath{\beta}$
  decay in the lhc era.
\newblock {\em Annalen der Physik}, 525(8-9):600--619, 2013.

\bibitem{1402-4896-2013-T152-014018}
N~Severijns and O~Naviliat-Cuncic.
\newblock Structure and symmetries of the weak interaction in nuclear beta
  decay.
\newblock {\em Physica Scripta}, 2013(T152):014018, 2013.

\bibitem{RevModPhys.87.1483}
K.~K. Vos, H.~W. Wilschut, and R.~G.~E. Timmermans.
\newblock Symmetry violations in nuclear and neutron $\ensuremath{\beta}$
  decay.
\newblock {\em Rev. Mod. Phys.}, 87:1483--1516, Dec 2015.

\bibitem{gonzalez2019new}
Martin Gonzalez-Alonso, Oscar Naviliat-Cuncic, and Nathal Severijns.
\newblock New physics searches in nuclear and neutron $\beta$ decay.
\newblock {\em Progress in Particle and Nuclear Physics}, 104:165--223, 2019.

\bibitem{Ohayon2018}
Ben Ohayon, Joel Chocron, Tsviki Hirsh, Ayala Glick-Magid, Yonatan Mishnayot,
  Ish Mukul, Hitesh Rahangdale, Sergei Vaintraub, Oded Heber, Doron Gazit, and
  Guy Ron.
\newblock Weak interaction studies at saraf.
\newblock {\em Hyperfine Interactions}, 239(1):57, Nov 2018.

\bibitem{Hoferichter:2015ipa}
Martin Hoferichter, Philipp Klos, and Achim Schwenk.
\newblock {Chiral power counting of one- and two-body currents in direct
  detection of dark matter}.
\newblock {\em Phys. Lett.}, B746:410--416, 2015.

\bibitem{Menendez:2012tm}
J.~Menendez, D.~Gazit, and A.~Schwenk.
\newblock {Spin-dependent WIMP scattering off nuclei}.
\newblock {\em Phys. Rev.}, D86:103511, 2012.

\bibitem{Klos:2013rwa}
P.~Klos, J.~Mene\'ndez, D.~Gazit, and A.~Schwenk.
\newblock {Large-scale nuclear structure calculations for spin-dependent WIMP
  scattering with chiral effective field theory currents}.
\newblock {\em Phys. Rev.}, D88(8):083516, 2013.
\newblock [Erratum: Phys. Rev.D89,no.2,029901(2014)].

\bibitem{PhysRevD.94.063505}
Martin Hoferichter, Philipp Klos, Javier Men\'endez, and Achim Schwenk.
\newblock Analysis strategies for general spin-independent wimp-nucleus
  scattering.
\newblock {\em Phys. Rev. D}, 94:063505, Sep 2016.

\bibitem{1475-7516-2013-02-004}
A.~Liam Fitzpatrick, Wick Haxton, Emanuel Katz, Nicholas Lubbers, and Yiming
  Xu.
\newblock The effective field theory of dark matter direct detection.
\newblock {\em Journal of Cosmology and Astroparticle Physics}, 2013(02):004,
  2013.

\bibitem{PhysRevC.89.065501}
Nikhil Anand, A.~Liam Fitzpatrick, and W.~C. Haxton.
\newblock Weakly interacting massive particle-nucleus elastic scattering
  response.
\newblock {\em Phys. Rev. C}, 89:065501, Jun 2014.

\bibitem{PhysRev.104.254}
T.~D. Lee and C.~N. Yang.
\newblock Question of parity conservation in weak interactions.
\newblock {\em Phys. Rev.}, 104:254--258, Oct 1956.

\bibitem{PhysRev.106.517}
J.~D. Jackson, S.~B. Treiman, and H.~W. Wyld.
\newblock Possible tests of time reversal invariance in beta decay.
\newblock {\em Phys. Rev.}, 106:517--521, May 1957.

\bibitem{WALECKA1975113}
J.D. Walecka.
\newblock Section 4 - semileptonic weak interactions in nuclei.
\newblock In Vernon~W. Hughes and C.S. Wu, editors, {\em Muon Physics, Volume
  II: Weak Interactions}, pages 113--218. Academic Press, 1975.

\bibitem{Edmonds:1974:AMQM}
A.~R. Edmonds.
\newblock {\em Angular Momentum in Quantum Mechanics}.
\newblock Princeton University Press, Princeton, NJ, 3rd printing, with
  corrections, 2nd edition, 1974.
\newblock Reprinted in 1996.

\bibitem{glick2021formalism}
Ayala Glick-Magid and Doron Gazit.
\newblock A formalism to assess the accuracy of nuclear-structure weak
  interaction effects in precision $\beta$-decay studies.
\newblock {\em J. Phys. G: Nucl. Part. Phys. (in press) arXiv:2107.10588},
  2021.

\bibitem{cirgiliano2019precision}
Vincenzo Cirgiliano, Alejandro Garcia, Doron Gazit, Oscar Naviliat-Cuncic, Guy
  Savard, and Albert Young.
\newblock Precision beta decay as a probe of new physics.
\newblock {\em arXiv:1907.02164}, 2019.

\bibitem{Cat__2007}
Oscar Cat{\`{a}} and Vicent Mateu.
\newblock Chiral perturbation theory with tensor sources.
\newblock {\em Journal of High Energy Physics}, 2007(09):078--078, sep 2007.

\bibitem{Cirigliano:2013xha}
Vincenzo Cirigliano, Susan Gardner, and Barry Holstein.
\newblock {Beta Decays and Non-Standard Interactions in the LHC Era}.
\newblock {\em Prog. Part. Nucl. Phys.}, 71:93--118, 2013.

\bibitem{Bhattacharya2016}
Tanmoy Bhattacharya, Vincenzo Cirigliano, Saul~D. Cohen, Rajan Gupta, Huey-Wen
  Lin, and Boram Yoon.
\newblock Axial, scalar, and tensor charges of the nucleon from $2+1+1$-flavor
  lattice qcd.
\newblock {\em Phys. Rev. D}, 94:054508, Sep 2016.

\bibitem{PhysRev.112.1375}
Steven Weinberg.
\newblock Charge symmetry of weak interactions.
\newblock {\em Phys. Rev.}, 112:1375--1379, Nov 1958.

\bibitem{Ademollo1964}
M.~Ademollo and R.~Gatto.
\newblock Nonrenormalization theorem for the strangeness-violating vector
  currents.
\newblock {\em Phys. Rev. Lett.}, 13:264--266, Aug 1964.

\bibitem{DONOGHUE1990243}
John~F. Donoghue and D.~Wyler.
\newblock Isospin breaking and the precise determination of vud.
\newblock {\em Physics Letters B}, 241(2):243 -- 248, 1990.

\bibitem{RevModPhys.90.015008}
Leendert Hayen, Nathal Severijns, Kazimierz Bodek, Dagmara Rozpedzik, and
  Xavier Mougeot.
\newblock High precision analytical description of the allowed
  $\ensuremath{\beta}$ spectrum shape.
\newblock {\em Rev. Mod. Phys.}, 90:015008, Mar 2018.

\bibitem{HAYEN2019152}
L.~Hayen and N.~Severijns.
\newblock Beta spectrum generator: High precision allowed $\beta$ spectrum
  shapes.
\newblock {\em Computer Physics Communications}, 240:152 -- 164, 2019.

\bibitem{Jackson1957}
J.D. Jackson, S.B. Treiman, and H.W. Wyld.
\newblock Coulomb corrections in allowed beta transitions.
\newblock {\em Nuclear Physics}, 4:206--212, aug 1957.

\bibitem{glickmagid2021nuclear}
Ayala Glick-Magid, Christian Forssén, Daniel Gazda, Doron Gazit, Peter
  Gysbers, and Petr Navrátil.
\newblock Nuclear ab initio calculations of 6he $\beta$-decay for beyond the
  standard model studies.
\newblock {\em Physics Letters B}, 832:137259, 2022.

\bibitem{PhysRevC.94.035503}
M.~Gonz\'alez-Alonso and O.~Naviliat-Cuncic.
\newblock Kinematic sensitivity to the fierz term of $\ensuremath{\beta}$-decay
  differential spectra.
\newblock {\em Phys. Rev. C}, 94:035503, Sep 2016.

\bibitem{Mishnayot-23Ne}
Yonatan Mishnayot, Ayala Glick-Magid, Hitesh Rahangdale, Guy Ron, Doron Gazit,
  Jason~T. Harke, Ben Ohayon, Aaron Gallant, Nicholas~D. Scielzo, Sergey
  Vaintraub, Tsviki Hirsch, Christian Forss\'en, Daniel Gazda, Peter Gysbers,
  Javier Men\'endez, Petr Navratil, Leonid Weissman, Arik Kreisel, Boaz Kaizer,
  Hodaya Dafna, and Maayan Buzaglo.
\newblock Constraining new physics with a new measurement of the
  $^{23}\text{Ne}$ branching ratio.
\newblock {\em arXiv:2107.14355}, 2021.

\bibitem{GLICKMAGID2017285}
Ayala Glick-Magid, Yonatan Mishnayot, Ish Mukul, Michael Hass, Sergey
  Vaintraub, Guy Ron, and Doron Gazit.
\newblock Beta spectrum of unique first-forbidden decays as a novel test for
  fundamental symmetries.
\newblock {\em Phys. Lett. B}, 767:285 -- 288, 2017.

\bibitem{refId0}
{Israel Mardor}, {Ofer Aviv}, {Marilena Avrigeanu}, {Dan Berkovits}, {Adi
  Dahan}, {Timo Dickel}, {Ilan Eliyahu}, {Moshe Gai}, {Inbal Gavish-Segev},
  {Shlomi Halfon}, {Michael Hass}, {Tsviki Hirsh}, {Boaz Kaiser}, {Daniel
  Kijel}, {Arik Kreisel}, {Yonatan Mishnayot}, {Ish Mukul}, {Ben Ohayon},
  {Michael Paul}, {Amichay Perry}, {Hitesh Rahangdale}, {Jacob Rodnizki}, {Guy
  Ron}, {Revital Sasson-Zukran}, {Asher Shor}, {Ido Silverman}, {Moshe
  Tessler}, {Sergey Vaintraub}, and {Leo Weissman}.
\newblock The soreq applied research accelerator facility (saraf): Overview,
  research programs and future plans.
\newblock {\em Eur. Phys. J. A}, 54(5):91, 2018.

\bibitem{Itzykson:1980:QTF}
Claude Itzykson and Jean~Bernard Zuber.
\newblock {\em Quantum Field Theory}.
\newblock International Series in Pure and Applied Physics. McGraw-Hill
  International Book Co., New York, 1980.
\newblock Reprinted by Dover, 2006.

\bibitem{Mendenhall2013}
M.~P. Mendenhall, R.~W. Pattie, Y.~Bagdasarova, D.~B. Berguno, L.~J. Broussard,
  R.~Carr, S.~Currie, X.~Ding, B.~W. Filippone, A.~Garc\'{\i}a, P.~Geltenbort,
  K.~P. Hickerson, J.~Hoagland, A.~T. Holley, R.~Hong, T.~M. Ito, A.~Knecht,
  C.-Y. Liu, J.~L. Liu, M.~Makela, R.~R. Mammei, J.~W. Martin, D.~Melconian,
  S.~D. Moore, C.~L. Morris, A.~P\'erez~Galv\'an, R.~Picker, M.~L. Pitt,
  B.~Plaster, J.~C. Ramsey, R.~Rios, A.~Saunders, S.~J. Seestrom, E.~I.
  Sharapov, W.~E. Sondheim, E.~Tatar, R.~B. Vogelaar, B.~VornDick, C.~Wrede,
  A.~R. Young, and B.~A. Zeck.
\newblock Precision measurement of the neutron $\ensuremath{\beta}$-decay
  asymmetry.
\newblock {\em Phys. Rev. C}, 87:032501, Mar 2013.

\bibitem{Mund2013}
D.~Mund, B.~M\"arkisch, M.~Deissenroth, J.~Krempel, M.~Schumann, H.~Abele,
  A.~Petoukhov, and T.~Soldner.
\newblock Determination of the weak axial vector coupling
  $\ensuremath{\lambda}\mathbf{=}{g}_{A}/{g}_{V}$ from a measurement of the
  $\ensuremath{\beta}$-asymmetry parameter $a$ in neutron beta decay.
\newblock {\em Phys. Rev. Lett.}, 110:172502, Apr 2013.

\end{thebibliography}


\end{document}